\documentclass[aps,twocolumn,nofootinbib]{revtex4}
\usepackage{graphicx}
\usepackage{amsmath}
\usepackage{amssymb}
\usepackage{mathtools}
\usepackage{hyperref}
\usepackage{bm}
\usepackage[T1]{fontenc}
\usepackage{upquote}
\usepackage{xcolor}
\usepackage{soul}

\DeclareMathOperator{\Ima}{im}

\newcommand{\rn}{{\rm n}}

\begin{document} 

\title{\bf Flows, scaling, and the control of moment hierarchies for
stochastic chemical reaction networks}

\author{Eric Smith}

\affiliation{Earth-Life Science Institute, Tokyo Institute of
Technology, 2-12-1-IE-1 Ookayama, Meguro-ku, Tokyo 152-8550, Japan}

\affiliation{Department of Biology, Georgia Institute of
Technology, 310 Ferst Drive NW, Atlanta, GA 30332, USA}

\affiliation{Santa Fe Institute, 1399 Hyde Park Road, Santa Fe, NM
87501, USA}

\affiliation{Ronin Institute, 127 Haddon Place, Montclair, NJ 07043,
USA} 

\author{Supriya Krishnamurthy}

\affiliation{Department of Physics, Stockholm University, SE- 106 91,
Stockholm, Sweden} 

\date{\today}
\begin{abstract}

  Stochastic chemical reaction networks (CRNs) are complex systems
  which combine the features of concurrent transformation of multiple
  variables in each elementary reaction event, and nonlinear relations
  between states and their rates of change.  Most general results
  concerning CRNs are limited to restricted cases where a topological
  characteristic known as \textit{deficiency} takes value 0 or 1,
  implying uniqueness and positivity of steady states and surprising,
  low-information forms for their associated probability
  distributions.  Here we derive equations of motion for fluctuation
  moments at all orders for stochastic CRNs at general deficiency.  We
  show, for the standard base-case of proportional sampling without
  replacement (which underlies the mass-action rate law), that the
  generator of the stochastic process acts on the hierarchy of
  factorial moments with a finite representation.  Whereas simulation
  of high-order moments for many-particle systems is costly, this
  representation reduces solution of moment hierarchies to a
  complexity comparable to solving a heat equation.  At steady states,
  moment hierarchies for finite CRNs interpolate between low-order and
  high-order scaling regimes, which may be approximated separately by
  distributions similar to those for deficiency-0 networks, and
  connected through matched asymptotic expansions.  In CRNs with
  multiple stable or metastable steady states, boundedness of
  high-order moments provides the starting condition for recursive
  solution downward to low-order moments, reversing the order usually
  used to solve moment hierarchies.  A basis for a subset of network
  flows defined by having the same mean-regressing property as the
  flows in deficiency-0 networks gives the leading contribution to
  low-order moments in CRNs at general deficiency, in a
  $1/n$-expansion in large particle numbers.  Our results give a
  physical picture of the different informational roles of
  mean-regressing and non-mean-regressing flows, and clarify the
  dynamical meaning of deficiency not only for first-moment conditions
  but for all orders in fluctuations.

\end{abstract}

\maketitle

\section{Introduction}

Random walks of multiple independent particles on ordinary graphs are
simple processes in several fundamental
senses~\cite{Lovasz:survey:93,Luo:spectral_embed:03,Lee:spectral:10}.
Each step involves a change of a single degree of freedom, and for the
base case of proportional sampling without replacement, the rate law
for fluxes is linear in concentrations.  Moreover, identification of
topological properties of ordinary graphs which determine
characteristics of random-walk dynamics can generally be carried out
in polynomial time~\cite{Lovasz:survey:93,Hordijk:autocat_alg:04}.

Just the opposite is true of the stochastic processes associated with
Chemical Reaction Networks (CRNs).  Each elementary reaction event
(for a general network) can involve the concurrent conversion of
multiple inputs into multiple outputs
~\cite{Danos:rule_based_modeling:08,Harmer:info_carriers:10}, making
the elementary network on which the reactions occur not an ordinary
graph, but a \textit{directed multi-hypergraph}
~\cite{Andersen:comp_rules:13,Andersen:generic_strat:14}.  For the
base case of proportional sampling without replacement over reactants,
the rate law becomes nonlinear, leading generically to possibilities
for complex dynamics and multiple (stable or metastable) steady
states~\cite{Feinberg:notes:79,Gunawardena:CRN_for_bio:03,Baez:QTSM:17}.
Moreover, identification of key topological properties such as
shortest reaction sequences connecting inputs to outputs or complete
sets of self-amplifying cycles, which affect the character of CRN
dynamics, are known to be NP-hard problems for
hypergraphs~\cite{Berge:hypergraphs:73, Andersen:NP_autocat:12}.

As a consequence, although stochastic processes with the essential
features of CRNs are ubiquitous in biochemistry~\cite{Metzler:BC:03},
systems biology~\cite{Palsson:systems_bio:06},
ecology~\cite{Smith:pop:11}, and
epidemiology~\cite{Allen:stoch_process:03}, and are thus of large
practical and theoretical interest, few results exist for these
systems~\cite{Feinberg:def_01:87,Craciun:multistability:05,%
  Craciun:bistability:06,Ji:PhD:11,Joshi:Survey:15} compared with the
large literature that exists for random walks on ordinary
graphs~\cite{Lovasz:survey:93}.  In addition to systems which clearly
have CRN structure because the underlying processes obey constraints
of stoichiometry (the source of concurrency), the CRN framework is
flexible enough to furnish a representation for systems of broad
interest to non-equilibrium thermodynamics such as the zero-range
process~\cite{Evans:zero_range:05}, where topological characteristics
that are known to lead to simple steady states for CRNs can be used to
sieve for exactly solvable cases.  Indeed, some of the key results
that are known for stochastic CRNs~\cite{Anderson:product_dist:10}
were motivated in part by earlier work of Kelly on a related class of
queuing networks~\cite{Kelly:rev_stoc_networks:79}.

\subsubsection*{Partitioning off the simple sub-architecture in the
core of complex CRNs}  

A representation scheme for CRNs was made standard by the work of
Feinberg~\cite{Feinberg:notes:79,Feinberg:def_01:87}, which separates
the structure of independent reaction events from the stoichiometric
relations that determine their action in the chemical state space.  We
will emphasize that, well beyond the use made by its original authors,
this decomposition defines the fundamental partition in CRN
architecture between a sub-system with the same simplicity as the
random walk on an ordinary graph, and the remainder of the CRN
constraints responsible for concurrency, nonlinearity, and their
resulting complexity.  Identifying the subsystem isomorphic to a
simple process is the key to decomposing the scaling behaviors in the
moment hierarchies of CRNs, and to distinguishing the roles of
different classes of flows in the dynamics and steady states.

The emphasis in the work of Feinberg, and in closely-related work by
Horn and Jackson~\cite{Horn:mass_action:72}, was the existence and
uniqueness of solutions to the mass-action rate equations with
strictly positive concentrations: a limited, deterministic, and static
problem.  We will show below that the Feinberg decomposition is even
more useful in the analysis of the stochastic processes associated
with CRNs, where in addition to exact results in deterministic limits,
it can serve as a foundation for systematic approximation methods in
the general case.  A host of results follow, including novel
representations of the generator of the stochastic process acting on
the moment hierarchy, duality relations of the kind explored in
Stochastic Thermodynamics~\cite{Seifert:stoch_thermo_rev:12}, and
scaling relations that suggest solution methods using matched
asymptotic expansions.

\subsubsection*{An expanded role for deficiency}

A new dimensional property termed \textit{deficiency} was introduced
by Feinberg~\cite{Feinberg:notes:79}, which was central to both the
results on existence and uniqueness of steady states, and to a limited
but important corollary about the form of some of the distributions
associated with such states.  Deficiency can be computed as a
topological index from the graphical structure associated with a
CRN~\cite{Feinberg:notes:79,Gunawardena:CRN_for_bio:03,Baez:QTSM:17},
but its importance comes from its meaning as a count of the
dimensionality of chemical flows that can proceed in a steady state
without being subject to mean-regression due to changes in chemical
concentrations.  When there are no such regression-free flows -- when
deficiency equals zero -- the network is guaranteed to have unique
strictly positive steady states at general parameters.  A remarkable
result due to Anderson, Craciun, and Kurtz
(ACK)~\cite{Anderson:product_dist:10} is that under the same
conditions, the steady-state distributions have a simple factorial
form under proportional sampling rules, and a class of related forms
under more general rules (\textit{e.~g.}, Michaelis kinetics), as long
as they sample chemical species independently of one another.

The concept expressed by deficiency remains key to organizing the
stochastic processes for general CRNs, even when their deficiency is
nonzero, and the clues for why this should be so are already latent in
the ACK theorem ~\cite{Anderson:product_dist:10}.  For the simple case
of proportional sampling without replacement (which gives rise to
mass-action kinetics, and which we will assume in the remainder of the
article), the ACK solutions are either products of Poisson
distributions, or hypersurfaces within such product-Poissons
constrained by particle conservation laws (with no loss of generality
for the claims below).  In a Poisson distribution, all higher moments
are universal functions of the mean value, so in a product-Poisson,
the entire moment hierarchy is controlled by the set of first-moments.
In a deficiency-zero network at steady state, the ACK theorem
effectively states that first-moment values carry all the
``information'' in the distribution.  When deficiency is nonzero, the
mean-regressing flows are no longer the \emph{exclusive} dynamical
entities, but as we will show they remain the \emph{dominant} entities
governing low-order moments, in an asymptotic expansion where the
small parameter is the order of any moment of the distribution
relative to its mean particle number.  The approach by which we will
construct this result also shows how to extract other scaling regimes
associated with the remaining flows in networks with non-zero
deficiency, and the way these control complementary asymptotic
expansions for high-order moments relative to mean particle numbers.

\subsubsection*{Coupling CRN theory to Doi operator algebras for
stochastic processes}

Our approach to the stochastic processes associated with CRNs grows
out of a set of linear-algebra methods due to Masao
Doi~\cite{Doi:SecQuant:76,Doi:RDQFT:76}, for the treatment of
generating functions for general discrete-state stochastic processes.
The Doi operator algebra provides a starting point for numerous
solution methods,\footnote{One of the best-known of these is the
  coherent-state expansion of generating functionals due to Luca
  Peliti~\cite{Peliti:PIBD:85,Peliti:AAZero:86}.  It is particularly
  useful for semiclassical approximations and other stationary-point
  methods, which we mention but do not pursue in depth here.} but its
simplifying effect is particularly elucidating for CRNs.  In the
classical mass-action theorems of Feinberg, the rate equations for
CRNs take an awkward and not-very-perspicuous form, in which
stoichiometry is expressed asymmetrically in nonlinear rate laws and
concurrency constraints on inputs and outputs.  In the Doi
representation of the full stochastic process, all formal asymmetry
between inputs and outputs disappears.  It can be seen that the
asymmetry of the Feinberg problem reflects the way the particular
projection operator to the first-moment equations of motion interacts
with the formally-symmetric generator of all fluctuations.  Within the
framework of the Doi algebra, we formulate the projection operator
that gives the equations of motion for arbitrary moments, and show how
the Feinberg equations generalize to a new representation of the
generator of the stochastic process acting on the moment hierarchy,
which directly expresses the scaling influence of different network
flows.  The underlying symmetry of the Doi representation of the
stochastic process generator remains, and can serve as a point of
departure for the derivation of duality relations for stochastic CRNs,
which we mention here but develop in a separate
publication~\cite{Smith:DP_duality:17}.

\subsubsection*{Organization of the presentation}

\noindent The presentation is organized as follows:
\smallskip 

In Sec.~\ref{sec:CRN_SP} we introduce the general concepts and
notation associated first with Chemical Reaction Networks, then with
general discrete-state stochastic processes, and finally for the
particular forms of stochastic processes associated with CRNs.  The
culmination of this section is Eq.~(\ref{eq:L_psi_from_A}), the
Liouville-form expression for the generator of the stochastic process
of a CRN motivated by the Feinberg decomposition, which is the basis
for all other results in the paper. 

Sec.~\ref{sec:dyn_hierarchies} introduces the factorial moments which
are the natural observables for CRNs with simple mass-action rate
laws, and shows that a dynamical equation for all orders of
fluctuations is closed and has a finite-order generator in this set of
moments.  The culmination of this section is the representation of the
generator in Eq.~(\ref{eq:Glauber_moment_fact_prod}).  

Sec.~\ref{sec:top_def_flows} then shows how topological
characteristics of the CRN are linked to dynamical properties of
flows, and reviews the Feinberg deficiency-0 theorem and the
associated Anderson-Craciun-Kurtz theorem.  We introduce what we term
the \textit{stoichiometric decomposition} of the Liouville operator in
Eq.~(\ref{eq:Ls_reps}), which separates two dynamically different
classes of mean-regressing and non-mean-regressing flows.

Sec.~\ref{sec:scaling} uses the representation of the generator on the
lattice of factorial moments to show how different combinations of
rate constants from a CRN govern scaling properties of moments in
different ranges of the moment order.  This section shows how matched
asymptotic expansions can be used to solve for steady-state moment
hierarchies recursively, and shows the (related) sense in which the
deficiency-0-like subset of flows in the stoichiometric decomposition
dominate low-order moments.  Sec.~\ref{sec:worked_examp} then provides
a sequence of worked examples of ascending complexity to introduce
each of the concepts above and show its effects in a solution.

We have chosen to develop all main results in their general forms in
Sections~\ref{sec:CRN_SP} -- \ref{sec:scaling}, in the interest of
economy and continuity of the argument, postponing examples to
Sec.~\ref{sec:worked_examp} where they may be directly compared.  The
simplest case and the starting point involves one species and
deficiency zero~(\ref{sec:g_1_to_2}), illustrating the way the ACK
theorem is recovered in the Doi algebra with a one-line proof.  We
then introduce nonzero deficiency while keeping the same mass-action
equations of motion~(\ref{sec:g_123_conn}), to show how the subspace
of flows resembling a zero-deficiency network can be extracted (but
also why, in the general case, this cannot be expressed as a
zero-deficiency sub-process of the full process), and how it controls
the scaling of low-order moments.  Next we hold deficiency fixed but
change the network topology to one in which autocatalytic feedback
produces multiple steady states in the mass-action
approximation~(\ref{sec:g_0123_unconn}).  This case introduces the
first non-trivial role for the asymptotic expansion and shows,
counter-intuitively, how a criterion of boundedness for asymptotically
\emph{high-order} moments anchors the recursion downward to specify
the low-order moments of the ergodic distribution over the two steady
states.  All effects up to this point are illustrated with
single-species networks.  At the end we introduce a two-species
network in which cross-catalysis replaces the single-species
autocatalysis~(\ref{sec:two_species_bistable}), yielding an equivalent
bistable classical system if the species are not distinguished.  This
case demonstrates a non-trivial use of the stochastic process
generator acting on the moment hierarchy, and shows how the
factorability of the ACK theorem for multi-species distributions is
lost at non-zero deficiency.

Readers who prefer to alternate general notation and instances are
encouraged to browse the examples in Sec.~\ref{sec:worked_examp} in
parallel with reading the formal development in the earlier sections.
A more direct track to the main results on recursions in the moment
hierarchy is also provided in~\cite{Krishnamurthy:CRN_moments:17}.

\section{Chemical Reaction Networks and their associated stochastic
  processes}
\label{sec:CRN_SP}

The next three sub-sections review the standard concepts for CRNs and
two aspects of stochastic-process algebras -- representations of the
generator and then the Doi operator formalism -- and introduce the
notation in which we will represent them in this paper.  In
Sec.~\ref{sec:CRN_SPs} these are then brought together to obtain the
Liouville-operator representation of the generator for a stochastic
process CRN that will be the basis for all further constructions.

\subsection{The elements of a CRN}
\label{sec:CRN_elements}

\begin{figure}[ht]
  \begin{center} 
  \includegraphics[scale=0.6]{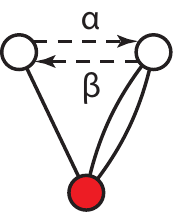}
  \caption{
  Graphical representation for a minimal CRN model and the first
  example in the sequence that will be developed in
  Sec.~\ref{sec:worked_examp}.  One chemical species, two complexes, a
  single linkage class, and deficiency zero.  These terms are defined
  in the remainder of the section.
    \label{fig:g_1_to_2} 
  }
  \end{center}
\end{figure}

Our decomposition of CRNs follows that of
Feinberg~\cite{Feinberg:notes:79}, but we use a more complete graphic
representation, illustrated for our simplest example process in
Fig.~\ref{fig:g_1_to_2}.  We will present CRNs using both this graphic
form, and a corresponding reaction-scheme form such as
\begin{align}
  {\rm A} 
& \xrightleftharpoons[\beta]{\alpha}
  2 {\rm A} .
\label{eq:A_AA_full_scheme}
\end{align}
The following list explains the fundamental division of Feinberg, Horn
and Jackson that separates the so-called \textit{complex
network}\footnote{Here the stress is on the first syllable --
{\tt \textquotesingle com,plex} network -- to be distinguished from references
to systems that are {\tt com\textquotesingle plex}.} from the
stoichiometric relations that interpret its action in the state space
of chemical species:

\begin{trivlist}

\item \textbf{The chemical species:} In examples we will denote
  explicit species names with single capital Roman letters such as
  ${\rm A}$.  Where a set of species is indicated, we index them with
  subscripts $p \in 1 , \ldots , P$, the number of distinct chemical
  species.  In network diagrams a species is denoted with a filled
  dot: \includegraphics[scale=0.65]{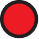}

\item \textbf{Complex:} a multiset of species, which is the input or
  output of a reaction.  It may be written as a sum rather than a set.
  Examples might be ${\rm A}$, $2 {\rm A}$ (or equivalently ${\rm A} +
  {\rm A}$).  In network diagrams, a complex is denoted by an open
  circle with one or more (labeled) dashed line-stubs indicating the
  reaction(s) in which it participates and (labeled) solid line-stubs
  indicating the participating species:
  \includegraphics[scale=0.65]{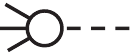} (In one-species
  or one-reaction networks, or after the network diagram has been
  assembled, we may suppress the labels to improve readability.)  In
  set notation, we index complexes with subscripts $i$ or $j$.

\item \textbf{(Directed) reaction:} an ordered pair of complexes with
  an associated rate constant.  For complexes indexed $i$ and $j$,
  respectively, the reaction from $i$ to $j$ would correspond to the
  ordered pair $\left( i , j \right)$, and the associated rate
  constant is denoted $k_{ji}$.  (For simple models, rate constants
  may be given simplifying labels such as $\alpha$.)

  An expression with one or more reaction is known as a
  \textit{reaction scheme}, such as
\begin{equation}
  {\rm A} 
  \overset{\alpha}{\rightharpoonup}
  2 {\rm A} . 
\label{eq:A_AA_half_scheme}
\end{equation}
In network diagrams, a reaction is represented with a dashed arrow
between the input and output complexes (optionally labeled with the
rate constant):
\includegraphics[scale=0.65]{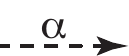} 

\item \textbf{Chemical Reaction Network (CRN):} a collection of
  reactions.  By default we adopt the finest-grained description in
  which all reactions are unidirectional; bi-directional reactions are
  indicated with pairs of directed arrows.  In reaction schemata we
  may also condense notation for bidirectional reactions, as above in
  Scheme~\ref{eq:A_AA_full_scheme}.

  Graphically a CRN is a well-formed doubly-bipartite network (two
  kinds of nodes and two kinds of links), in which all reactions
  terminate in two complexes, and all line-stubs from a complex are
  filled by the appropriate reactions or links to chemical
  species\footnote{Strictly speaking, a CRN corresponds to a
  \textit{directed multi-hypergraph}, in which the reactions
  correspond to directed hyperedges, and the input and output
  complexes are their vertex sets.  The graphic depiction used here
  and elsewhere is a bipartite representation of the underlying
  hypergraph.  The computational complexity of numerous search and
  optimization problems on hypergraphs, which are simple on ordinary
  graphs, is one consequence of the concurrency of inputs and outputs
  on a hyperedge.}, as in Fig.~\ref{fig:g_1_to_2}. 

\item \textbf{Adjacency/rate matrix:} In Feinberg's representation of
  CRNs~\cite{Feinberg:notes:79,Feinberg:def_01:87}, only the complexes
  and the reactions are denoted explicitly, and they form an ordinary,
  directed graph.  In a stochastic formulation, reactions can occur as
  independent events on the links, analogous to the steps in a simple
  random walk.  Complexes are treated as if they have
  \textit{activities}, and the set of rate constants map these
  activities to reaction rates.  In this way, both stoichiometric
  constraints and the determination of activities of complexes are
  cordoned off as separate information from the rate- and
  connectivity-structure of the complex-network.  The latter is given
  by an ordinary \textit{adjacency/rate matrix}, identical in form to
  the \textit{graph Laplacian} for a simple random walk.  Following
  the notation made standard by Feinberg, we denote this matrix by
  $\mathbb{A}$.

  Arranging complexes in a column vector indexed by $i$, let $w_i$ be
  the indicator function that is nonzero on complex $i$ only (so the
  $j$th component ${\left( w_i \right)}_j \equiv {\delta}_{ij}$), and
  $w_i^T$ its transpose which can act as a projection operator.  Then
  the adjacency/rate matrix can be written as a sum of dyadics,
  arranged in a variety of ways.  Two of these we call the
  \textit{reaction representation} and the \textit{complex
  representation}, written
\begin{align}
  \mathbb{A}
& = 
  \sum_{\left( i , j \right)}
  \left( w_j - w_i \right)
  k_{ji}
  w_i^T 
& \mbox{reaction rep.}
\nonumber \\
& = 
  \sum_i 
  w_i 
  \sum_j
  \left( 
    k_{ji} w_j^T - k_{ij} w_i^T
  \right)
& \mbox{complex rep.}
\label{eq:A_two_reps}
\end{align}
In expressions such as the canonical mass-action rate law -- which
will appear as Eq.~(\ref{eq:Glauber_1st_moment}) below, after its
context has been properly introduced -- the row vectors $w^T$ select
the activities determining reaction rates, and the column vectors $w$
identify the net flux into and out of complexes.  The reaction
representation is often more intuitive for constructing generators of
the stochastic process, while the complex representation, in
accumulating all flows into or out of a complex, more directly
reflects the cause of mean-regression that underlies the concept of
deficiency.

\item \textbf{Stoichiometric matrix:} In order to connect complexes to
  species, treat $\rn \equiv \left[ {\rn}_p \right]$ as a column
  vector, and introduce a matrix $Y$ with rows $y_p \equiv {\left[
      y_p^i \right]}^T$.  $y_p^i$ are the \textit{stoichiometric
    coefficients} indicating the number of instances of species $p$ in
  complex $i$.  In graphs such as Fig.~\ref{fig:g_1_to_2}, this is the
  number of solid lines from the species node $p$ to the complex node
  $i$.  The product $Y \mathbb{A}$ connects fluxes at complexes (the
  row index on $\mathbb{A}$) to fluxes at species (the row index on $Y
  \mathbb{A}$).

\item \textbf{Interpretation of complex activities:} Complexes are not
  the same as chemical species, and their (virtual) activities in the
  Feinberg complex-graph must be fixed in terms of the (actual)
  activities of the species.  We will call any such dependence an
  \textit{interpretation} of the complexes.\footnote{It is an
  interesting question, which we leave for other work, whether the
  random walk on the complex network can be represented formally in
  terms of a linear algebra of pseudo-particles, which are formal
  proxies for the products of operators representing real particles of
  the chemical species.}

  To give an interpretation of complex activities in terms of species
  activities that is convenient to use with the adjacency/rate matrix,
  let ${\Psi}_Y \equiv \left[ {\Psi}_Y^i \right]$ be a column vector
  with components given by 
\begin{align}
  {\Psi}_Y^i \! \left( \rn \right) 
& \equiv 
  \prod_p 
  \frac{
    {\rn}_p ! 
  }{
    \left( {\rn}_p - y^i_p \right) ! 
  }
\label{eq:Psi_only_i_def}
\end{align}
${\Psi}_Y$ defines the activity products corresponding to proportional
sampling without replacement on the discrete indices ${\rn}_p$.  We
will return below to the relation between discrete sampling and
mass-action rate laws, after we have introduced notation and concepts
for the stochastic process that governs mass action and all higher
moments.  

\end{trivlist}

\subsection{Representations of a stochastic process}
\label{sec:SP_reps}

A certain standard machinery underlies all discrete-state stochastic
processes, including those associated with CRNs.  Before deriving the
particular forms for the network decomposition of
Sec.~\ref{sec:CRN_elements}, we introduce general notation and
constructions in this section.

The point we wish to emphasize is that, while the stochastic process
for a CRN has a uniquely-defined generator, that generator may have
many representations, depending on whether we solve the stochastic
process for its probability density, or the moment-generating
functional of that density, or the hierarchy of moments evaluated
directly.

When the activities of complexes are defined by proportional sampling
without replacement as in Eq.~(\ref{eq:Psi_only_i_def}) (the simplest
case, corresponding to ideal gases or solutions), the moments that
appear naturally in all rate equations are what we term
\textit{factorial moments}, and it is for these that the equations of
motion take the most compact form.

The essential components in the stochastic-process description are
then the following: 
\begin{trivlist}

\item \textbf{Probability density function and transfer matrix:}
  States of the CRN are indexed by values of the vector $\rn$, and
  reactions are treated as instantaneous changes of
  state.\footnote{This level of coarse-graining in the description of
  reaction events is the standard assumption also in Stochastic
  Thermodynamics~\cite{Seifert:stoch_thermo_rev:12}.}  Our starting
  point in describing the stochastic process is a probability density
  function ${\rho}_{\rn}$ indexed on the values of $\rn$.

  The density $\rho$ evolves on a time coordinate $\tau$ under a
  \textit{master equation}
\begin{align}
  \frac{\partial \rho}{\partial \tau} 
& = 
  {\rm T} \rho 
\nonumber \\ 
\mbox{shorthand for } \quad
  \frac{\partial {\rho}_{\rn}}{\partial \tau} 
& = 
  \sum_{{\rn}^{\prime}}
  {\rm T}_{\rn {\rn}^{\prime}} 
  {\rho}_{{\rn}^{\prime}} . 
\label{eq:ME_genform}
\end{align}
The matrix ${\rm T} \equiv \left[ {\rm T}_{\rn {\rn}^{\prime}}
\right]$ is called the \textit{transfer matrix}, and is one
representation of the generator of the stochastic process.

\item \textbf{Moment-generating function and Liouville operator:} The
  \textit{moment-generating function} is formed from ${\rho}_{\rn}$
  with the introduction of a vector $z \equiv \left[ z_p \right]$ of
  complex coefficients, as the Laplace transform
\begin{equation}
  \phi \! 
  \left( z \right) \equiv 
  \sum_{\rn}
  \left( 
    \prod_p 
    z_p^{{\rn}_p}
  \right) 
  {\rho}_{\rn} . 
\label{eq:gen_fn_multi_arg}
\end{equation}
The generating function evolves under a \textit{Liouville equation} of
the form 
\begin{align}
  \frac{\partial \phi}{\partial \tau} 
& = 
  - \mathcal{L} \phi 
\nonumber \\ 
  \mbox{shorthand for } \quad
  \frac{\partial}{\partial \tau} 
  \phi \! \left( z \right)
& = 
  - \mathcal{L} \! 
  \left( z , \frac{\partial}{\partial z} \right)
  \phi \! \left( z \right) . 
\label{eq:Liouville_eq_multi_arg}
\end{align}
$\mathcal{L}$ is called the \textit{Liouville operator}.  Its
form is defined from the transfer matrix in Eq.~(\ref{eq:ME_genform}),
and it provides an alternative representation of the generator of the
stochastic process, acting on Laplace transforms.

\item \textbf{Expectations and moments and their time-dependence:} The
  expectation of an arbitrary function $O \! \left( \rn \right)$ (for
  \textit{``Observable''}) of the components of $\rn$, in the
  background $\rho$, is denoted
\begin{equation}
  \left< 
    O \! \left( \rn \right)
  \right> \equiv 
  \sum_{\rn}
  O \! \left( \rn \right)
  {\rho}_{\rn}
\label{eq:exp_denote}
\end{equation}
Since $\rho$ may be a continuous-valued quantity whereas $\rn$ is
discrete, we introduce a particular short-hand with math-Italic
font for the first moment
\begin{equation}
  n \equiv 
  \left< \rn \right> = 
  \sum_{\rn} \rn 
  {\rho}_{\rn} ,
\label{eq:mean_notation}
\end{equation}
which may vary continuously if $\rho$ does.  The mass-action rate
equations are expressed entirely in terms of $n$.  Depending on the
form of the underlying distribution $\rho$ and whether higher-order
correlations can be expressed as functions of $n$, the mass-action
equations may be exact or they may involve a (generally-unregulated)
approximation known as the \textit{mean-field approximation}.

$n$ may also be obtained from the generating function as
\begin{equation}
  n = 
  {
    \left. 
      \frac{\partial}{\partial z}
      \log \phi \! \left( z \right) 
    \right|
  }_{z \equiv 1} , 
\label{eq:mean_from_MGF}
\end{equation}
and expectations of more complex observables can be built up
by acting on $\phi$ appropriately with higher-order derivatives in
$z$.  

The time-dependence of $n$, which is the object of classical
first-moment equations or \textit{chemical rate equations}, can be
obtained by acting on either $\rho$ or $\phi$ with its corresponding
generator, by Equations~(\ref{eq:ME_genform})
or~(\ref{eq:Liouville_eq_multi_arg}).

\item \textbf{Factorial moments for CRNs:} In many applications, the
  natural moments for which to study dynamics are either ordinary
  powers ${\rn}^k$, obtained by acting on $\phi$ with higher
  derivatives in $z$, or cumulants~\cite{vanKampen:Stoch_Proc:07},
  obtained by acting on $\log \phi$ with higher derivatives in $z$.
  For CRNs with reaction rate laws corresponding to proportional
  sampling without replacement, however, the simplest dynamical
  relations are obtained for the factorial moments, for which we
  therefore introduce a specific notation.  For a single component
  ${\rn}_p$ and power $k_p$,
\begin{align}
  {\rn}_p^{\underline{k_p}} 
& \equiv 
  \frac{
    {\rn}_p !
  }{
    \left( {\rn}_p - k_p \right) ! 
  }
& ; \; k_p \le {\rn}_p 
\nonumber \\ 
& \equiv 
 0 
& ; \; k_p > {\rn}_p .
\label{eq:factorial_moment_not}
\end{align}
(We adopt the convention in the second line of
Eq.~(\ref{eq:factorial_moment_not}) for $k_p > {\rn}_p$ because it
allows a simplification in sum notations below.)  We will use the form
of Eq.~(\ref{eq:factorial_moment_not}) as a general formula for
truncated factorials, allowing other integer-valued arguments such as
stoichiometric coefficients to take the place of ${\rn}_p$ as the
argument.

\item \textbf{Moment hierarchy and a new representation for the
generator:} For a vector $k \equiv \left[ k_p \right]$ of powers, we
introduce the \textit{factorial moment hierarchy} indexed by $k$, as
the expectation
\begin{equation}
  {\Phi}_k \equiv 
  \left<
    \prod_p
    {\rn}_p^{\underline{k_p}}
  \right> . 
\label{eq:Phi_def}
\end{equation}
We show in Sec.~\ref{sec:dyn_hierarchies} that $\Phi$ evolves in time
as
\begin{align}
  \frac{\partial \Phi}{\partial \tau} 
& = 
  \Lambda \Phi 
\nonumber \\ 
\mbox{shorthand for } \quad
  \frac{\partial {\Phi}_k}{\partial \tau} 
& = 
  \sum_{k^{\prime}}
  {\Lambda}_{k k^{\prime}} 
  {\Phi}_{k^{\prime}} . 
\label{eq:Phi_time_evol}
\end{align}
The matrix $\Lambda \equiv \left[ {\Lambda}_{k k^{\prime}} \right]$
has finitely many nonzero entries determined by the stoichiometric
coefficients, and its form may be derived from the Liouville operator
$\mathcal{L}$.  $\Lambda$ provides yet a third representation of the
generator of the stochastic process, which is particularly well-suited
to the study of CRNs because it exposes different scaling regimes
controlled by combinations of the rate parameters corresponding to
different flows.  We develop the implications of scaling in
Sec.~\ref{sec:scaling}.

\end{trivlist}

\subsection{The Doi operator algebra for a discrete-state stochastic
process} 
\label{sec:Doi_reps}

For most of its purposes as a generating function, it is not necessary
that $\phi \! \left( z \right)$ be an analytic function of a vector
$z$ of complex-valued arguments.  Often only the formal power series
in $z$, and its algebra with the derivative $\partial / \partial z$,
is required.  

The abstraction of the linear algebra of generating functions in terms
of formal raising and lowering operators follows a procedure due to
Masao Doi~\cite{Doi:SecQuant:76,Doi:RDQFT:76}.  We have elaborated the
details of the mapping and its interpretation extensively
elsewhere~\cite{Smith:Signaling:11,Smith:LDP_SEA:11,Smith:evo_games:15},
and here we only summarize the notation, which by now is
standard.\footnote{Indeed, most treatments open directly with the Doi
algebra~\cite{Mattis:RDQFT:98,Cardy:FTNEqSM:99}.  We have used the
two-step introduction by way of conventional analytic generating
functions because it clarifies the meaning of some terms in the Doi
algebra that can be obscure when presented without introduction.}

The Doi algebra denotes the argument-variables and their derivatives
as abstract \textit{raising} and \textit{lowering} operators,
\begin{align}
  z_p 
& \rightarrow a^{\dagger}_p
& 
  \frac{\partial}{\partial z_p}
& \rightarrow 
  a_p , 
\label{eq:a_adag_defs}
\end{align}
because partial differentiation then imposes on these operators the
conventional commutation algebra 
\begin{align}
  \left[ 
    a_p , a^{\dagger}_q 
  \right] = 
  {\delta}_{pq} , 
\label{eq:comm_relns}
\end{align}
where ${\delta}_{pq}$ is the Kronecker $\delta$.  

Generating functions, which are polynomials multiplying the number 1,
are written as the action of the raising operators on a formal
\textit{right-hand null state}, while the projection operator that
takes the trace of a generating function with an integral is written
as a \textit{left-hand null state}:
\begin{align}
  1 
& \rightarrow 
  \left| 0 \right)
& 
  \int d^P \! z \, 
  {\delta}^P \! \left( z \right) 
& \rightarrow 
  \left( 0 \right|
\label{eq:null_states}
\end{align}
where ${\delta}^P \! \left( z \right)$ is the Dirac $\delta$ in $P$
dimensions, and the inner product of the null states is normalized:
$\left( 0 \mid 0 \right) = 1$. 

The basis for generating functions is the set of \textit{number
states} which are elementary monomials.  For any vector $\rn$, 
\begin{align}
  \prod_{p = 1}^P
  z_p^{{\rn}_p} \times 
  1 
& \rightarrow 
  \prod_{p = 1}^P
  {
    a_p^{\dagger} 
  }^{{\rn}_p} 
  \left| 0 \right) \equiv 
  \left| \rn \right) ; 
\label{eq:number_states}
\end{align}
number states are eigenstates of the set of \textit{number operators}
$a^{\dagger}_p a_p$:  
\begin{align}
  a^{\dagger}_p a_p 
  \left| \rn \right) =
  {\rn}_p
  \left| \rn \right) .
\label{eq:num_states_eigs}
\end{align}
In particular for use with CRNs, we note the role of lowering
operators in extracting the truncated factorials of number arguments:
for any non-negative integer $k$,
\begin{align}
  a_p^k  
  \left| \rn \right)
& = 
  {\rn}_p^{\underline{k}}
  \left| \rn - k_p \right) 
\nonumber \\ 
  {a^{\dagger}_p}^k a_p^k 
  \left| \rn \right)
& = 
  {\rn}_p^{\underline{k}}
  \left| \rn \right) .
\label{eq:trunc_fact_extract}
\end{align}
where $\left| \rn - k_p \right)$ is the number state with $k$
subtracted from ${\rn}_p$ and all ${\rn}_q$ for $q \neq p$ unchanged. 

With these steps the generating function becomes a vector in a linear
space: 
\begin{align}
  \phi \! \left( z \right) = 
  \sum_{\rn}
  \prod_{p = 1}^P
  z_p^{{\rn}_p} 
  {\rho}_{\rn}
& \rightarrow 
  \sum_{\rn}
  {\rho}_{\rn}
  \left| \rn \right) \equiv 
  \left| \phi \right) . 
\label{eq:genfun_to_state}
\end{align}
All number states are normalized with respect to the \textit{Glauber
inner product}, defined by 
\begin{align}
  \left( 0 \right|
  e^{\sum_p a_p}
  \left| \rn \right) = 1 , \quad
  \forall \rn , 
\label{eq:Glauber_inn_prod}
\end{align}
and the Glauber inner product with a generating function is simply the
trace of the underlying probability density: 
\begin{align}
  \left( 0 \right|
  e^{\sum_p a_p}
  \left| \phi \right) = 
  \sum_{\rn}
  {\rho}_{\rn} = 1 . 
\label{eq:Glauber_is_trace}
\end{align}
The above conventions define the standard representation in which we
will work with generating functions in the remainder of this article. 

\subsection{The stochastic process for a CRN}
\label{sec:CRN_SPs}

We now apply the above operator formalism to the particular forms of
transfer matrices and generating functions produced by CRNs, and show
how the reaction and complex representations decompose the resulting
representations of the generators.  

Corresponding to the column vector ${\Psi}_Y$ of activities from
Eq.~(\ref{eq:Psi_only_i_def}), we introduce a second column vector
${\psi}_Y \equiv \left[ {\psi}_Y^i \right]$ which takes as its
argument the Doi lowering operators $a_p$ that extract the truncated
factorials of ${\rn}_p$ according to
Eq.~(\ref{eq:trunc_fact_extract}).  Making use of the
notation~(\ref{eq:factorial_moment_not}) for these truncated
factorials, we may write the coefficients in these two vectors as
\begin{align}
  {\Psi}_Y^i \! \left( \rn \right) 
& \equiv 
  \prod_p 
  {\rn}_p^{\underline{y_p^i}} 
\nonumber \\
  {\psi}_Y^i \! \left( a \right) 
& \equiv 
  \prod_p 
  a_p^{y^i_p}
\label{eq:Psi_psi_i_def}
\end{align}

There is a corresponding row vector for the adjoints:
${\psi}_Y^{\dagger} \equiv {\left[ {{\psi}^{\dagger}}_Y^i \right]}^T$
is a row vector of components defined on the complex indices $i$,
where the notation means
\begin{equation}
  {{\psi}^{\dagger}}_Y^i \! 
  \left( a^{\dagger} \right) \equiv 
  \prod_p 
  {a^{\dagger}}_p^{y^i_p}
\label{eq:psi_dag_i_def}
\end{equation}
(As we did with the notation~(\ref{eq:factorial_moment_not}) for
factorial moments, we take the
definitions~(\ref{eq:Psi_psi_i_def},\ref{eq:psi_dag_i_def}) for $\psi$
and ${\psi}^{\dagger}$ as general forms, in which other arguments
besides $a$ and $a^{\dagger}$, such as the first-moment value $n$ can
appear.  This is convenient when expressing the approximation made in
the mass-action rate law, and exhibiting its relation to the exact
equation of motion as an operator expression.)

We now use the above notations to group the sample numbers and index
shifts that describe the concurrent conversion of reactants into
products in the elementary reaction events of a CRN.  We begin with
the transfer matrix, and then show the simplifications afforded by
working with the Liouville operator. 

For proportional sampling without replacement, the number dependence
of the probability for the reaction $\left( i, j \right)$ from state
$\rn$ is simply given by the component ${\Psi}_Y^i \! \left( \rn
\right)$ from Eq.~(\ref{eq:Psi_psi_i_def}).  The way the master
equation acts on indices is slightly more complicated: For the
delivery of probability into state $\rn$, the master equation must
sample both $\rho$ and ${\Psi}_Y^i$ at a value shifted from $\rn$ by
the stoichiometric coefficients that are consumed at complex $i$ minus
those that are produced at complex $j$.  The simplest way to express
such shifts is to let the vector of \textit{shift operators}
$e^{\partial / \partial \rn}$ (with the exponential evaluated
component-wise on $\rn$) serve as an argument to ${\psi}_Y^i$, thus:
\begin{equation}
  {\psi}_Y^i \! 
  \left( 
    e^{\partial / \partial \rn}
  \right) \equiv 
  \prod_p 
  e^{y^i_p \partial / \partial {\rn}_p} = 
  e^{
    {y^i}^T \partial / \partial \rn 
  } ; 
\label{eq:psi_of_shift_def}
\end{equation}
with this convention for both the rate constant and the shift
operator, the matrix ${\rm T}$ from Eq.~(\ref{eq:ME_genform}) can be
written
\begin{align}
  {\rm T} 
& = 
  {\psi}_Y^{\dagger} \! \left( e^{-\partial / \partial n} \right) 
  \mathbb{A}
  \left[ 
    {\psi}_Y \! \left( e^{\partial / \partial n} \right) \cdot 
    {\Psi}_Y \! \left( \rn \right) 
  \right]
\nonumber \\
& = 
  \sum_{\left( i , j \right)}
  \left[
    {\psi}_Y^j \! \left( e^{-\partial / \partial n} \right) - 
    {\psi}_Y^i \! \left( e^{-\partial / \partial n} \right) 
  \right]
  k_{ji}
  {\psi}_Y^i \! \left( e^{\partial / \partial n} \right) 
  {\Psi}_Y^i \! \left( \rn \right) .
\label{eq:T_psi_from_A}
\end{align}
Here the dot-product ($\cdot$) between ${\psi}_Y \! \left( e^{\partial
/ \partial n} \right)$ and ${\Psi}_Y \! \left( \rn \right)$ in the
first line indicates that these two vectors are to be multiplied
component-wise with respect to the complex index $i$, so that their
product is then extracted by the indicator functions $w_i^T$ in
$\mathbb{A}$ from Eq.~(\ref{eq:A_two_reps}).  The cross-term in
${\psi}_Y^j$ and ${\psi}_Y^i {\Psi}_Y^i$, made explicit in the second
line of Eq.~(\ref{eq:T_psi_from_A}), performs the required index shift
on $\rn$ in both $\rho$ and ${\Psi}_Y$ to account for the particles
lost from the system's state through complex $i$ and those gained by
the system's state through complex $j$.  The other cross-term, with
${\psi}_Y^i \!  \left( e^{-\partial / \partial n}
\right)$, simply cancels the shift operators in ${\psi}_Y^i \!  \left(
  e^{\partial / \partial n} \right)$ and represents the loss of
probability from state $\rn$ with rate ${\Psi}_Y^i \! \left( \rn
\right)$.

Working with the Liouville operator from
Eq.~(\ref{eq:Liouville_eq_multi_arg}) is much more straightforward,
because the lowering operators $a_p$ both extract sample numbers and
shift indices according to Eq.~(\ref{eq:trunc_fact_extract}), so these
do not need to be separately tracked as they are in the transfer
matrix.  Using the exact
definitions~(\ref{eq:Psi_psi_i_def},\ref{eq:psi_dag_i_def}) with the
arguments $a$ and $a^{\dagger}$ implicit, $\mathcal{L}$ takes the form
\begin{align}
  - \mathcal{L} 
& = 
  {\psi}^{\dagger}_Y 
  \mathbb{A}
  {\psi}_Y
\label{eq:L_psi_from_A}
\end{align}
\textbf{Eq.~(\ref{eq:L_psi_from_A}) is one of the central equations of
  this paper, and underlies many of the simplifications we present
  here.}
In this expression, all the formal asymmetry of the standard
first-moment rate equations from the CRN literature has disappeared,
and particle consumption and creation are now treated symmetrically.
This is the first of many simplifications gained by working with the
Laplace transform and the Doi operator algebra.


\section{The dynamics of moment hierarchies}
\label{sec:dyn_hierarchies}

From the foregoing constructions we can directly compute the equations
of motion for arbitrary moments of the density ${\rho}_{\rn}$.  These
equations are finitely generated if we work in a basis of factorial
moments, the demonstration of which is the main result of this
section. 

\subsection{Dynamics of factorial moments for a single species}
\label{sec:fact_mom_inn_prod}

For a non-negative integer $k$, the operator that extracts the
truncated factorial ${\rn}_p^{\underline{k}}$ from number-states,
which are the basis for a general state vector $\left| \phi \right)$,
is $a_p^{k}$.  Therefore the expectation of ${\rn}_p^{\underline{k}}$
in the state $\left| \phi \right)$ is given by
\begin{equation}
  \left< 
    {\rn}_p^{\underline{k}}
  \right> = 
  \left( 0 \right|
  e^{\sum_q a_q}
  a_p^{k}
  \left| \phi \right) , 
\label{eq:n_fact_mom_def}
\end{equation}
and from Eq.~(\ref{eq:Liouville_eq_multi_arg}) and the
form~(\ref{eq:L_psi_from_A}) for $\mathcal{L}$, its
time-dependence is given by
\begin{align}
  \frac{\partial}{\partial \tau}
  \left< 
    {\rn}_p^{\underline{k}}
  \right> 
& = 
  \left( 0 \right| 
  e^{\sum_q a_q}
  a_p^k 
  \left( - \mathcal{L} \right)
  \left| \phi \right) 
\nonumber \\ 
& = 
  \left( 0 \right| 
  e^{\sum_q a_q}
  a_p^k 
  {\psi}^{\dagger}_Y \! \left( a^{\dagger} \right) 
  \mathbb{A}  
  {\psi}_Y \! \left( a \right) 
  \left| \phi \right) . 
\label{eq:Glauber_moment_fact}
\end{align}

In order to obtain a recursion relation for the time dependence of
$\left< {\rn}_p^{\underline{k}} \right>$ in terms of the values of
other factorial moments, we must commute the product of lowering
operators $a_p^k$ through all powers of raising operators, which are
gathered in the coefficients of the row-vector ${\psi}^{\dagger}_Y$.
The result of the commutation is a finite series with descending
powers of $a_p$ and $a_p^{\dagger}$.  For positive integers $k$ and
$y$, the evaluation of operator products of powers of raising and
lowering operators is given by\footnote{The proof is by induction.  If
  $k \ge y$, start with an elementary evaluation of $a_p^k
  {a^{\dagger}}_p$ and then induct on $y$.  If $k \le y$, start with
  an elementary evaluation of $a_p {a^{\dagger}}_p^{y}$ and induct on
  $k$.}
\begin{align}
  a_p^k
  {a^{\dagger}}_p^{y} 
& = 
  \sum_{j = 0}^{\min \left( k , y \right)}
  \frac{
    k! y!
  }{
    j! \left( k-j \right) ! \left( y-j \right) ! 
  } \, 
  {a^{\dagger}_p}^{y - j}
  a_p^{k-j} 
\nonumber \\ 
& = 
  \sum_{j = 0}^k
  \left(
    \begin{array}{c}
      k \\
      j 
    \end{array}
  \right)
  y^{\underline{j}} \, 
  {a^{\dagger}_p}^{y - j}
  a_p^{k-j} . 
\label{eq:OP_commute}
\end{align}
The first line emphasizes the symmetric roles of $k$ and $y$ in the
combinatorial coefficient.  In the second line we have used the
definition~(\ref{eq:factorial_moment_not}) applied to
$y^{\underline{j}}$ (rather than ${\rn}^{\underline{j}}$) to simplify
the index of summation in the case that $k > y$.

Now we may expand the evaluation appearing in
Eq.~(\ref{eq:Glauber_moment_fact}), using the
sum~(\ref{eq:OP_commute})
\begin{align}
  \left( 0 \right|
  e^{\sum_q a_q} 
  a_p^k 
  {{\psi}^{\dagger}}_Y^i \! \left( a^{\dagger} \right) 
& = 
  \left( 0 \right|
  \sum_{j = 0}^k
  \left(
    \begin{array}{c}
      k \\ j 
    \end{array}
  \right)
  {
    \left( y_p^i \right) 
  }^{\underline{j}} \,  
  e^{\sum_q a_q}
  a_p^{k - j}
\nonumber \\ 
& = 
  \sum_{j = 0}^k
  \left(
    \begin{array}{c}
      k \\ j 
    \end{array}
  \right)
  {
    \left( y_p^i \right) 
  }^{\underline{j}} \,  
  \left( 0 \right|
  e^{\sum_q a_q}
  a_p^{k - j} , 
\label{eq:np_psii_comm_exp}
\end{align}
as an operator identity acting on general states.  Here we have used
the property of the Doi operator algebra that commutation through the
exponential $e^{\sum_q a_q}$ shifts all $a_p^{\dagger} \rightarrow
\left( a_p^{\dagger} + 1 \right)$, after which all factors of
$a_p^{\dagger}$ annihilate the right ground state $\left( 0 \right|$.
Thus we have eliminated all factors of $a^{\dagger}$, along the way
extracting from ${\psi}_Y^{\dagger}$ the factorial moments ${ \left(
    y_p^i \right) }^{\underline{j}}$ of the stoichiometric
coefficients.

The powers of the lowering operator $a_p^{k - j}$ in
Eq.~(\ref{eq:np_psii_comm_exp}) are the same form as terms already in
${\psi}_Y$, so we can absorb them into ${\psi}_Y$ by shifting the
stoichiometric coefficients in row $p$, which we denote as 
\begin{align}
  \left( 0 \right| 
  e^{\sum_q a_q}
  a_p^{k-j} 
  {\psi}_Y \! \left( a \right) 
  \left| \phi \right) 
& = 
  \left( 0 \right| 
  e^{\sum_q a_q}
  {\psi}_{Y + {\left( k-j \right)}_p} \! \left( a \right) 
  \left| \phi \right) 
\nonumber \\ 
& =  
  \left< 
    {\Psi}_{Y + {\left( k-j \right)}_p} \! \left( \rn \right) 
  \right>
\label{eq:shifted_psi_Psi_exp}
\end{align}
$Y + {\left( k-j \right)}_p$ is the matrix in which the ${}_p^i$
component is $y_p^i + k - j$, $\forall i$, and $y_q^i$ is unchanged
for $q \neq p$.

From these evaluations we can re-express
Eq.~(\ref{eq:Glauber_moment_fact}) as 
\begin{align}
  \frac{\partial}{\partial \tau}
  \left< 
    {\rn}_p^{\underline{k}}
  \right> 
& = 
  \sum_{j = 0}^k
  \left(
    \begin{array}{c}
      k \\ j 
    \end{array}
  \right)
  Y_p^{\underline{j}} \,  
  \mathbb{A}  
  \left< 
    {\Psi}_{Y + {\left( k-j \right)}_p} \! \left( \rn \right) 
  \right> 
\nonumber \\ 
& = 
  \sum_{j = 1}^k
  \left(
    \begin{array}{c}
      k \\ j 
    \end{array}
  \right)
  Y_p^{\underline{j}} \,  
  \mathbb{A}  
  \left< 
    {\Psi}_{Y + {\left( k-j \right)}_p} \! \left( \rn \right) 
  \right> 
\nonumber \\
& \equiv 
  \sum_{j = 1}^k
  \left(
    \begin{array}{c}
      k \\ j 
    \end{array}
  \right)
  Y_p^{\underline{j}} \,  
  e^{
    \left( k - j \right) \, \partial / \partial Y_p
  }
  \mathbb{A}  
  \left< 
    {\Psi}_{Y} \! \left( \rn \right) 
  \right>
\label{eq:Glauber_moment_fact_assum}
\end{align}
Here $Y_p^{\underline{j}}$ is a row vector in which the $i$th
component is the factorial moment ${ \left( y_p^i \right)
}^{\underline{j}}$.  Note that $Y_p^{\underline{0}}$ is the row vector
of 1s, $Y_p^{\underline{1}} = Y_p$, the $p$th row of $Y$, etc.  The
first line of Eq.~(\ref{eq:Glauber_moment_fact_assum}) contains the
full sum over $j$ from Eq.~(\ref{eq:np_psii_comm_exp}), and the second
line uses the fact that $1^T \mathbb{A} \equiv 0$ to eliminate the $j
= 0$ term.  In the third line, we have expressed the shift of
coefficients in the $p$th row of $Y$ again using an exponential shift
operator denoted $e^{\partial / \partial Y_p}$, which acts on all
components $y_p^i$.  This will be a convenient notation for working
with moments involving multiple species.

\subsection{The generator of the stochastic process acting on the
  moment hierarchy}
\label{sec:dyn_mom_hierarchy}

For multiple species, we generalize the notation to an integer-valued
vector of powers $k \equiv \left[ k_p \right]$, and arrange summation
indices similarly in vectors $j \equiv \left[ j_p \right]$.  Recalling
the definition~(\ref{eq:Phi_def}) of the factorial moment hierarchy
$\Phi$, we write its time derivative as
\begin{widetext}
\begin{align}
  \frac{\partial}{\partial \tau} {\Phi}_k \equiv 
  \frac{\partial}{\partial \tau}
  \left< 
    \prod_p 
    {\rn}_p^{\underline{k_p}}
  \right> 
& = 
  \dot{\prod_p}
  \left[ 
    \sum_{j_p = 0}^{k_p}
    \left(
      \begin{array}{c}
        k_p \\ j_p 
      \end{array}
    \right)
    Y_p^{\underline{j_p}} \,  
    e^{
      {\left( k - j \right)}_p \, \partial / \partial Y_p
    }
  \right]
  \mathbb{A}  
  \left< 
    {\Psi}_{Y} \! \left( \rn \right) 
  \right>
\nonumber \\ 
  \mbox{expands to } 
& = 
  \sum_{j_1 = 0}^{k_1}
  \left(
    \begin{array}{c}
      k_1 \\ j_1 
    \end{array}
  \right) \ldots 
  \sum_{j_P = 0}^{k_P}
  \left(
    \begin{array}{c}
      k_P \\ j_P 
    \end{array}
  \right)  
  \left[ 
    \dot{\prod_p}
    Y_p^{\underline{j_p}} \,  
  \right]
  \mathbb{A}  
  \left< 
    {\Psi}_{Y + \left( k - j \right)} \! \left( \rn \right) 
  \right> 
\nonumber \\ 
& \equiv 
  \sum_{k^{\prime}}
  {\Lambda}_{k k^{\prime}} 
  {\Phi}_{k^{\prime}} . 
\label{eq:Glauber_moment_fact_prod}
\end{align}
\end{widetext}
The notation ${\dot{\prod}}_p$ denotes a product over species $p$
within each index $i$ of the row vectors $Y_p^{\underline{j_p}}$.
(Note that in the sums over $j_p$, we must now retain the $j_p = 0$
entries, because even if one index $Y_p^{\underline{j_p}} = {\left[ 1
\right]}^T$, there may be others in the sum where $j_{p^{\prime}} \neq
0$, and the product $\left( {\dot{\prod}}_p Y_p^{\underline{j_p}}
\right) \mathbb{A}$ is only assured to vanish when all $j_p = 0$.)
Now the shift operator $e^{ {\left( k - j \right)}_p \, \partial /
\partial Y_p }$ in the first line of
Eq.~(\ref{eq:Glauber_moment_fact_prod}) offsets all coefficients
$y_p^i$ in the row $p$ by ${\left( k - j \right)}_p$, so we write the
matrix $Y$ in the second line with its rows shifted uniformly by the
entries in the column vector $\left( k - j \right)$.

The third line of Eq.~(\ref{eq:Glauber_moment_fact_prod}) gives the
definition of the matrix $\Lambda$ introduced in
Eq.~(\ref{eq:Phi_time_evol}).  Each entry in the matrix ${\Psi}_{Y +
  \left( k - j \right)} \! \left( \rn \right)$ is itself a truncated
factorial, so the expression is closed on $\Phi$.  Although the orders
$k_p$ may be arbitrarily large, the matrix $\Lambda$ has only
finitely-many nonzero entries, limited by the largest values of $j_p$
for which the rows $Y_p^{\underline{j_p}}$ have non-vanishing entries
(recalling the definition~(\ref{eq:factorial_moment_not}) for
truncated factorials).

\textbf{Eq.~(\ref{eq:Glauber_moment_fact_prod}) is the main result
  with which we will work in this paper.}  It contains the rate
equations for the species numbers $n$ studied by Feinberg, Horn and
Jackson, and extends these to give a compact representation for the
dynamics of all higher-order moments as well.  In
Sec.~\ref{sec:worked_examp} we illustrate graphic methods for
representing $\Lambda$.

\section{Network topology, deficiency, and the classification and role
of different network flows}
\label{sec:top_def_flows}

The results up to this point are true for a general CRN.  Thus they
say nothing directly about the topological properties that may afford
simplifications such as uniqueness and positivity of steady states or
factorability of distributions.  In this section we shift to a
consideration of topology and its implications, including the concept
of deficiency and the Feinberg~\cite{Feinberg:def_01:87} and
Anderson-Craciun-Kurtz (ACK) theorems~\cite{Anderson:product_dist:10}

\subsection{Deficiency, and a basis to decompose network flows}

The connectivity of the adjacency matrix $\mathbb{A}$, together with
the stoichiometric matrix $Y$, determines the linear subspace of
$\rn$-values that can be accessed through any flow on the network,
called the \textit{stoichiometric subspace}.  The number and character
of steady states depends on whether the $\rn$-dependence of the
mass-action rate equations within this subspace admits a Lyapunov
function~\cite{Horn:mass_action:72,Baez:QTSM:17}.  That, in turn,
depends on whether all flows that transport net matter into or out of
any complex must also transport some net matter into or out of some
chemical species, thus increasing its chemical potential in a
direction that opposes the flow.  If so, then all the flows are
mean-regressing and strictly positive steady states are unique.  If
not, there are net fluxes at the complexes that do not lead to net
fluxes of species, and for these there is no force leading to
mean-regression.  In the latter case, multiple steady states or
non-strictly positive steady states\footnote{These would be boundary
  solutions where some concentrations equal zero.} cannot be ruled
out~\cite{Craciun:multistability:05,Craciun:bistability:06}.

The character of the dynamical steady states therefore depends on the
relative dimension of $\ker \mathbb{A}$ (the flows that absorb or emit
no material at the complexes) and $\ker Y \mathbb{A}$ (the flows that
absorb or emit no material at the chemical species), within the
stoichiometric subspace.  A sketch of the demonstration that this is a
topological characteristic, following
Feinberg~\cite{Feinberg:notes:79,Feinberg:def_01:87}, Horn and
Jackson~\cite{Horn:mass_action:72}, follows:

For any CRN, the stoichiometric subspace corresponds to 
\begin{align}
  S \equiv \Ima \left( Y \mathbb{A} \right)
\label{eq:S_def}
\end{align}
The dimension of $S$, and of the subspace of flows through complexes
that do not produce motions within $S$, are denoted
\begin{align}
  s 
& \equiv 
  \dim \left( S \right)
\nonumber \\
  \delta
& \equiv 
  \dim 
  \left( 
    \ker \left( Y \right) \cap \Ima \left( \mathbb{A} \right) 
  \right) . 
\label{eq:s_delt_defs}
\end{align}
$\delta$ is the \textit{deficiency} of the CRN.  

The following relations hold (as identities) among dimensions in $Y
\mathbb{A}$ and $\mathbb{A}$:
\begin{align}
  \dim \left( \Ima \left( \mathbb{A} \right) \right) 
& = 
  \dim \left( \Ima \left( Y \mathbb{A} \right) \right) + 
  \dim 
  \left( 
    \ker Y \cap \Ima \left( \mathbb{A} \right) 
  \right) 
\nonumber \\ 
& = 
  s + \delta . 
\label{eq:Im_A_dim}
\end{align}
If $C$ is the total number of complexes, then it follows that 
\begin{align}
  C 
& = 
  \dim \left( \Ima \left( \mathbb{A} \right) \right) + 
  \dim \left( \ker \left( \mathbb{A} \right) \right)
\nonumber \\ 
& = 
  s + \delta + 
  \dim \left( \ker \left( \mathbb{A} \right) \right) 
\label{eq:A_ident}
\end{align}

The expression for $\dim \left( \ker \left( \mathbb{A}
\right) \right)$ is simple, and follows from the fact that
$\mathbb{A}$ functions as an ordinary graph Laplacian on each
connected component of the complexes.  The argument involves the
following observations:
\begin{trivlist}

\item \textbf{Weak reversibility and linkage classes:} Connected
  components in the simple graph that includes only complexes and
  reactions are termed \textit{linkage classes} in the CRN literature.
  \textit{Weak reversibility} is the condition that any node in a
  linkage class can be reached from any other by some sequence of
  reactions.  The subset of complexes in a linkage class, which can be
  reached starting from any complex, and which subsequently are never
  exited, is called a \textit{strong terminal linkage class}.  Weak
  reversibility of the whole CRN ensures that each linkage class is a
  strong terminal linkage class, and (as in
  Gunawardena~\cite{Gunawardena:CRN_for_bio:03}) we will limit to this
  case for simplicity.\footnote{The more general case differs only by
  superficial book-keeping to exclude complexes that are exited
  permanently.}

\item \textbf{The counting rule for deficiency:} Weakly reversible
  processes are ergodic on each linkage class, so by the
  Perron-Frobenius theorem or an equivalent
  argument~\cite{Baez:QTSM:17}, there is one basis vector for $\ker
  \left( \mathbb{A} \right)$ for each linkage class.  Let $l$ denote
  the number of linkage classes.  Then $l = \dim \left( \ker \left(
  \mathbb{A} \right) \right)$, and the Feinberg counting result that
\begin{equation}
  \delta = C - s - l
\label{eq:Feinberg_counting}
\end{equation}
follows from Eq.~(\ref{eq:A_ident}).

\item \textbf{Complex-balanced steady states:} Flows in $\ker \left( A
  \right)$ are termed \textit{complex-balanced}, because they require
  no net transport of flux to or from any complex to the species that
  make it up.  All steady states must (tautologically) be in $\ker
  \left( Y \mathbb{A} \right)$.  If $\delta = 0$ the two spaces have
  the same dimension and thus are the same.  In other words, the
  steady-state condition that there be no sources or sinks at species
  nodes entails the condition that there be no sources or sinks at
  complexes.  In this case a convexity
  argument~\cite{Feinberg:notes:79,Baez:QTSM:17} implies that there is a
  unique steady state in the positive orthant for any value of the
  rate constants.

\item \textbf{Nonzero deficiency:} If $\delta > 0$ there are
  species-balancing flows that are not complex-balancing, and $\dim
  \left( \ker \left( \mathbb{A} \right) \right)$ is larger than the
  number of constraints from $\partial n / \partial \tau = 0$ (which
  is only $s$).  Steady states then generally exist out of the
  subspace of $\ker \left( \mathbb{A} \right)$.

\end{trivlist}

\subsubsection{Using mean-regression as a basis to decompose flows
beyond the ($\delta = 0$)-condition}

We noted above that a basis for $\ker \left( A \right)$ has one vector
on each linkage class.  At the level of individual linkage classes,
these are profiles in $\Psi$ proportional to the maximal eigenvectors
of the Perron-Frobenius theorem for simple diffusion under
$\mathbb{A}$.  To characterize the remainder of the space of
activities on the complexes, we require a basis that decomposes the
pre-image of the stoichiometric subspace from the remaining flows.

\begin{trivlist}

\item Let ${\left\{ e_{\alpha} \right\}}_{\alpha = 1}^s$ be a basis
  for ${\ker \left( Y \mathbb{A} \right)}^{\perp} \subseteq
  {\mathbb{R}}^C$.  Changes in $\Psi$ along these directions lead to
  changes in flows $\partial n / \partial \tau$ within $S$, which we
  term \textit{$s$-flows}.

\item Let ${\left\{ {\tilde{e}}_{\beta} \right\}}_{\beta =
    1}^{\delta}$ be a basis for $\ker \left( Y \mathbb{A} \right) /
  \ker \left( \mathbb{A} \right)$.  Changes in $\Psi$ along these
  directions lead to changes by species-balanced but not
  complex-balanced flows, which do not alter $\partial n / \partial
  \tau$ and which we term \textit{$\delta$-flows}.

\item If follows that jointly $\left\{ {\left\{ e_{\alpha}
      \right\}}_{\alpha = 1}^s, {\left\{ {\tilde{e}}_{\beta}
      \right\}}_{\beta = 1}^{\delta} \right\}$ form a basis for ${\ker
    \left( \mathbb{A} \right)}^{\perp} \subseteq {\mathbb{R}}^C$.

\end{trivlist}

This basis leads to a decomposition of $\mathbb{A}$ and therefore of
$\mathcal{L}$, different from either the reaction or the complex
representations in Eq.~(\ref{eq:A_two_reps}), which we might term the
``stoichiometric'' representation:
\begin{align}
  - \mathcal{L} 
& = 
  {\psi}^{\dagger}_Y 
  \mathbb{A}
  \left\{ 
    \sum_{\alpha = 1}^s
    e_{\alpha} e_{\alpha}^T + 
    \sum_{\beta = 1}^{\delta}
    {\tilde{e}}_{\beta} {\tilde{e}}_{\beta}^T 
  \right\} 
  {\psi}_Y
& \mbox{stoich.~rep.}
\label{eq:Ls_reps}
\end{align}
The $s$-flows and $\delta$-flows distinguished in the stoichiometric
representation turn out to make dimensionally different contributions
to the moment equations~(\ref{eq:Glauber_moment_fact_prod}).  The
original Feinberg result, expressed in this language as saying that
the $s$-flows completely determine the first-moment equations of
motion -- and that these are the \emph{only} such flows for CRNs of
deficiency zero -- extends to a claim that the $s$-flows
\emph{dominate} the behavior of moments at all orders lower than the
mean particle numbers, for general CRNs where they are not the only
flows.  We explore the consequences of these scaling dimensions in
Sec.~\ref{sec:scaling}.

\subsection{The Feinberg deficiency-0 theorem and the
  Anderson-Craciun-Kurtz theorem}
\label{sec:def0_and_ACK}

Feinberg's Deficiency-0 theorem was originally
framed~\cite{Feinberg:notes:79} as a result in the convex analysis of
the mass-action rate law, which is generally assumed to be a
mean-field approximation.  The Anderson-Craciun-Kurtz
theorem~\cite{Anderson:product_dist:10} uses the Feinberg existence
proof for solutions to the mass-action rate law, but in proving that
the underlying distributions have product-Poisson form,\footnote{under
the sampling model assumed throughout this paper} it actually
strengthens Feinberg's original theorem: because factorial moments of
Poisson distributions have exactly the relation to the first moment
assumed in mean-field formulations, the Feinberg solution to the
mass-action rate equations need not be framed as a result in
(unregulated) mean-field approximation, but can rather be seen as an
exact result.

The Doi operator algebra, together with the stoichiometric
decomposition in Eq.~(\ref{eq:Ls_reps}), provides an elegant way to
both prove the ACK theorem and see that it makes the Feinberg theorem
exact when $\delta = 0$, and also see why and how these results no
longer hold when $\delta > 0$.

We begin with the set of cases $\sum_p k_p = 1$ of
Eq.~(\ref{eq:Glauber_moment_fact_assum}), in which the sum on $j$
contains the single term $j = k$; these are the equations of motion for
the set of first moments.  For species $p$:\footnote{Note how, from
the symmetric form of $\mathcal{L}$ in Eq.~(\ref{eq:L_psi_from_A}),
the asymmetry of the usual rate equations has resulted from the
projection onto the dynamics of a particular moment.}
\begin{align}
  \frac{\partial}{\partial \tau}
  \left< 
    {\rn}_p
  \right> 
& = 
  Y_p \,  
  \mathbb{A}  
  \left( 0 \right| 
  e^{\sum_q a_q}
  {\psi}_Y \! \left( a \right) 
  \left| \phi \right) 
\nonumber \\ 
& = 
  Y_p \,  
  \mathbb{A}  
  \left< 
    {\Psi}_Y \! \left( \rn \right) 
  \right> 
\nonumber \\
& = 
  Y_p  
  \mathbb{A} 
  \sum_{\alpha = 1}^s
  e_{\alpha} e_{\alpha}^T 
  \left< 
    {\Psi}_{Y} \! \left( \rn \right) 
  \right> . 
\label{eq:Glauber_1st_moment}
\end{align}
So far we work at general deficiency, but because $Y \mathbb{A}
{\tilde{e}}_{\beta} \equiv 0 , \forall \beta$ by construction, only
the basis elements corresponding to $s$-flows are non-zero.

Eq.~(\ref{eq:Glauber_1st_moment}) is exact for this stochastic
process, and it is almost the same as the standard expression for the
mass-action rate law, except that it involves an expectation of the
observables ${\Psi}_Y \! \left( \rn \right)$, which may include
higher-order correlations in $\rn$.  Arbitrarily ignoring these
correlations, and replacing $\left< {\Psi}_{Y} \! \left( \rn \right)
\right>$ with ${\psi}_{Y} \! \left( \left< \rn \right> \right) $,
defines the mean-field approximation.

The one case where the mean-field form is exact is when the state
$\left| \phi \right)$ equals some \textit{coherent state} $\left| \xi
\right)$.  The coherent states are the generating functions of Poisson
distributions, constructed in the Doi algebra as
\begin{align}
  \left| \xi \right) 
& \equiv 
  e^{
    \left( a^{\dagger} - 1 \right) \xi
  } 
  \left| 0 \right) 
\nonumber \\
& \leftrightarrow 
  e^{
    {\left( z - 1 \right)}^T \xi
  } \cdot 1 = 
  e^{- 1^T \xi}
  \sum_{{\rn}_1} \ldots  
  \sum_{{\rn}_P} 
  \prod_p 
  \frac{z_p^{{\rn}_p} {\xi}_p^{{\rn}_p}}{{\rn}_p !} . 
\label{eq:coh_state_def}
\end{align}
$\xi \equiv \left[ {\xi}_p \right]$ is a vector of the mean particle
numbers ${\xi}_p = n_p$ for the $P$ chemical species.

Coherent states are eigenstates of the Doi lowering operator, and thus
\begin{equation}
  {\psi}_Y \! \left( a \right)
  \left| \xi \right) = 
  {\psi}_Y \! \left( \xi \right)
  \left| \xi \right) . 
\label{eq:psi_coherent_eigs}
\end{equation}
giving the mean-field form as an exact result:
\begin{align}
  \left< 
    {\Psi}_{Y} \! \left( \rn \right) 
  \right> 
& = 
  \left( 0 \right| 
  e^{\sum_q a_q}
  {\psi}_Y \! \left( a \right) 
  \left| \xi \right) 
\nonumber \\ 
& = 
  {\psi}_{Y} \! \left( \xi \right) 
\nonumber \\ 
& = 
  {\psi}_{Y} \! \left( \left< \rn \right> \right) . 
\label{eq:s_projectors_k1}
\end{align}
Using the basis $\left\{ e_{\alpha} \right\}$ to handle the counting
of dimensions in the stoichiometric subspace (not all linear
combinations of $n_p$ are necessarily dynamic in a particular CRN),
the condition that $\partial n / \partial \tau = 0$ in
Eq.~(\ref{eq:Glauber_1st_moment}) becomes
\begin{equation}
  e_{\alpha}^T 
  {\psi}_Y \! \left( \xi \right) = 0 \; ; \qquad
  \forall \alpha \in 1 , \ldots , s . 
\label{eq:projectors_cond}
\end{equation}
These are the set of equations proved by Feinberg to have a unique,
strictly-positive solution when $\delta = 0$, for all non-degenerate
values of the rate constants.  Thus, \emph{if} the distribution
$\left| \phi \right)$ is a coherent state, the Feinberg result is an
exact solution.

The coherent state identified by Eq.~(\ref{eq:projectors_cond}) is the
ACK solution \emph{if it is a solution at all}.  However, whereas any
mean-field equation can be solved for some coherent-state parameter,
the corresponding state is only a solution to the whole moment
hierarchy if it also leads to stasis of all higher moments.  To see
why this is assured at $\delta = 0$ and not otherwise, we insert the
stoichiometric decomposition~(\ref{eq:Ls_reps}) into the general
moment hierarchy~(\ref{eq:Glauber_moment_fact_prod}) to obtain:
\begin{widetext}
\begin{align}
  \frac{\partial}{\partial \tau}
  \left< 
    \prod_p 
    {\rn}_p^{\underline{k_p}}
  \right> 
& = 
  \sum_{j_1 = 0}^{k_1}
  \left(
    \begin{array}{c}
      k_1 \\ j_1 
    \end{array}
  \right) \ldots 
  \sum_{j_P = 0}^{k_P}
  \left(
    \begin{array}{c}
      k_P \\ j_P 
    \end{array}
  \right)  
  \left[ 
    \dot{\prod_p}
    Y_p^{\underline{j_p}} \,  
  \right]
  \mathbb{A}  
  \left\{ 
    \sum_{\alpha = 1}^s
    e_{\alpha} e_{\alpha}^T + 
    \sum_{\beta = 1}^{\delta}
    {\tilde{e}}_{\beta} {\tilde{e}}_{\beta}^T 
  \right\} 
  \left< 
    {\Psi}_{Y + \left( k - j \right)} \! \left( \rn \right) 
  \right> 
\label{eq:Glauber_moment_fact_stoich}
\end{align}
\end{widetext}
On a coherent state, the vector of factorial moments $\left< {\Psi}_{Y
    + \left( k - j \right)} \! \left( \rn \right) \right>$ differs
from the value $\left< {\Psi}_Y \! \left( \rn \right) \right>$ for the
first-moment condition only by an overall factor $\prod_p
{\xi}_p^{{\left( k-j \right)}_p}$.  If $\delta = 0$, only $s$-flows
are present in the decomposition, and the set of projections $\left\{
  e_{\alpha}^T \left< {\Psi}_{Y + \left( k - j \right)} \! \left( \rn
    \right) \right> \right\} = \prod_p {\xi}_p^{{\left( k-j
    \right)}_p} \times \left\{ e_{\alpha}^T \left< {\Psi}_Y \! \left(
      \rn \right) \right> \right\}$ vanishes exactly when
Eq.~(\ref{eq:projectors_cond}) holds, for all values of $k$ and $j$.
Although the truncated factorials of stoichiometric coefficients
$\left[ \dot{\prod_p} Y_p^{\underline{j_p}} \, \right]$ generally
differ from $Y$, this changes only the weight of the inner products
with the $e_{\alpha}$ and thus the strength with which each $s$-flow
contributes to the rate equation away from the steady state.
\textbf{Hence the coherent-state solution identified by the
  first-moment condition~(\ref{eq:projectors_cond}) is a steady-state
  solution for the whole moment hierarchy, proving the ACK theorem.}

If $\delta > 0$, then $\delta$-flows also exist in the
sum~(\ref{eq:Glauber_moment_fact_stoich}).  However, whereas $Y
\mathbb{A} {\tilde{e}}_{\beta} \equiv 0 , \forall \beta$, $\left[
  {\dot{\prod}}_p Y_p^{\underline{j_p}} \, \right]$ will not generally
project out $\mathbb{A} {\tilde{e}}_{\beta}$ in the higher-order terms
$\sum_p j_p > 1$, except possibly in special cases of fine-tuning of
the rate parameters.  Therefore, in general, $k$-dependent linear
combinations of ${\tilde{e}}_{\beta}^T \left< {\Psi}_{Y + \left( k - j
    \right)} \!  \left( \rn \right) \right>$ and $e_{\alpha}^T \left<
  {\Psi}_{Y + \left( k - j \right)} \! \left( \rn \right) \right>$
will be required to vanish at a steady state, obviating any simple
Poisson solution.

\section{Scaling regimes, matched asymptotic expansions, and the
controlling role of mean-regressing flows} 
\label{sec:scaling}

The series expansion in factorials of the stoichiometry from
Eq.~(\ref{eq:Glauber_moment_fact_prod}), together with the
interpretation of each term as a projection operator for some
product-Poisson distribution that we used to prove the ACK theorem in
Sec.~\ref{sec:def0_and_ACK}, provides a way to associate different
regions in the lattice of factorial moments with control by different
subsets of the network flows.  Both a set of practical solution
methods for steady states, based on matched asymptotic expansions, and
the concept of approximating a complex distribution locally by a
product-Poisson distribution, follow when we recognize that different
terms in the series expansion are associated with different scaling
behaviors because they capture distinct combinations of the rate
constants from the network.

The next three sub-sections cover the following topics in order:
\begin{trivlist}

\item \textbf{Descaling:} We first introduce the descaling of the
moment hierarchy with coherent-state parameters.  Since, for Poisson
distributions these are the only scale parameters, descaling turns the
moment hierarchy for a coherent state into a vector of
1s.\footnote{This is true as long as the descaling is done with the
coherent state's own $\xi$ values.  More generally the moment
hierarchy becomes a geometric progression.}

\item \textbf{Matched asymptotic expansions:} Generalizing the
Feinberg steady-state condition to all orders, as we did above to
prove the ACK theorem, allows us to use the condition $\Lambda \Phi =
0$ as a recursion relation on $k$ to solve the moment hierarchy, much
like the solution of any Laplace equation, with $\Lambda$ serving as
the Laplacian on the lattice of moments.  When we do this with the
descaled moment hierarchy, it becomes easy to show that recursion
upward in any component of $k$ produces a convergent power-series
expansion on the range $k / \xi \ll 1$ (where $\xi$ stands for
whichever particle number corresponds to the component of $k$ being
incremented), whereas recursion downward in $k$ is convergent for $k /
\xi \gg 1$.  This suggests a general method of solution for moment
hierarchies using matched asymptotic expansions, where the matching
conditions are imposed in the region $k \sim \xi$. 

\item \textbf{$1/n$-expansion about Poisson backgrounds:} The same
small-parameter recursion that controls the asymptotic expansion for
$k / \xi \ll 1$ also shows a sense in which the mean-regressing flows
(the $s$-flows) define a leading-Poisson approximation to low-order
moments for a general CRN.  This is true even when the projection onto
the basis $\left\{ e_{\alpha} \right\}$ does not define a
zero-deficiency sub-network of the original CRN.  We can construct a
linear combination of the coherent-state solutions to the first-moment
steady-state conditions for which the remainder term that must be
added to obtain an exact solution makes a contribution that is
$\mathcal{O} \! \left( k / \xi \right)$ smaller than the contribution
of the Poisson backgrounds for the low-order moments (those with $k /
\xi \ll 1$).  

\end{trivlist}

\subsection{Descaling the moment equations with coherent-state
parameters}

Just as $Y_p$ projects out the ACK product-Poisson for $\delta = 0$
networks, each of the projectors $ \left[ {\dot{\prod}}_p
Y_p^{\underline{j_p}} \, \right]$ in the moment-recursion
equation~(\ref{eq:Glauber_moment_fact_prod}) projects out some
product-Poisson distribution if values for the corresponding
coherent-state parameters can be found.  (If they are not unique, it
can project out more than one such solution.)  We may choose to
reference exact solutions for ${\Phi}_k$ to locally-chosen Poisson
distributions in different regions of $k$ corresponding to different
terms $j \equiv \left[ j_p \right]$, and let the recursion equations
solve for the (smaller) deviations from these reference-Poissons.

For any vector $\xi \equiv \left[ {\xi}_p \right]$ of mean values, we
may descale the activities ${\Psi}_Y^i$ from
Eq.~(\ref{eq:Psi_psi_i_def}) as
\begin{equation}
  {\hat{\Psi}}_Y^i \! \left( \rn \right) \equiv 
  \prod_p 
  \frac{
    {\rn}_p^{\underline{y_p^i}} 
  }{
    {\xi}_p^{y_p^i}
  }
\label{eq:Psi_from_fact_mom_hat}
\end{equation}
If we normalize the ${\Psi}_Y$ vectors in this way, a corresponding
counter-normalization of the adjacency matrix can be defined as
\begin{equation}
  {\hat{\mathbb{A}}}_{ji} \equiv 
  {\mathbb{A}}_{ji} 
  \prod_p 
  {\xi}_p^{y_p^i} .
\label{eq:A_hat_def}
\end{equation}
In cases where we wish to use the stoichiometric
representation~(\ref{eq:Ls_reps}), a similar descaling of the
corresponding projection vectors is
\begin{align}
  {
    \left( {\hat{e}}_{\alpha} \right) 
  }_i 
& \equiv 
  \frac{
    1 
  }{
    \prod_p {\xi}_p^{y_p^i}
  }
  {
    \left( e_{\alpha} \right) 
  }_i 
& 
  {
    \left( {\hat{e}}_{\alpha}^T \right) 
  }_i 
& \equiv 
  {
    \left( e_{\alpha}^T \right) 
  }_i 
  \prod_p {\xi}_p^{y_p^i}
\nonumber \\
  {
    \left( {\hat{\tilde{e}}}_{\beta} \right) 
  }_i 
& \equiv 
  \frac{
    1 
  }{
    \prod_p {\xi}_p^{y_p^i}
  }
  {
    \left( {\tilde{e}}_{\beta} \right) 
  }_i 
& 
  {
    \left( {\hat{\tilde{e}}}_{\beta}^T \right) 
  }_i 
& \equiv 
  {
    \left( {\tilde{e}}_{\beta}^T \right) 
  }_i 
  \prod_p {\xi}_p^{y_p^i} , 
\label{eq:descaled_basis_vecs}
\end{align}
for all $\alpha$ and $\beta$.  All these are chosen so that
\begin{align}
  \mathbb{A} 
  \left< {\Psi}_Y \! \left( \rn \right) \right> 
& \equiv 
  \hat{\mathbb{A}} 
  \left< {\hat{\Psi}}_Y \! \left( \rn \right) \right> 
\nonumber \\ 
& \equiv 
  \hat{\mathbb{A}} 
  \left\{ 
    \sum_{\alpha = 1}^s
    {\hat{e}}_{\alpha} {\hat{e}}_{\alpha}^T + 
    \sum_{\beta = 1}^{\delta}
    {\hat{\tilde{e}}}_{\beta} {\hat{\tilde{e}}}_{\beta}^T 
  \right\} 
  \left< {\hat{\Psi}}_Y \! \left( \rn \right) \right> 
\label{eq:A_Psi_hats_cons}
\end{align}

Applying the scale
transformations~(\ref{eq:Psi_from_fact_mom_hat},\ref{eq:A_hat_def}) to
the equations of motion~(\ref{eq:Glauber_moment_fact_prod}) for the
moment hierarchy $\Phi$ gives the equation for a descaled hierarchy
$\hat{\Phi}$, in which the leading geometric dependence on $\xi$ has
been factored out:
\begin{widetext}
\begin{align}
  \frac{\partial}{\partial \tau} {\hat{\Phi}}_k \equiv 
  \frac{\partial}{\partial \tau}
  \left< 
    \prod_p 
    \frac{
      {\rn}_p^{\underline{k_p}}
    }{
      {\xi}_p^{k_p}
    }
  \right> 
& = 
  \sum_{j_1 = 0}^{k_1}
  \left(
    \begin{array}{c}
      k_1 \\ j_1 
    \end{array}
  \right) \ldots 
  \sum_{j_P = 0}^{k_P}
  \left(
    \begin{array}{c}
      k_P \\ j_P 
    \end{array}
  \right)  
  \left[ 
    \dot{\prod_p}
    \frac{
      Y_p^{\underline{j_p}} 
    }{
      {\xi}_p^{j_p}
    } \,  
  \right]
  \hat{\mathbb{A}}
  \left< 
    {\hat{\Psi}}_{Y + \left( k-j \right)} \! \left( \rn \right) 
  \right>
\nonumber \\ 
& \equiv 
  \sum_{k^{\prime}}
  {\hat{\Lambda}}_{k k^{\prime}} 
  {\hat{\Phi}}_{k^{\prime}} . 
\label{eq:Glauber_moment_fact_hats_alt}
\end{align}
\end{widetext}
Following Feinberg but extending his consideration to all moments, we
try to construct steady states, for which
Eq.~(\ref{eq:Glauber_moment_fact_hats_alt}) can be used as a recursion
relation among the moments of ${\hat{\Phi}}_k$.

\subsection{Matched asymptotic expansions for the steady-state
condition}
\label{sec:match_AE}

The terms that govern the behavior of recursions in the components of
$k$, if Eq.~(\ref{eq:Glauber_moment_fact_hats_alt}) is used to
(exactly or approximately) solve for ${\hat{\Phi}}_k$, are
combinations of the form 
\begin{equation}
  \left(
    \begin{array}{c}
      k \\ j 
    \end{array}
  \right)
  \frac{
    Y^{\underline{j}} 
  }{
    {\xi}^j
  } = 
  \frac{
    k! 
  }{
    \left( k-j \right) !\, 
    {\xi}^j
  }
  \frac{
    Y^{\underline{j}} 
  }{
    j ! 
  } . 
\label{eq:gen_scaling_term}
\end{equation}
(with $k$, $j$, $Y$, and $\xi$ carrying indices for each $p$, which we
suppress to reduce clutter).  The ratios $Y^{\underline{j}} / j!$ are
fixed parameters of $\hat{\Lambda}$ and in any case only finite in
number.  The ratios that govern scaling behavior across the moment
hierarchy are the terms $k! / \left[ \left( k-j \right) !\, {\xi}^j
\right]$.  Because the stoichiometric coefficients and therefore the
limits in the sums over $j$ are finite, it is possible to consider a
range of typical particle number $n \sim \xi \gg \max \left( j
\right)$, in which $k! / \left[
\left( k-j \right) !\, {\xi}^j \right] \sim {\left( k / \xi
\right)}^j$ in the ranges that govern the transition between scaling
regions.  These scale factors govern the stability of asymptotic
expansions as follows: 

For any fixed value of $k$, increasing $j$ in the
sum~(\ref{eq:Glauber_moment_fact_hats_alt}) \emph{lowers} the order of
all moments in $\left< {\hat{\Psi}}_{Y + \left( k-j
\right)} \! \left( \rn \right) \right>$, at the same time multiplying
the corresponding term by a coefficient $\sim {\left( k / \xi
\right)}^j$.  Let $\underline{j}$ be the smallest value at which
${\dot{\prod}}_p \left( Y_p^{\underline{j_p}} / {\xi}_p^{j_p} \right)
\hat{\mathbb{A}}$ does not vanish.  (In general, this occurs when
$\sum_q j_q = 1$, so exactly one of the terms $Y_p^{\underline{j_p}} =
Y_p$ and $Y_q^{\underline{j_q}} = 1$ for all other $q \neq p$.)  

To extend the recursion upward by one order, we must increment $k$
while holding $j$ fixed at $\underline{j}$.  The new moments appearing
at order $k$ are referred to those at the immediately preceding order
in the recursion by higher-order terms $j > \underline{j}$ in the sum
at the current $k$.  The relative magnitude of the preceding terms to
the new terms scales as $\sim {\left( k / \xi \right)}^{j -
  \underline{j}}$.  For $k \ll \xi$, successively higher-order terms
are expressed as sums of lower-order terms with positive powers of
$\left( k / \xi \right)$, consistent with both a non-zero radius of
convergence, and with damping-out of uncertainties in the initial
conditions of the recursion.  (The latter property is important for an
asymptotic expansion to provide a robust solution algorithm.)

For $k \gg \xi$ the opposite is true: the \emph{lower-order} terms
must be solved as functions of the higher-order terms, which are
multiplied by positive powers of $\left( \xi / k \right)$.  Thus in
this range the downward recursion is consistent with a non-zero radius
of convergence, and damps out uncertainties in the starting conditions
assumed at large $k$.

This argument is the basis for a solution in terms of matched
asymptotic expansions, where stable recursions are carried out
starting respectively from $k = 0$ (up-going) and from asymptotically
large $k$ (down-going), and matching conditions are imposed in the
overlap region $k \sim \xi$, which are marginally stable for both
series.  An interesting feature of this solution is that, for CRNs
where the mean-field approximation predicts multiple steady states,
there may still be unique large-$k$ asymptotic behaviors required to
ensure boundedness of moments at all orders.  In such cases, \emph{it
is the downward recursion from large $k$ that anchors the solution to
the moment hierarchy}.  This is a counter-intuitive result given the
conventional mean-field approach to moment closure, which attempts to
anchor all higher-order moments in solutions to the first moments, but
as a consequence cannot obtain the ergodic sum over multiple steady
states which is a property of the exact all-orders solution. 

We do not offer a formal proof that these asymptotic expansions can be
consistently performed for all CRNs and all dimensionalities of the
moment hierarchy, which is an exercise beyond the scope of the current
paper.  However, in one dimension,
the recursion is elementary to define, and for higher-dimensional systems
we offer examples of decompositions of the solution for which
numerical simulation suggests that a similar expansion can be used.

\subsection{Leading Poisson approximations to nonzero-deficiency CRNs}
\label{sec:leading_Poiss_arg}

The above analysis of the scaling of terms in an asymptotic expansion
for solutions to $\hat{\Lambda} \hat{\Phi} = 0$ has an immediate
corollary: in general steady-state solutions, a basis of
product-Poisson distributions associated with $s$-flows dominates the
low-order moments.  The kernel of the argument is that, although the
projection operators associated with both $s$- and $\delta$-flows
share in the same scaling, at the lowest order where the upward-going
recursion begins, the $\delta$-flow contributions are projected out --
recall that their absence from the first-moment conditions is their
defining feature -- therefore only $s$-flow contributions serve as
seeds for the polynomial expansion in $\left( k / \xi \right)$.  We
now demonstrate that relation:

The steady-state condition for
Eq.~(\ref{eq:Glauber_moment_fact_hats_alt}), with the stoichiometric
decomposition inserted from Eq.~(\ref{eq:Ls_reps}), becomes
\begin{widetext}
\begin{align}
  0 
& = 
  \sum_{j_1 = 0}^{k_1}
  \left(
    \begin{array}{c}
      k_1 \\ j_1 
    \end{array}
  \right) \ldots 
  \sum_{j_P = 0}^{k_P}
  \left(
    \begin{array}{c}
      k_P \\ j_P 
    \end{array}
  \right)  
  \left[ 
    \dot{\prod_p}
    \frac{
      Y_p^{\underline{j_p}} 
    }{
      {\xi}_p^{j_p}
    } \,  
  \right]
  \hat{\mathbb{A}}
  \left\{ 
    \sum_{\alpha = 1}^s
    {\hat{e}}_{\alpha} {\hat{e}}_{\alpha}^T + 
    \sum_{\beta = 1}^{\delta}
    {\hat{\tilde{e}}}_{\beta} {\hat{\tilde{e}}}_{\beta}^T 
  \right\} 
  \left< 
    {\hat{\Psi}}_{Y + \left( k-j \right)} \! \left( \rn \right) 
  \right>
\label{eq:Glauber_moment_fact_hats_stoich}
\end{align}
\end{widetext}
The lowest-$k$ conditions that the moment hierarchy must satisfy are the
first-moment conditions, which are the set of terms $\sum_q k_q = 1$
and $j = k$, for which exactly one $Y_p^{\underline{j_p}} = Y_p$ and
$Y_q^{\underline{j_q}} = 1$ for all other $q \neq p$, as noted in the
previous section.

Write the state vector $\left| \phi \right)$ for a general solution to
Eq.~(\ref{eq:Glauber_moment_fact_hats_stoich}) as a sum
\begin{equation}
  \left| \phi \right) = 
  \sum_{\gamma}
  c_{\gamma}
  \left| {\xi}^{\left( \gamma \right)} \right) + 
  \left| {\phi}^{\prime} \right) ,
\label{eq:state_division}
\end{equation}
in which $\left\{ {\xi}^{\left( \gamma \right)} \right\}$ is the set
of all mean-field solutions to the first-moment steady state
conditions, and $c_{\gamma}$ are coefficients to be determined. By
construction $e_{\alpha}^T \sum_{\gamma} c_{\gamma} {\psi}_{Y} \! 
\left( {\xi}^{\left( \gamma \right)} \right) = 0$, $\forall
\alpha$.\footnote{Note that $\left| {\phi}^{\prime} \right)$, as the
generating function for a difference of distributions, will not
generally be derived from any distribution with all positive values.}
Refer to the corresponding expectations as
\begin{align}
  \left( 0 \right| 
  e^{\sum_q a_q}
  {\psi}_Y \! \left( a \right) 
  c_{\gamma}
  \left| {\xi}^{\left( \gamma \right)} \right) 
& \equiv 
  {
    \left< 
      {\Psi}_Y \! \left( \rn \right) 
    \right>
  }^{\left( \gamma \right)}
\nonumber \\
  \left( 0 \right| 
  e^{\sum_q a_q}
  {\psi}_Y \! \left( a \right) 
  \left| {\phi}^{\prime} \right) 
& \equiv 
  {
    \left< 
      {\Psi}_Y \! \left( \rn \right) 
    \right>
  }^{\prime} , 
\label{eq:exp_pert_not}
\end{align}
and likewise for higher-order moments and descaled moments
$\hat{\Psi}$. 

Under the decomposition~(\ref{eq:state_division}), the steady-state
condition~(\ref{eq:Glauber_moment_fact_hats_stoich}) becomes
\begin{widetext}
\begin{align}
& 
  \sum_{j_1 = 0}^{k_1}
  \left(
    \begin{array}{c}
      k_1 \\ j_1 
    \end{array}
  \right) \ldots 
  \sum_{j_P = 0}^{k_P}
  \left(
    \begin{array}{c}
      k_P \\ j_P 
    \end{array}
  \right)  
  \left[ 
    \dot{\prod_p}
    \frac{
      Y_p^{\underline{j_p}} 
    }{
      {\xi}_p^{j_p}
    } \,  
  \right]
  \hat{\mathbb{A}}
  \left\{ 
    \sum_{\alpha = 1}^s
    {\hat{e}}_{\alpha} {\hat{e}}_{\alpha}^T + 
    \sum_{\beta = 1}^{\delta}
    {\hat{\tilde{e}}}_{\beta} {\hat{\tilde{e}}}_{\beta}^T 
  \right\} 
  {
    \left< 
      {\hat{\Psi}}_{Y + \left( k-j \right)} \! \left( \rn \right) 
    \right>
  }^{\prime}
\nonumber \\
& = 
  - \sum_{\gamma}
  \sum_{j_1 = 0}^{k_1}
  \left(
    \begin{array}{c}
      k_1 \\ j_1 
    \end{array}
  \right) \ldots 
  \sum_{j_P = 0}^{k_P}
  \left(
    \begin{array}{c}
      k_P \\ j_P 
    \end{array}
  \right)  
  \left[ 
    \dot{\prod_p}
    \frac{
      Y_p^{\underline{j_p}} 
    }{
      {\xi}_p^{j_p}
    } \,  
  \right]
  \hat{\mathbb{A}}
  \sum_{\beta = 1}^{\delta}
  {\hat{\tilde{e}}}_{\beta} {\hat{\tilde{e}}}_{\beta}^T 
  {
    \left< 
      {\hat{\Psi}}_{Y + \left( k-j \right)} \! \left( \rn \right) 
    \right>
  }^{\left( \gamma \right)} . 
\label{eq:Glauber_moment_fact_hats_primes}
\end{align}
\end{widetext}
If there is a unique steady-state solution, and if the descaled moment
hierarchy $\hat{\Phi}$ is descaled with this $\xi$, then by
construction it will be the case that ${ \left< {\hat{\Psi}}_{Y +
      \left( k-j \right)} \!  \left( \rn \right) \right> }^{\left(
    \gamma \right)} = c_{\gamma} \left[ 1 \right]$ (the vector of all
1s), $\forall k, j$ on the right-hand side of
Eq.~(\ref{eq:Glauber_moment_fact_hats_primes}), which contains a
single term in the sum on $\gamma$.\footnote{Note that, even in the
  case of a unique steady state, we cannot presume that $c_{\gamma} =
  1$ unless $\left< {\Psi}_Y \right>$ includes a term proportional to
  ${\Phi}_0 \equiv \left< 1 \right>$, because the first-moment
  condition does not otherwise fix the normalization of the geometric
  sequence within the total distribution.}  In the more general case,
we can choose the descaling parameters so that ${ \left<
    {\hat{\Psi}}_{Y + \left( k-j \right)} \! \left( \rn \right)
  \right> }^{\left( \gamma \right)} = c_{\gamma} \left[ 1 \right]$ for
a particular $\gamma$ of our choice. (Generally this means descaling
whichever term makes the largest contribution on the right-hand side,
to remove $k$- and $j$- dependence in that term.)  The argument that
the term $\left| {\phi}^{\prime} \right)$ is sub-leading is then made
in two steps:
\begin{enumerate}

\item For some set of coefficients $\left\{ c_{\gamma} \right\}$ we
can ensure that $e_{\alpha}^T {\left< {\Psi}_{Y} \!  \left( \rn
\right) \right>}^{\prime} = 0$.
This is because the set of steady-state solutions for ${\xi}^{\left(
\gamma \right)}$ form a basis for the set of all solutions to the
first-moment steady-state conditions.  In general (even if there is
only one solution for $\xi$), the required coefficients $\left\{
c_{\gamma} \right\}$ may need to be determined by matching conditions
to a large-$k$ asymptotic expansion.\footnote{The mono-stable solution
in Sec.~\ref{sec:g_123_conn} illustrates the need for a non-trivial
normalization in the case of a unique solution, and the bistable
solution in Sec.~\ref{sec:g_0123_unconn} illustrates the case of
solution for a linear combination of Poisson backgrounds.}  (Note that
the values of ${\tilde{e}}_{\beta}^T {\left< {\Psi}_{Y} \!
\left( \rn \right) \right>}^{\prime}$ are unconstrained at order
$\sum_p k_p = 1$, and must be determined as part of the recursion on
$k$.)

\item For $\sum_p k_p > 1$, the leading-order dependence on ${ \left<
{\hat{\Psi}}_{Y + \left( k-j \right)} \! \left( \rn \right) \right>
}^{\left( \gamma \right)}$ on the right-hand side of
Eq.~(\ref{eq:Glauber_moment_fact_hats_primes}) comes when $\sum_p j_p
= 2$, and by Eq.(\ref{eq:gen_scaling_term}) this term is $\mathcal{O}
\! \left[ {\left( k / \xi \right)}^2 { \left< {\hat{\Psi}}_{Y + \left(
k-j \right)} \! \left( \rn \right) \right> }^{\left( \gamma \right)}
\right]$.  The leading term on the left-hand side arises where $\sum_p
j_p = 1$, involves only the $s$-flows, and is $\mathcal{O} \! \left[
\left( k / \xi \right) {\left< {\Psi}_{Y + \left( k-j \right)} \!
\left( \rn \right) \right>}^{\prime} \right]$.  Hence we
conclude that, at the level of counting naive scaling dimensions,
there is a perturbative expansion in small $\left( k / \xi \right)$
about a sum of coherent states, in which ${\left< {\Psi}_{Y + \left(
k-j \right)} \!  \left( \rn \right) \right>}^{\prime} \sim \mathcal{O}
\! \left[ \left( k / \xi \right) { \left< {\hat{\Psi}}_{Y + \left( k-j
\right)} \! \left( \rn \right) \right> }^{\left( \gamma \right)} \right]$.
The reason it is meaningful to make such a scaling comparison, when
the $j$ values used to estimate the powers of scale factors are
different on the left-hand and right-hand sides of
Eq.~(\ref{eq:Glauber_moment_fact_hats_primes}), is that we have been
free to choose the de-scaling parameter for $\hat{\Phi}$ so that for
whichever $\gamma$ gives the largest contribution, ${ \left<
{\hat{\Psi}}_{Y + \left( k-j \right)} \! \left( \rn \right) \right>
}^{\left( \gamma \right)} = c_{\gamma} \left[ 1 \right]$, removing
$k$- and $j$-dependence from the source on the right-hand side of the
equation. 

\end{enumerate}
This completes the argument.

\section{Worked examples}
\label{sec:worked_examp}

We now demonstrate the above results for moment hierarchies in a
cascade of examples.  Each successive example increases the generality
of the problem and introduces a new feature of general solutions of
CRN's.
Explicit constructions involving transfer matrices or Liouville
operators are given in the main text where they first occur, and the
corresponding forms that differ only by elaboration for later examples
are removed to App.~\ref{sec:app_CRN_algebra}.

\subsection{A CRN with 1-species, 2 states and no conserved quantities}
\label{sec:g_1_to_2}

This model is a minimal non-trivial form for a CRN, showing how
uniqueness of positive steady states follows from deficiency-0, and
exhibiting the proof of the ACK theorem in terms of coherent-state
projection operators from Sec.~\ref{sec:def0_and_ACK}.  The CRN is
given by the graph introduced in Fig.~\ref{fig:g_1_to_2} with the
associated reaction scheme~(\ref{eq:A_AA_full_scheme}).  Its
mean-field rate equation is
\begin{align}
  \frac{\partial n}{\partial \tau}
& = 
  \alpha n - 
  \beta n^2 . 
\label{eq:rate_g_12}
\end{align}
The master equation, illustrating the
decomposition~(\ref{eq:T_psi_from_A}) for the transfer matrix, is
\begin{equation}
  \frac{\partial {\rho}_n}{\partial \tau} = 
  \left[
    \left( e^{-\partial / \partial n} -1 \right) 
    \alpha n + 
    \left( e^{\partial / \partial n} -1 \right) 
    \beta n \left( n-1 \right)
  \right]
  {\rho}_n . 
\label{eq:ME_g_1to2}
\end{equation}

For the transfer matrix~(\ref{eq:ME_g_1to2}), the Liouville operator is
\begin{align}
  \mathcal{L} 
& = 
  \left( 1 - a^{\dagger} \right) 
  \left(
    \alpha a^{\dagger} a - 
    \beta a^{\dagger} a^2
  \right)
\nonumber \\ 
& = 
  \left( 1 - a^{\dagger} \right) 
  \left( a^{\dagger} \! a \right) 
  \left(
    \alpha - 
    \beta a 
  \right) . 
\label{eq:L_1_to_2}
\end{align}
The first line is a direct translation of the reaction-representation
from Eq.~(\ref{eq:A_two_reps}) for the conversion of particles in each
unidirectional reaction.  The second line extracts the overall factor
of the projection operator $\left( \alpha - \beta a \right)$ that
vanishes on a coherent state with parameter $\xi = \alpha / \beta$ in
all moment equations, which is the proof of the ACK theorem given in
Sec.~\ref{sec:def0_and_ACK}.

\subsection{A CRN with 1-species, 3 states and
    no conserved quantities
} 
\label{sec:g_123_conn}

\begin{figure}[ht]
  \begin{center} 
  \includegraphics[scale=0.6]{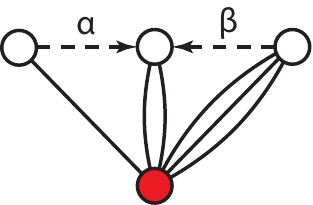}
  \caption{
  An additional state added relative to the model of
  Fig.~\ref{fig:g_1_to_2}.  This model is not weakly reversible.
    \label{fig:g_1_and_3_to_2} 
  }
  \end{center}
\end{figure}

\begin{figure}[ht]
  \begin{center} 
  \includegraphics[scale=0.6]{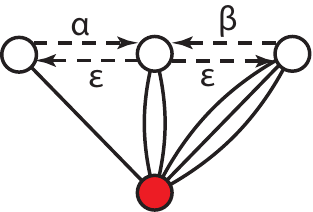}
  \caption{
  A variant one-species model in which the complex-graph is weakly
  reversible.  The steady-state concentrations are the same as those 
  from Fig.~\ref{fig:g_1_and_3_to_2}, which appears as a regular limit
  at $\epsilon \rightarrow 0$.  Because $\delta = 1$, the distribution
  at the steady state is no longer Poisson.
    \label{fig:g_123_conn} 
  }
  \end{center}
\end{figure}

In this section we introduce a one-parameter family of models which
have the same rate equation over the entire family.  Fig.
\ref{fig:g_1_and_3_to_2} depicts a limiting member which lacks weak
reversibility.  Although the limiting case is formally outside the
scope of the assumptions in the rest of the paper, it is useful to
highlight the role of $\delta$-flows in driving solutions away from
the Poisson form associated with the ACK theorem.  Weak reversibility
may be established without changing the rate equation by adding two
reactions to the graph of Fig.~\ref{fig:g_1_and_3_to_2}, to obtain the
family (over the rate parameter $\epsilon$) of graphs shown in
Fig.~\ref{fig:g_123_conn}, which we analyze below.

For this model, we demonstrate the stoichiometric decomposition of the
Liouville operator from Eq.~(\ref{eq:Ls_reps}), and descaling of the
moment recursion equation.  For the parameters we will use in
simulations, a single, downward-going asymptotic expansion (obtained
in \cite{Krishnamurthy:CRN_moments:17}) is sufficient to solve the
entire moment hierarchy to arbitrary precision, starting from an
analytically derived large-$k$ limiting form.  However, we will also
demonstrate the upward/downward matched asymptotic expansion to
illustrate the stability properties of the recursion in small-$k$ and
large-$k$ ranges.  Exact solutions to moment hierarchies of this kind
are only possible for birth-death~\cite{vanKampen:Stoch_Proc:07} type
CRNS with one species, of which this class of models is an example, or
for $\delta=0$ CRN's, whereas the asymptotic expansions have a much
wider applicability, as we demonstrate in later sections.
 

The rate equation for the CRN's in both Fig. ~\ref{fig:g_1_and_3_to_2}
and Fig.~\ref{fig:g_123_conn} is
\begin{align}
  \frac{\partial n}{\partial \tau}
& = 
  \alpha n - 
  \beta n^3 . 
\label{eq:rate_g_123}
\end{align}
which differs from Eq.~(\ref{eq:rate_g_12}) only in changing the
activities that govern particle creation and destruction.

The simplest CRN graph with rate equation~(\ref{eq:rate_g_123}) is
shown in Fig.~\ref{fig:g_1_and_3_to_2}, with associated reaction
scheme
\begin{align}
  {\rm A} 
& \overset{\alpha}{\rightharpoonup} 
  2 {\rm A} & 
  2 {\rm A} 
& \overset{\beta}{\leftharpoondown} 
  3 {\rm A} . 
\label{eq:A_AA_AAA_NR_scheme}
\end{align}
The model introduces competing auto-catalysis at two orders: particle
creation occurs in proportion to the density of existing particles,
while particle destruction occurs in proportion to the cube of the
density.  Because the CRN in Fig.~\ref{fig:g_1_and_3_to_2} has only
three complexes, however, the rate equation has only two non-trivial
roots, and therefore cannot support multiple positive steady states.
(It does, however, have the marginally stable steady state $n \equiv
0$.)

The CRN of Fig.~\ref{fig:g_123_conn} with reaction
scheme
\begin{align}
  {\rm A} 
& \xrightleftharpoons[\epsilon]{\alpha}
  2 {\rm A} & 
  2 {\rm A} 
& \xrightleftharpoons[\beta]{\epsilon}
  3 {\rm A} . 
\label{eq:A_AA_AAA_R_scheme}
\end{align}
These may be checked from Eq.~(\ref{eq:Feinberg_counting}) to have
deficiency $\delta = 1$.

The master equation is provided in Eq.~(\ref{eq:ME_g_123_R}), and the
associated Liouville operator is
\begin{align}
  \mathcal{L} 
& = 
  \left( 1 - a^{\dagger} \right) 
  \left[
    \alpha a^{\dagger} a - 
    \epsilon \left( 1 - a^{\dagger} \right) 
    a^{\dagger} a^2 - 
    \beta {a^{\dagger}}^2 a^3 
  \right]
\nonumber \\ 
& = 
  \left( 1 - a^{\dagger} \right) 
  \left( a^{\dagger} \! a \right) 
  \left[
    \left( 
      \alpha - 
      \epsilon a 
    \right) + 
    \left( a^{\dagger} \! a - 1 \right) 
    \left( 
      \epsilon - 
      \beta a 
    \right) 
  \right] . 
\label{eq:L_123_R}
\end{align}
The limit $\epsilon \rightarrow 0$ is degenerate with the graph of
Fig.~\ref{fig:g_1_and_3_to_2}.  The important feature of this
Liouville operator is that the two projection terms, $\left( \alpha -
\epsilon a \right)$ and $\left( \epsilon - \beta a \right)$, are now
multiplied by distinct non-trivial operators (respectively $1$ and
$\left( a^{\dagger} \! a - 1 \right)$, and cannot both be made to
vanish independently at a single Poisson solution at general values of
$\epsilon$.  This is the way in which deficiency-1 (attained by adding
a complex within a linkage class) moves the CRN outside the scope of
the ACK theorem.

\subsubsection{The stoichiometric decomposition}

We can use the property suggested by the concept of deficiency -- the
categorization of flows as mean-regressing versus non-mean-regressing
-- to further clarify how the non-independence of the projection
terms in the Liouville operator~(\ref{eq:L_123_R}) results in
deviations from Poisson steady-state form.  

The three decompositions of the Liouville operator~(\ref{eq:L_123_R}),
from Eq.~(\ref{eq:A_two_reps}) and Eq.~(\ref{eq:Ls_reps}), are given by
\begin{widetext}
\begin{align}
  \mathcal{L} 
& = 
  \begin{array}{c}
    \left[ 
      \begin{array}{ccc}
        a^{\dagger} & {a^{\dagger}}^2 & {a^{\dagger}}^3  
      \end{array}
    \right] \\ 
    \phantom{} \\ 
    \phantom{}
  \end{array} 
  \left\{ 
    \left[
      \begin{array}{r}
        1 \\
        - 1 \\
        0 
      \end{array}
    \right] 
    \begin{array}{c}
      \left[ 
        \begin{array}{ccc}
          \alpha & -\epsilon & 0  
        \end{array}
      \right] \\ 
      \phantom{} \\ 
      \phantom{}
    \end{array} + 
    \left[
      \begin{array}{r}
        0 \\
        1 \\
        - 1 
      \end{array}
    \right] 
    \begin{array}{c}
      \left[ 
        \begin{array}{ccc}
          0 & \epsilon & - \beta 
        \end{array}
      \right] \\ 
      \phantom{} \\ 
      \phantom{}
    \end{array}
  \right\}
  \left[
    \begin{array}{c}
      a^1 \\
      a^2 \\
      a^3
    \end{array}
  \right] 
\nonumber \\
& = 
  \begin{array}{c}
    \left[ 
      \begin{array}{ccc}
        a^{\dagger} & {a^{\dagger}}^2 & {a^{\dagger}}^3  
      \end{array}
    \right] \\ 
    \phantom{} \\ 
    \phantom{}
  \end{array} 
  \left\{ 
    \left[
      \begin{array}{r}
        1 \\
        0 \\
        0 
      \end{array}
    \right] 
    \begin{array}{c}
      \left[ 
        \begin{array}{ccc}
          \alpha & - \epsilon & 0  
        \end{array}
      \right] \\ 
      \phantom{} \\ 
      \phantom{}
    \end{array} + 
    \left[
      \begin{array}{r}
        0 \\
        1 \\
        0 
      \end{array}
    \right] 
    \begin{array}{c}
      \left[ 
        \begin{array}{ccc}
          - \alpha & 2 \epsilon & - \beta 
        \end{array}
      \right] \\ 
      \phantom{} \\ 
      \phantom{}
    \end{array} + 
    \left[
      \begin{array}{r}
        0 \\
        0 \\
        1 
      \end{array}
    \right] 
    \begin{array}{c}
      \left[ 
        \begin{array}{ccc}
          0 & - \epsilon & \beta 
        \end{array}
      \right] \\ 
      \phantom{} \\ 
      \phantom{}
    \end{array}
  \right\}
  \left[
    \begin{array}{c}
      a^1 \\
      a^2 \\
      a^3
    \end{array}
  \right] 
\nonumber \\
& = 
  \begin{array}{c}
    \left[ 
      \begin{array}{ccc}
        a^{\dagger} & {a^{\dagger}}^2 & {a^{\dagger}}^3  
      \end{array}
    \right] \\ 
    \phantom{} \\ 
    \phantom{}
  \end{array} 
  \frac{1}{{\alpha}^2 + {\beta}^2}
  \left\{ 
    \left[
      \begin{array}{c}
        {\alpha}^2 \\
        - {\alpha}^2 + {\beta}^2 \\
        - {\beta}^2 
      \end{array}
    \right] 
    \begin{array}{c}
      \left[ 
        \begin{array}{ccc}
          \alpha & 0 & -\beta  
        \end{array}
      \right] \\ 
      \phantom{} \\ 
      \phantom{}
    \end{array} + 
    \left[
      \begin{array}{r}
        1 \\
        - 2 \\
        1 
      \end{array}
    \right] 
    \begin{array}{c}
      \left[ 
        \begin{array}{ccc}
          \alpha {\beta}^2 & 
          - \epsilon \left( {\alpha}^2 + {\beta}^2 \right) & 
          \beta  {\alpha}^2
        \end{array}
      \right] \\ 
      \phantom{} \\ 
      \phantom{}
    \end{array}
  \right\}
  \left[
    \begin{array}{c}
      a^1 \\
      a^2 \\
      a^3
    \end{array}
  \right] 
\nonumber \\
\label{eq:Lv_123_R}
\end{align}
\end{widetext}
The top line is the reaction representation (since the reactions are
bi-directional, we have combined both departure terms in the row
vectors).  The middle line is the complex representation.  In this
representation it is clear why, if $\epsilon \rightarrow 0$, the
reaction cannot be complex-balanced: all terms in a given row vector
have the same sign, so any positive density produces non-zero flows at
some complexes.  The bottom line is the stoichiometric representation.
The first dyadic corresponds to the $s$-flow $\mathbb{A} e e^T$, and
the second dyadic corresponds to the $\delta$-flow $\mathbb{A}
\tilde{e} {\tilde{e}}^T$.

Note that the diagonalization of the $s$-flow couples activity in the
complex $a$ to changes of probability across complexes
${a^{\dagger}}^2$ and ${a^{\dagger}}^3$, and vice-versa with activity
at $a^3$ and change of probability across complexes $a^{\dagger}$ and
${a^{\dagger}}^2$.  Thus, despite the similarity in form to the
deficiency-0 projector in the Liouville operator from
Eq.~(\ref{eq:L_1_to_2}), the $s$-flow projection in
Eq.~(\ref{eq:Lv_123_R}) cannot be written as a stand-alone Liouville
operator from a deficiency-0 sub-network of the current network.

To illustrate the way in which different combinations of $s$- and
$\delta$-flows control the scaling of ${\Phi}_k$ in different regions,
we note the forms of projection operators at different orders in the
sum~(\ref{eq:Glauber_moment_fact_prod}):
\begin{align}
  Y = 
  Y^{\underline{1}} 
& = 
  \left[ 
    \begin{array}{ccc}
      1 & 2 & 3 
    \end{array}
  \right]
& 
  Y^{\underline{1}} \mathbb{A}
& = 
  \left[ 
    \begin{array}{ccc}
      \alpha & 0 & -\beta 
    \end{array}
  \right]
\nonumber \\ 
  Y^{\underline{2}} 
& = 
  \left[ 
    \begin{array}{ccc}
      0 & 2 & 6 
    \end{array}
  \right]
& 
  \frac{
    Y^{\underline{2}} \mathbb{A}
  }{
    2! 
  }
& = 
  \left[ 
    \begin{array}{ccc}
      \alpha & \epsilon & -2 \beta 
    \end{array}
  \right]
\nonumber \\ 
  Y^{\underline{3}} 
& = 
  \left[ 
    \begin{array}{ccc}
      0 & 0 & 6 
    \end{array}
  \right]
& 
  \frac{
    Y^{\underline{3}} \mathbb{A}
  }{
    6! 
  }
& = 
  \left[ 
    \begin{array}{ccc}
      0 & \epsilon & -\beta  
    \end{array}
  \right]
\label{eq:123_y_form}
\end{align}
The lowest-order term $Y^{\underline{1}} \mathbb{A}$ projects out
solutions ${\Phi}_{k+2} / {\Phi}_k = \alpha / \beta$, while the
highest-order term $Y^{\underline{3}} \mathbb{A}$ projects out the
solution ${\Phi}_{k+2} / {\Phi}_{k+1} = \epsilon / \beta$.  These turn
out to be the two limiting moment ratios, respectively, in the limits
$k = 1$ (the moment recursion formula has no term at $k = 0$) and $k
\rightarrow \infty$, as we now demonstrate.

\subsubsection{Scaling behavior of the rate equation used as a
recursion relation}

The Poisson background in the expansion~(\ref{eq:state_division}),
projected out by $Y^{\underline{1}} \mathbb{A}$ and ensuring vanishing
of the $s$-flow contribution to the moment dynamics at each order $k$,
is given by ${\psi}_Y = {\left[
\begin{array}{ccc} \xi & {\xi}^2 & {\xi}^3 \end{array} \right]}^T$,
where the mean number $\xi$ satisfies
\begin{equation}
  {\xi}^2 = 
  \frac{\alpha}{\beta} . 
\label{eq:123_mean}
\end{equation}

This CRN has a unique steady state, so the terms appearing in
Eq.~(\ref{eq:Glauber_moment_fact_hats_primes}) are:
\begin{widetext}
\begin{align}
  k 
  \frac{Y}{\xi} 
  \hat{\mathbb{A}}
  \hat{e} {\hat{e}}^T 
  {
    \left< 
      {\hat{\Psi}}_{Y + \left( k-1 \right)} \! \left( \rn \right) 
    \right>
  }^{\prime} 
& = 
  \alpha k 
  \left(
    {\hat{\Phi}}_k^{\prime} - 
    {\hat{\Phi}}_{k+2}^{\prime}
  \right) 
\nonumber \\ 
  \sum_{j = 2}^k
  \left(
    \begin{array}{c}
      k \\ j 
    \end{array}
  \right)
  \frac{
    Y^{\underline{j}} 
  }{
    {\xi}^j
  }
  \hat{\mathbb{A}} 
  \left\{ 
    \hat{e} {\hat{e}}^T + 
    {\hat{\tilde{e}}} {\hat{\tilde{e}}}^T 
  \right\}
  {
    \left< 
      {\hat{\Psi}}_{Y + \left( k-j \right)} \! \left( \rn \right) 
    \right>
  }^{\prime} 
& = 
  \frac{\alpha}{\xi}
  \frac{k!}{\left( k-2 \right) !}
  \left\{ 
    \left(
      {\hat{\Phi}}_{k-1}^{\prime} - 
      {\hat{\Phi}}_{k+1}^{\prime}
    \right) + 
    \left( 
      \frac{\epsilon}{\sqrt{\alpha \beta}}
      {\hat{\Phi}}_k^{\prime} - 
      {\hat{\Phi}}_{k+1}^{\prime}
    \right) 
  \right.
\nonumber \\
& \qquad \qquad \qquad 
  \mbox{} + 
  \left. 
    \left(
      \frac{k-2}{\xi}
    \right)
    \left( 
      \frac{\epsilon}{\sqrt{\alpha \beta}}
      {\hat{\Phi}}_{k-1}^{\prime} - 
      {\hat{\Phi}}_k^{\prime}
    \right) 
  \right\}
\nonumber \\ 
  c_0 
  \sum_{j = 2}^k
  \left(
    \begin{array}{c}
      k \\ j 
    \end{array}
  \right)
  \frac{
    Y^{\underline{j}}
  }{
    {\xi}^j
  }
  \hat{\mathbb{A}} 
  {\hat{\tilde{e}}} {\hat{\tilde{e}}}^T 
  {\hat{\psi}}_Y \! \left( \xi \right) 
& = 
  c_0 
  \frac{\alpha}{\xi}
  \frac{k!}{\left( k-2 \right) !}
  \left(
    1 + \frac{k-2}{\xi}
  \right)
  \left(
    \frac{\epsilon}{\sqrt{\alpha \beta}} - 1 
  \right) 
\label{eq:123_projectors}
\end{align}
\end{widetext}
As the numerical evaluations below will show, this is a model in
which, despite uniqueness of the Poisson solution matching the first
and third moments, the overall normalization of the moment hierarchy
is not anchored in the lowest term ${\Phi}_0$, so a relative
normalization $c_0$ for the ${\psi}_Y \! \left( \xi \right)$
contribution is undetermined.  \emph{In this way, the role of the
  Poisson background in an exact solution of the moment hierarchy is
  different from a mean-field approximation.}  MFT would require
${\left< \rn \right>}^2 = \alpha / \beta$ in place of
Eq.~(\ref{eq:123_mean}), which requires ${\left< \rn \right>}^2 =
{\left( c_0 \xi \right)}^2$.  The freedom for $\xi$ to differ from
$\left< \rn \right>$ by the normalization $c_0$ is necessary, because
even for this simple network, the MFT prediction for the mean is not
valid.

The exact recursion relation for the deviations from Poisson moments
is then the finite sum
\begin{align}
  {\hat{\Phi}}_k^{\prime} - 
  {\hat{\Phi}}_{k+2}^{\prime}
& + 
  \frac{k-1}{\xi}
  \left\{ 
    \left(
      {\hat{\Phi}}_{k-1}^{\prime} - 
      {\hat{\Phi}}_{k+1}^{\prime}
    \right) + 
    \left( 
      \frac{\epsilon}{\sqrt{\alpha \beta}}
      {\hat{\Phi}}_k^{\prime} - 
      {\hat{\Phi}}_{k+1}^{\prime}
    \right) 
  \right.
\nonumber \\
& \qquad \qquad 
  \mbox{} + 
  \left. 
    \left(
      \frac{k-2}{\xi}
    \right)
    \left( 
      \frac{\epsilon}{\sqrt{\alpha \beta}}
      {\hat{\Phi}}_{k-1}^{\prime} - 
      {\hat{\Phi}}_k^{\prime}
    \right) 
  \right\}
\nonumber \\ 
& = 
  - c_0 
  \frac{k-1}{\xi}
  \left(
    1 + \frac{k-2}{\xi}
  \right)
  \left(
    \frac{\epsilon}{\sqrt{\alpha \beta}} - 1 
  \right) 
\label{eq:123_eval}
\end{align}

For parameters that produce suitably small mean particle number, the
recursion relation implied by Eq.~(\ref{eq:123_eval}) may be solved
for all $k$ to any desired precision, from the upper asymptotic
behavior alone, as shown in~\cite{Krishnamurthy:CRN_moments:17}.  We
are interested here, however, in understanding the small-$k$ and
large-$k$ behavior of the moments.  To this effect, we can see
directly both aspects of the scaling presented in
Sec.~\ref{sec:scaling}.  As an asymptotic expansion, the recursion
relation specifies the higher-order difference $\left(
  {\hat{\Phi}}_{k+2} - {\hat{\Phi}}_k\right)$ as a power series in
$\left( k / \xi \right)$ with coefficients from values and differences
of $\Phi$ at lower $k$ indices.

The seed for the expansion at orders ${\hat{\Phi}}_4^{\prime}$ and
higher from the Poisson background is the $k$-independent value $c_0
\left( \epsilon / \sqrt{\alpha \beta} - 1 \right)$, multiplied by a
polynomial of $\mathcal{O} \! \left( k / \xi \right)$.  Note, however,
that the terms ${\hat{\Phi}}_2^{\prime}$ and $c_0$, which are
undetermined by the first-moment condition $e^T \left< {\Psi}_Y \! 
\left( \rn \right) \right> = 0$, enter the recursion relation according to
the scaling of the overall asymptotic expansion, and are permitted to
be $\mathcal{O} \! \left( 1 \right)$ relative to $\left( \epsilon /
\sqrt{\alpha \beta} - 1 \right)$.

Finally, we observe that for a fine-tuned value of the rate parameters
$\epsilon = \sqrt{\alpha \beta}$, the correction term can be made to
vanish, and the ACK-like solution projected to zero by the $s$-flow
term $\mathbb{A} e e^T$ in Eq.~(\ref{eq:Lv_123_R}) becomes a
steady-state solution, even though for this CRN $\delta = 1$.

Next we illustrate how the asymptotic expansion with a requirement of
boundedness at large $k$ anchors the moment hierarchy at all orders.
We seed the downward-going asymptotic expansion with the leading
non-constant approximation to the recursion relation around the
limiting ratio projected out by $Y^{\underline{3}} \mathbb{A}$ from
Eq.~(\ref{eq:123_y_form}), which has the form
\begin{align}
  \frac{
    \left< 
      {\rn}^{\underline{k+1}}
    \right> 
  }{
    \left< 
      {\rn}^{\underline{k}}
    \right> 
  }
& \approx 
  \left( \frac{\epsilon}{\beta} \right)
  \left[ 
    1 - 
    \frac{
      {\epsilon}^2 - \alpha \beta
    }{
      \epsilon \beta 
    }
    \frac{1}{k-1}
  \right] .
\label{eq:123_large_k_rats}
\end{align}
The corresponding leading-order approximation for large-$k$ moments
may be written about any reference value $k_0$ as 
\begin{align}
  \left< 
    {\rn}^{\underline{k}}
  \right> 
& \approx 
  \mathcal{N}
  {
    \left( \frac{\epsilon}{\beta} \right)
  }^k
  \left[ 
    1 - 
    \frac{
      {\epsilon}^2 - \alpha \beta
    }{
      \epsilon \beta 
    }
    \log 
    \left( 
      \frac{k-1}{k_0}
    \right)
  \right]
\nonumber \\
& \approx 
  \mathcal{N}
  {
    \left( \frac{\epsilon}{\beta} \right)
  }^k
  {
    \left( 
      \frac{k_0}{k-1}
    \right)
  }^{
    \left( {\epsilon}^2 - \alpha \beta \right) / 
    \left( \epsilon \beta \right) 
  } , 
\label{eq:123_large_k_sol}
\end{align}
where $\mathcal{N}$  is an overall normalization to be determined.

\begin{figure*}[ht]
  \begin{center} 
  \includegraphics[scale=0.29]{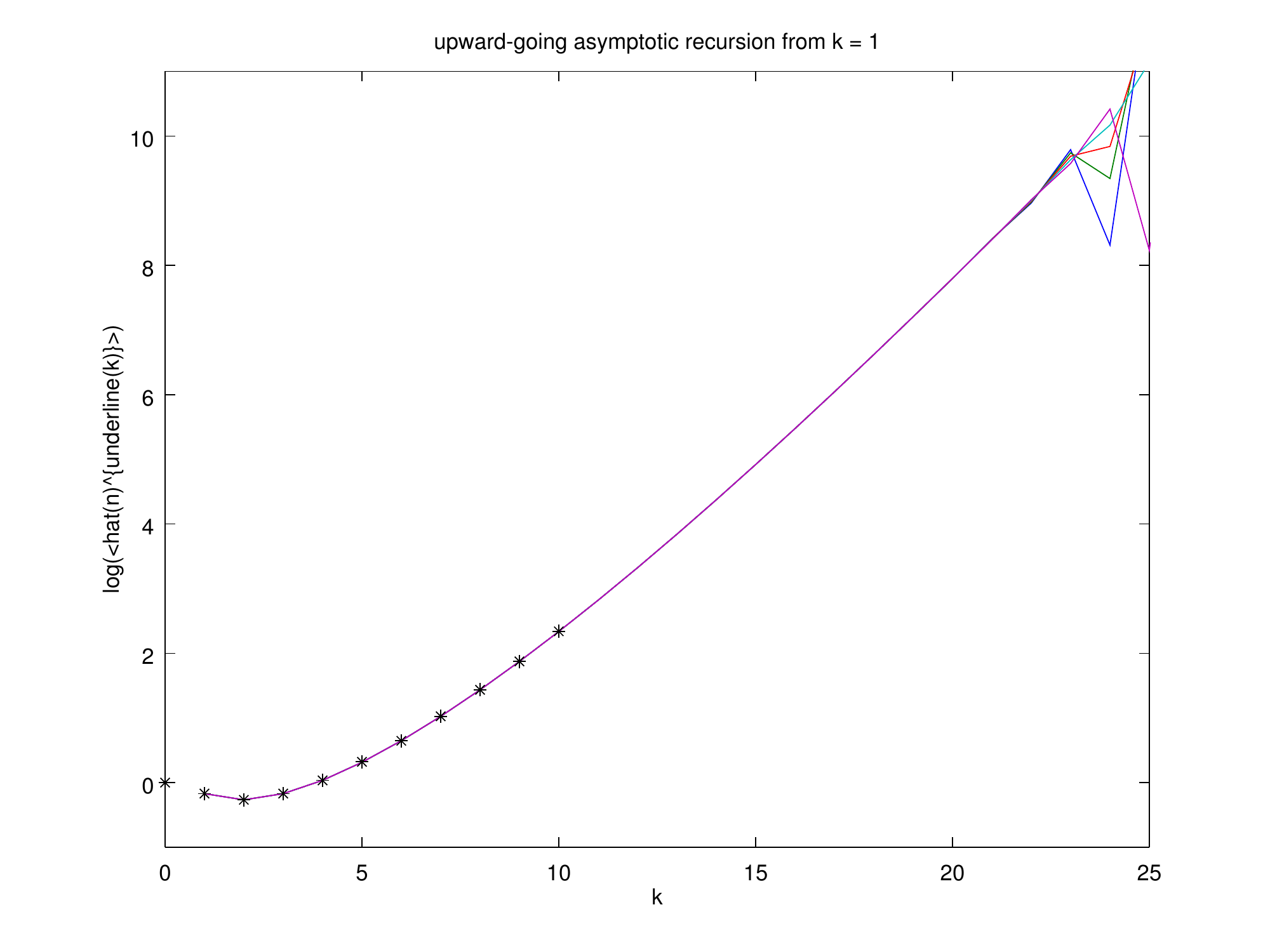} 
  \includegraphics[scale=0.29]{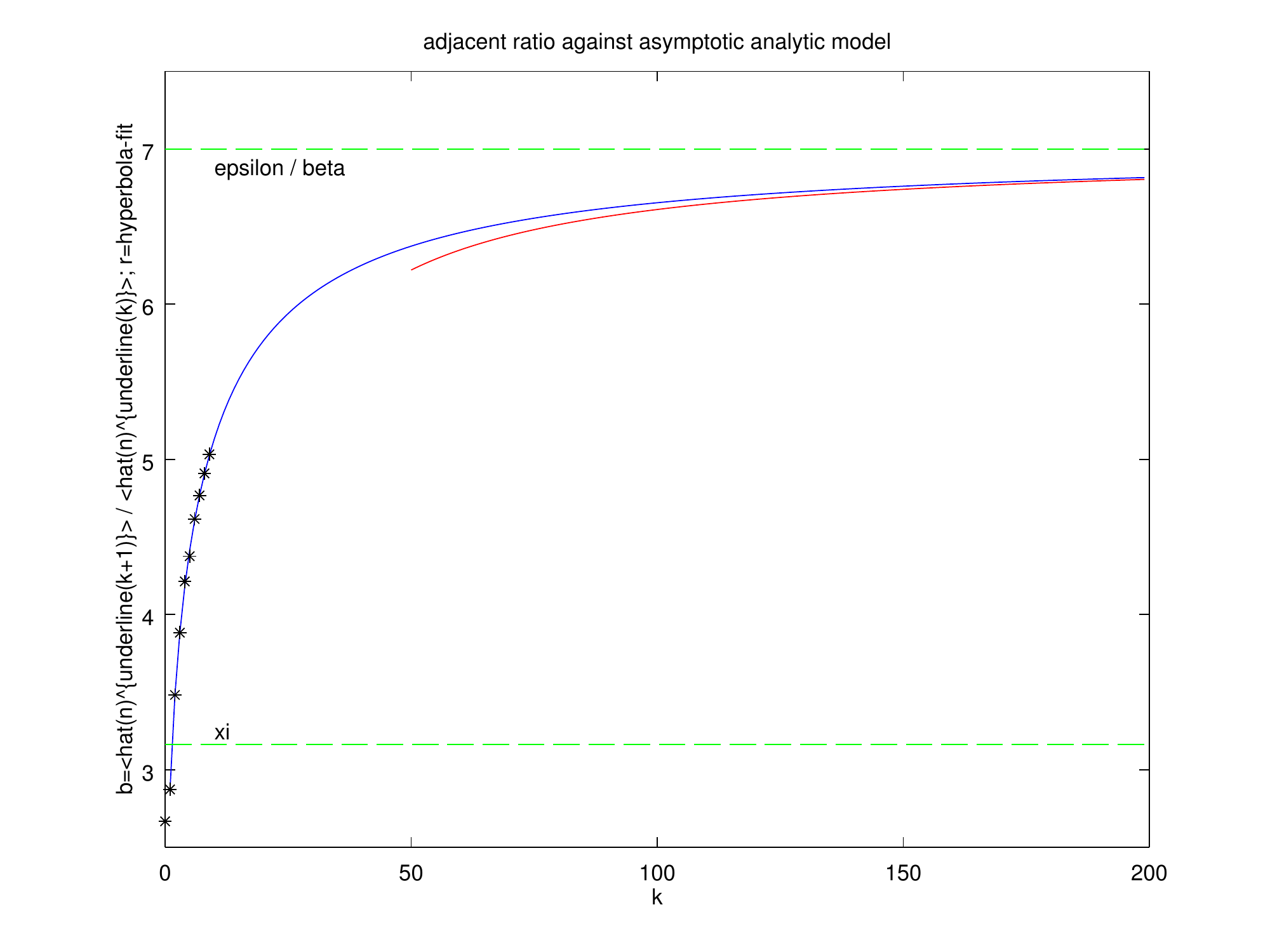} 
  \includegraphics[scale=0.29]{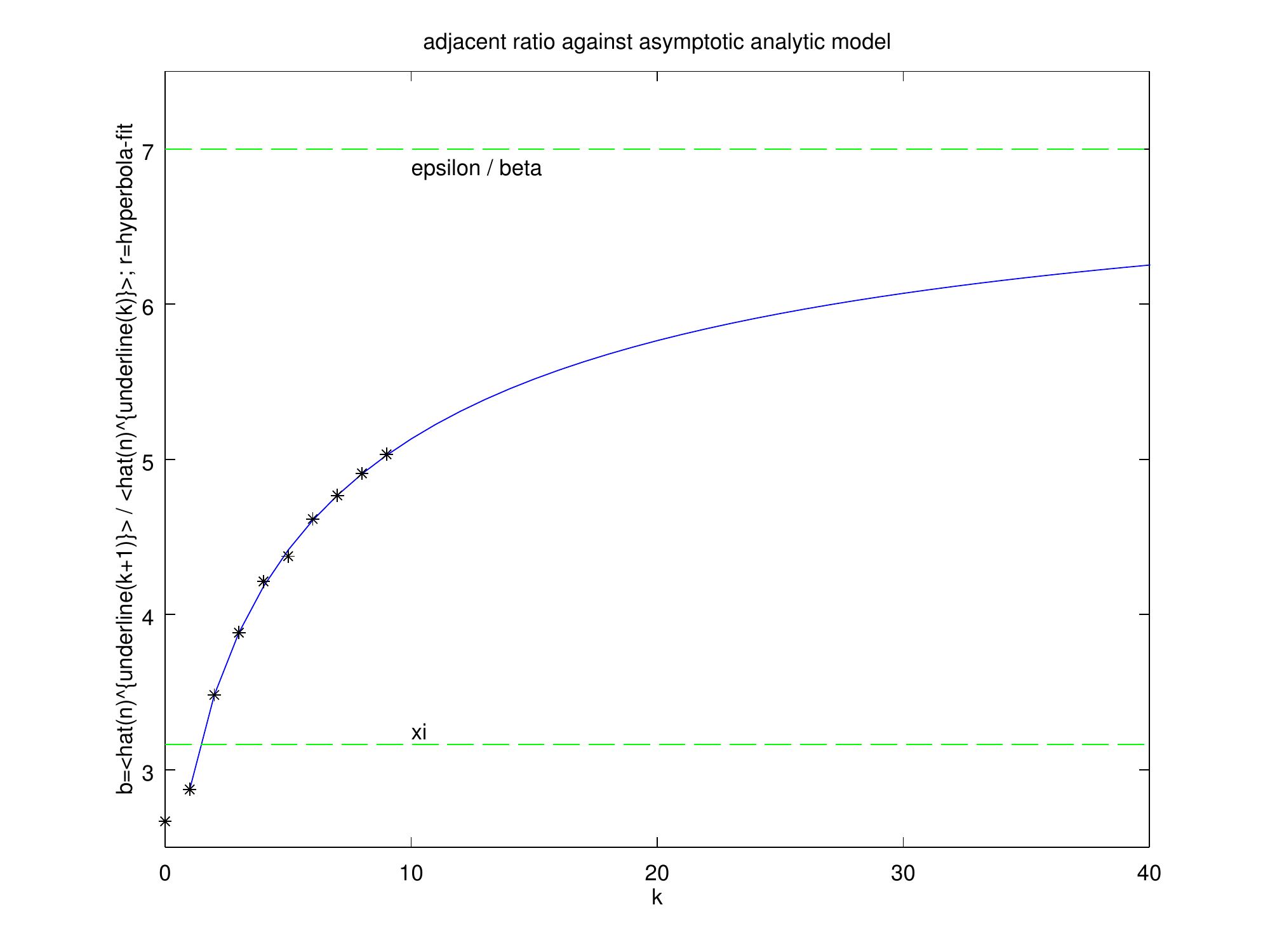} 
  \caption{
  Asymptotic expansions for moments and moment ratios for the model of
  Fig.~\ref{fig:g_123_conn}.  First plot shows an asymptotic expansion
  for $\log {\hat{\Phi}}_k$ descaled with $\xi = \sqrt{\alpha /
    \beta}$ under the recursion relation~(\ref{eq:123_eval}), upward
  from $k = 1$.  Five traces are generated by starting with $\left<
    \rn \right> \equiv \left< {\rn}^3 \right>$ fixed, and $\left<
    {\rn}^2 \right>$ values spaced by $1 \times {10}^{-10}$ around the
  stable value.  The group of trajectories become uncontrollably
  divergent by $k = 25$.  Black asterisks are evaluations of the
  corresponding moments from a Gillespie simulation, and the value of
  $\left< \rn \right>$ was used to supply the un-specified
  normalization $c_0 \approx 2.67$ in Eq.~(\ref{eq:123_eval}) for the
  recursion series.  Second and third panels, which differ only in the
  plotted range, show ratios ${\hat{\Phi}}_{k+1} / {\hat{\Phi}}_k$, 
  (for which $c_0$ appears only in the lowest term $\left< \rn \right> / 1$)
  computed by recursion downward from $k = 200$ with the starting
  approximation~(\ref{eq:123_large_k_rats}) (red curve in the second
  panel), with Gillespie simulation results overlaid.  The Poisson
  expectation $\xi$ from MFT, and the large-$k$ asymptotic limit
  $\epsilon / \beta$, are shown as dashed lines for reference. 
  \label{fig:n_hats_asymps_contigs} 
  } 
  \end{center}
\end{figure*}

Fig.~\ref{fig:n_hats_asymps_contigs} shows a comparison of the
numerical regression from Eq.~(\ref{eq:123_eval}), both for an
upward-going recursion from $k = 1$, and for a downward-going
recursion with the large-$k$ asymptotic
seed~(\ref{eq:123_large_k_sol}), to estimates of the first 10 moments
from a Gillespie simulation of the underlying process.  The parameters
used in the demonstration are: $\alpha = 100$, $\beta = 10$; $\epsilon
= 70$.  So the relevant parameters are $\xi = \sqrt{\alpha / \beta} =
\sqrt{10}$; $\epsilon / \beta = 7$; $\epsilon / \sqrt{\alpha \beta} =
\epsilon / \beta \xi \approx 2.2136$.  We have normalized the constant
$c_0$, which is unspecified by the recursion relation to the simulated
mean $\left< \rn \right>$, and find close agreement to all other the
moment ratios.

\subsection{A CRN with 1-species, 4 states, 2 linkage classes and no
  conserved quantities}
\label{sec:g_0123_unconn}

\begin{figure}[ht]
  \begin{center} 
  \includegraphics[scale=0.6]{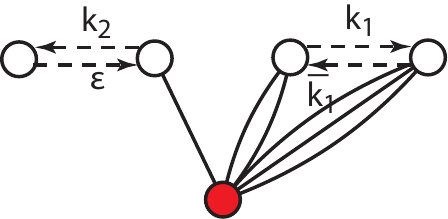}
  \caption{
  This network, for some rate constants, can have two non-equilibrium
  steady states in the mean-field approximation.  Even when this is
  the case, however, because particle number is finite, the stochastic
  system always has only one ergodically-sampled long-term steady
  distribution.  Some sub-graphs are common with
  Fig.~\ref{fig:g_123_conn}, but the complex graph has two linkage
  classes, again giving $\delta = 1$.
    \label{fig:g_0123_unconn} 
  }
  \end{center}
\end{figure}

We can preserve the number of $s$-flows and $\delta$-flows from the
previous model, but introduce the possibility for multi-stability, by
increasing both the number of complexes and the number of linkage
classes by one.  The resulting graph for a minimal model with this
elaboration is shown in Fig.~\ref{fig:g_0123_unconn}, and its reaction
scheme is given by
\begin{align}
  \varnothing
& \xrightleftharpoons[k_2]{\epsilon}
  {\rm A} & 
  2 {\rm A} 
& \xrightleftharpoons[{\bar{k}}_1]{k_1}
  3 {\rm A} .
\label{eq:cubic_1spec_scheme}
\end{align}
The corresponding rate equation is
\begin{align}
  \frac{\partial n}{\partial \tau}
& = 
  \epsilon - 
  k_2 n + 
  k_1 n^2 - 
  {\bar{k}}_1 n^3 . 
\label{eq:rate_g_0123}
\end{align}
We have labeled the reaction rate constants in this model to reflect a
set of cases that are often important in biochemistry and industrial
synthesis: a reaction that is (directly or indirectly) self-catalyzed
by feedback through the synthetic network is the main channel for
production and decay of the product (rate constants $k_1$ and
${\bar{k}}_1$), compared to an uncatalyzed pathway that has nonzero
but small rate ($\epsilon$), while a significant rate ($k_2$) remains
for spontaneous decay of the product.

Quadratic-order autocatalysis in this CRN comes from the same pair of
reactions as it does in Fig.~\ref{fig:g_123_conn}.  The addition of a
fourth state to the complex graph creates a cubic first-moment rate
equation and thus the possibility for multiple steady states.

For the graph of Fig.~\ref{fig:g_0123_unconn}, the Liouville operator
is
\begin{align}
  \mathcal{L} 
& = 
  \left( 1 - a^{\dagger} \right) 
  \left[
    \epsilon - 
    k_2 a + 
    k_1 {a^{\dagger}}^2 a^2 - 
    {\bar{k}}_1 {a^{\dagger}}^2 a^3 
  \right]
\nonumber \\ 
& = 
  \left( 1 - a^{\dagger} \right) 
  \left[
    \left( 
      \epsilon - 
      k_2 a 
    \right) + 
    \left( a^{\dagger} \! a \right) 
    \left( a^{\dagger} \! a - 1 \right) 
    \left( 
      k_1  - 
      {\bar{k}}_1 a 
    \right) 
  \right] . 
\label{eq:L_0123}
\end{align}
The stoichiometric decomposition of this operator is similar to that
from the previous model, and is given in
Eq.~(\ref{eq:Lv_one_species_bistable}).

The factorials $Y^{\underline{j}}$ and projection operators
$Y^{\underline{j}} \mathbb{A}$ appearing in
Eq.~(\ref{eq:Glauber_moment_fact_prod}) are given by 
\begin{align}
  Y = 
  Y^{\underline{1}} 
& = 
  \left[ 
    \begin{array}{cccc}
      0 & 1 & 2 & 3 
    \end{array}
  \right]
& 
  Y^{\underline{1}} \mathbb{A}
& = 
  \left[ 
    \begin{array}{cccc}
      \epsilon & -k_2 & k_1 & -{\bar{k}}_1
    \end{array}
  \right]
\nonumber \\ 
  Y^{\underline{2}} 
& = 
  \left[ 
    \begin{array}{cccc}
      0 & 0 & 2 & 6 
    \end{array}
  \right]
& 
  \frac{
  Y^{\underline{2}} \mathbb{A}
  }{
    2 ! 
  }
& = 
  2 
  \left[ 
    \begin{array}{cccc}
      0 & 0 & k_1 & -{\bar{k}}_1
    \end{array}
  \right]
\nonumber \\ 
  Y^{\underline{3}} 
& = 
  \left[ 
    \begin{array}{cccc}
      0 & 0 & 0 & 6 
    \end{array}
  \right]
& 
  \frac{
  Y^{\underline{3}} \mathbb{A}
  }{
    3 ! 
  }
& = 
  \left[ 
    \begin{array}{cccc}
      0 & 0 & k_1 & -{\bar{k}}_1
    \end{array}
  \right]
\label{eq:0123_yA_form}
\end{align}
The lowest-order (in $k / \xi$) projector in
Eq.~(\ref{eq:0123_yA_form}) is $Y^{\underline{1}} \mathbb{A}$, which
is the projection operator corresponding to the $s$-flow in
Eq.~(\ref{eq:Lv_one_species_bistable}) in the appendix.  If required
to vanish on a coherent state, it gives
\begin{equation}
  \epsilon - k_2 \xi + k_1 {\xi}^2 -{\bar{k}}_1 {\xi}^3 = 
  0 
\label{eq:0123_y1A_vanishing}
\end{equation}
For appropriate parameter choices, this may have either a unique
stable solution, or it may have three solutions, two stable and one
between them that is unstable.  Unlike the model of
Fig.~\ref{fig:g_123_conn}, both ${\Phi}_0$ and ${\Phi}_2$ have
non-zero coefficients in $\left< {\Psi}_Y \! \left( \rn \right)
\right>$, so the normalization of the mean is fixed relative to
${\Phi}_0 \equiv 1$.  Solutions of the form~(\ref{eq:state_division})
must satisfy $\sum_{\gamma} c_{\gamma} = 1$, and therefore we may set
${\left< {\Psi}_Y \! \left( \rn \right) \right>}^{\prime} \equiv 0$.

In this model (contrasted with the result in
Eq.~(\ref{eq:123_y_form})), both of the projection operators
$Y^{\underline{2}} \mathbb{A}$ and $Y^{\underline{3}} \mathbb{A}$
cancel the same ratio ${\Phi}_{k+2} / {\Phi}_{k+1} = k_1 / {\bar{k}}_1
\equiv K_1$, so there are only two scaling behaviors expressed in the
model, respectively at $k \rightarrow 0$ and $k \rightarrow \infty$. 

This CRN, also being a birth-death type process, can be solved exactly
\cite{Anderson:Lyapunov:15} for the steady state.  As for the 3-state
model, this moment hierarchy may also be solved by recursion from an
upper asymptotic limit that is derivable analytically (though again
the numerical calculation is stable only for sufficiently small mean
particle numbers.)  If we define $ \left<{\rn}^{\underline k}\right> /
\left<{\rn}^{\underline {k-1}}\right> \equiv R_k$, then the set of
$R_k$ must obey the recursion
\begin{widetext}
\begin{equation}
  R_k =  
  \frac{
      \epsilon + 
      \left( k-1 \right) \left( k-2 \right) k_1 
  }{
    k_2 - k_1 R_{k+1} +
    {\bar{k}}_1 R_{k+2} R_{k+1} - 
    \left( k-1 \right) 
    \left( 
      2k_1 - 2 {\bar{k}}_1 R_{k+1} 
    \right) + {\bar{k}}_1 
    \left( k-1 \right) \left( k-2 \right)
  } . 
\label{eq:rec_123}  
\end{equation}
\end{widetext}

When we solve the recursion~(\ref{eq:rec_123}) numerically, directly
in terms of moments ${\Phi}_k$, we begin with a more refined large-$k$
approximate form than the first-order approximation used as a seed in
Eq.~(\ref{eq:123_large_k_rats}).  The second-order leading
non-constant approximation, corresponding to the
form~(\ref{eq:123_large_k_sol}) given for the previous model, is
\begin{align}
  {\Phi}_k
& \approx 
  \mathcal{N}
  {
    \left( \frac{k_1}{{\bar{k}}_1} \right)
  }^k
  \left[
    1 + 
    \frac{\eta}{k-1} + 
    \frac{
      \eta
      \left(\eta / 2 - K_1 \right)
    }{
      k \left( k-1 \right)
    } 
  \right. 
\nonumber \\
& \qquad \qquad \qquad
  \mbox{} + 
  \left. 
    \mathcal{O} \! 
    \left( \frac{1}{k^3} \right)
  \right] , 
\label{eq:0123_large_k_sol}
\end{align}
where $\mathcal{N}$ is an arbitrary normalization to be fixed by
${\Phi}_0 = 1$, and 
\begin{equation}
  \eta \equiv 
  \frac{k_2}{{\bar{k}}_1} - 
  \frac{\epsilon}{k_1} . 
\label{eq:c_erg_def}
\end{equation}

The large-$k$ asymptotic behavior of $R_k$ in Eq.~(\ref{eq:rec_123})
can likewise be solved for an expansion in $1/k$ about the leading
fixed point, in the same manner as Eq.~(\ref{eq:123_large_k_rats}) for
the previous model.  In this case, the leading departure is
$\mathcal{O} \! \left( 1 / k^2 \right)$ rather than $\mathcal{O} \!
\left( 1 / k \right)$ as in the 3-complex model.  In general, it can
be shown that that the first departure from whatever fixed point is
dictated by the leading large-$k$ projection operator in
Eq.~(\ref{eq:Glauber_moment_fact_prod}) is determined by the highest
power of $k$ appearing in the expansion in the sum.  This order
corresponds to the largest stoichiometric coefficients for that
component of $k$ appearing in the CRN.

\subsubsection{Bistability in MFT and handling mixtures of Poisson
basis elements around $k = 0$}

The handling of multi-stability in CRNs with $\delta > 0$ introduces
several new interesting properties, both within the moment recursion
relations and in their relation to mean-field theory.  First, MFT will
generally predict multi-stability for Poisson solutions, whether or
not the mean particle number is large enough that trajectories in the
stochastic process actually generate a multi-modal density of particle
numbers.  The meaning of MFT solutions in relation to the analytic
structure of representations of the generating function is an
interesting topic from which we briefly draw results below, but mostly
refer to other
developments~\cite{Cardy:Instantons:78,Smith:Signaling:11,Smith:LDP_SEA:11,%
Smith:evo_games:15} (see also~\cite{Coleman:AoS:85}, Ch.~7).  Second
and more important, the moment relations are exact, and we therefore
expect them to possess unique solutions corresponding to the ergodic
distribution, even when mean particle number is large enough that the
MFT representation of multistability corresponds to a true
incipient\footnote{We say ``incipient'' because in conventional usage,
breaking of ergodicity is an asymptotic property, here in the scaling
variable $\left< \rn \right>$ as $\left<
\rn \right> \rightarrow \infty$.  Formally, ergodicity breaking can
still be considered well-defined even in finite-size systems, to the
extent that it is associated with stationary paths in a semiclassical
approximation that are essential singularities with respect to the
asymptotic expansion in fluctuations~\cite{Cardy:Instantons:78}.} 
breaking of ergodicity.  The expansion in high-order moments becomes
an important if cryptic representation of the trajectories responsible
for first-passages between domains.

We illustrate some of these properties for the case of bistability
with a numerical example at parameters: $\epsilon = 36$, $k_2 = 49$;
$k_1 = 14$; ${\bar{k}}_1 = 1$.  The two stable solutions to
Eq.~(\ref{eq:0123_y1A_vanishing}) are ${\xi}^{\left( 1 \right)} = 1$
and ${\xi}^{\left( 3 \right)} = 9$, and an unstable solution exists at
${\xi}^{\left( 2 \right)} = 4$.

\begin{figure}[ht]
  \begin{center} 
  \includegraphics[scale=0.4]{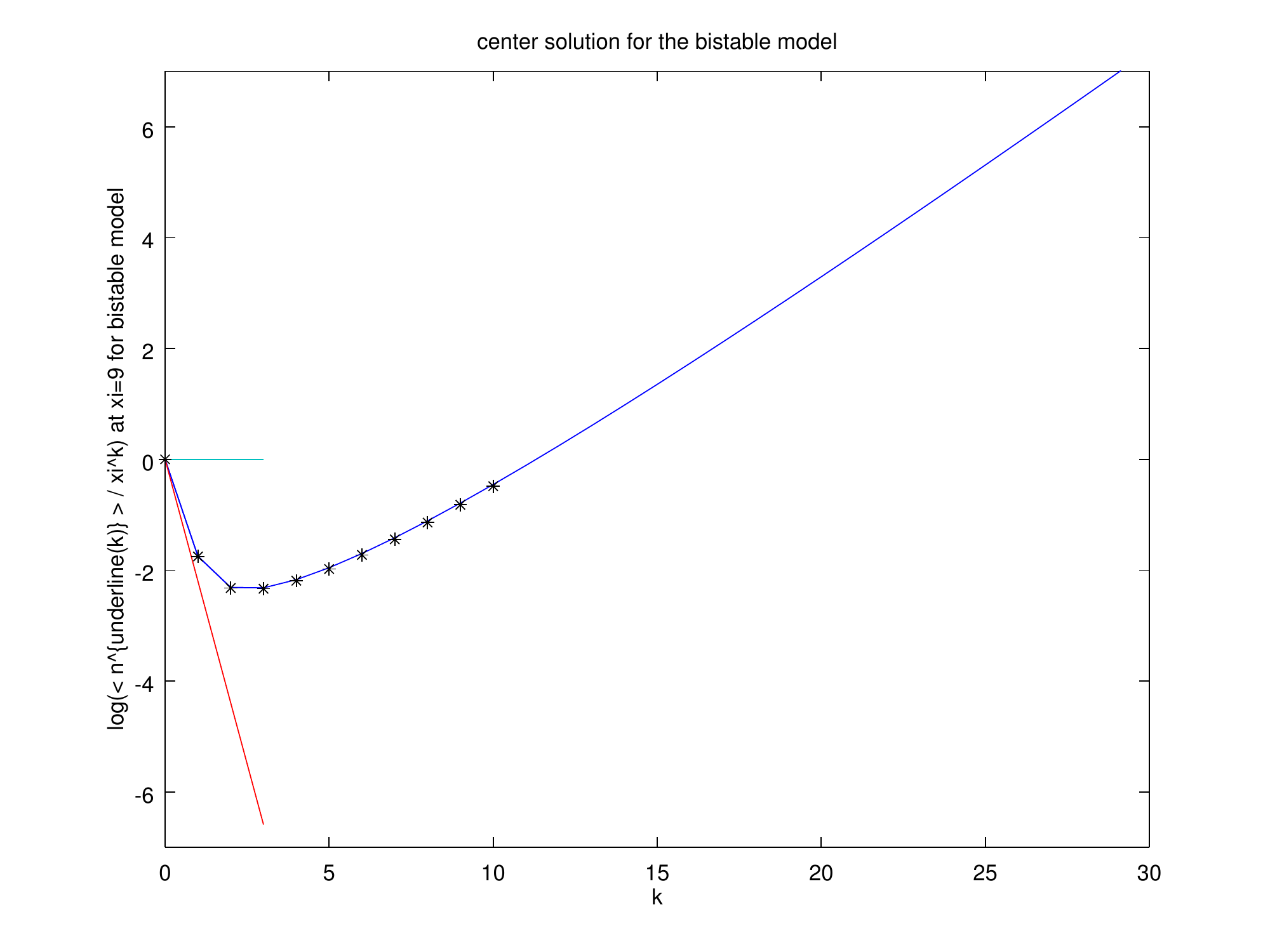} 
  \caption{
  The solution $\log {\hat{\Phi}}_k$ to
  the recursion relations~(\ref{eq:Glauber_moment_fact_assum}),
  descaled with ${\xi}^{\left( 3 \right)} = 9$, extended downward from
  $k = 200$ using the asymptotic
  approximation~(\ref{eq:0123_large_k_sol}).  The Poisson basis
  elements for ${\xi}^{\left( 1 \right)} = 1$ and ${\xi}^{\left( 3
  \right)} = 9$ corresponding to the stable solutions in MFT are
  shown, respectively, in red and green, for reference.  The values
  ${\hat{\Phi}}_k$ for $k \in 0, \ldots,
  3$ match an expansion~(\ref{eq:state_division}) with the
  coefficients~(\ref{eq:0123_mixture_coeffs}) and ${\left< {\Psi}_Y \!
  \left( \rn \right) \right>}^{\prime} \equiv 0$.  Symbols are from a
  direct Gillespie simulation. 
  \label{fig:0123_mid_soln} 
  } 
  \end{center}
\end{figure}

Fig.~\ref{fig:0123_mid_soln} shows the recursive solution for $\left<
{\hat{\rn}}^{\underline{k}} \right>$, descaled with ${\xi}^{\left( 3
\right)} = 9$, starting from the large-$k$ asymptotic
form~(\ref{eq:0123_large_k_sol}).  The solution exactly matches the
expansion~(\ref{eq:state_division}), with the coefficients given by
\begin{align}
  c_1
& \approx 
  0.987
&
  c_2 
& \approx 
  - 0.092
& 
  c_3
& \approx
  0.105 . 
\label{eq:0123_mixture_coeffs}
\end{align}

Mean-field theory suggests no natural interpretation of the
mixture~(\ref{eq:0123_mixture_coeffs}) with a negative coefficient on
an unstable solution.  What would normally be done instead in MFT is
to express the mean 
$\left< \rn \right> \approx 1.56$ directly as a mixture of the two
MFT-stable values ${\xi}^{\left( 1 \right)} = 1$ and ${\xi}^{\left( 3
\right)} = 9$ with a mixing coefficient
\begin{align}
  c_{\rm eff} \equiv 
  \frac{
    \left< {\rn} \right> - 1
  }{
    9 - 1 
  } \approx 
  0.071 . 
\label{eq:eff_mixture_coeffs}
\end{align}
We will use the phenomenological
description~(\ref{eq:eff_mixture_coeffs}) to understand qualitatively
how MFT and stationary-point expansions relate to the exact solution
of the moment hierarchy.

\subsubsection{Interpretation with a Kramers approximation for
  first-passage times}

\begin{figure}[ht]
  \begin{center} 
  \includegraphics[scale=0.375]{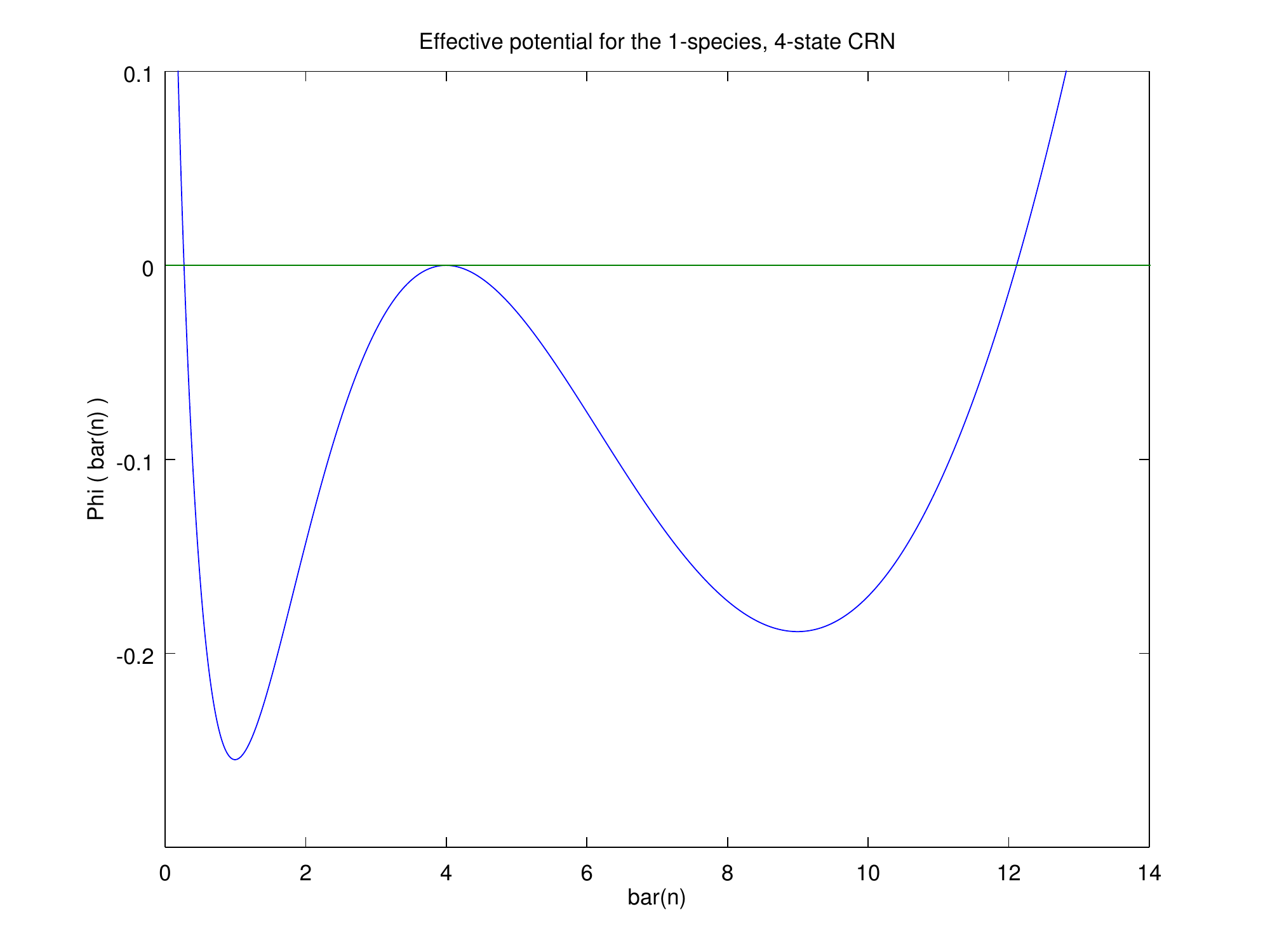} 
  \caption{
  The effective potential $\Xi \! \left( \bar{n} \right)$ from
  Eq.~(\ref{eq:0123_Kramers_probs_form}).
  \label{fig:0123_eff_pot} 
  } 
  \end{center}
\end{figure}

The interpretation of the ergodic solution in terms of a sum over
na{\"{\i}}ve mean-field backgrounds can be compared to a
stationary-point expansion using the method of instantons, which is
developed
in~\cite{Coleman:AoS:85,Smith:Signaling:11,Smith:evo_games:15}.
Stationary-point locations and probabilities are governed by the
minima of a non-equilibrium effective potential, which we have
computed for this particular network in~\cite{Smith:geo13:16} (Ch.7),
and which takes the form\footnote{In Ref.~\cite{Smith:geo13:16}, we
use the notation $\Phi \! \left( \bar{n} \right)$ for the effective
potential, which we change here to $\Xi \! \left( \bar{n} \right)$
to avoid a collision with the notation for the moment hierarchy.}
\begin{equation}
  \Xi \! \left( \bar{n} \right) = 
  \int_4^{\bar{\rn}} d n \,  
  \log \left(
  \frac{
    k_2  n + 
    {\bar{k}}_1 n^3
  }{
    \epsilon + 
    k_1 n^2
  }
  \right) . 
\label{eq:0123_Kramers_probs_form}
\end{equation}
The extrema of the effective potential are exactly the values of the
Poisson parameters ${\xi}^{\left( \gamma \right)}$.  (Here we
arbitrarily set the zero of the effective potential to $n = 4$, the
saddle point.)  A plot of the effective potential versus $n$ is shown
in Fig.~\ref{fig:0123_eff_pot}.

The probability to occupy either minimum may be approximated by the
Kramers formula~\cite{Sjogren:Kramers:16} derived from $\Xi$,
\begin{equation}
  p_{\bar{n}} \propto
  \sqrt{{\Xi}^{\prime \prime} \! \left( \bar{n} \right)}
  e^{- \Xi \! \left( \bar{n} \right)} , 
\label{eq:0123_p_occ_Kramers}
\end{equation}
which follows from a semiclassical approximation to the escape rates
by the non-trivial stationary trajectories known as
\textit{instantons}.  Fig.~\ref{fig:0123_eff_pot} shows that for these
parameters, the minima of $\Xi$ are $\approx -0.25$ and $\approx
-0.19$, respectively at $\bar{n} = 1$ and $\bar{n} = 9$, so the
Kramers approximation is not expected to be quantitatively accurate.
The corresponding second derivatives ${\Xi}^{\prime
\prime}$ take values $\approx 0.07$ and $\approx 0.02$.

\begin{figure}[ht]
  \begin{center} 
  \includegraphics[scale=0.3]{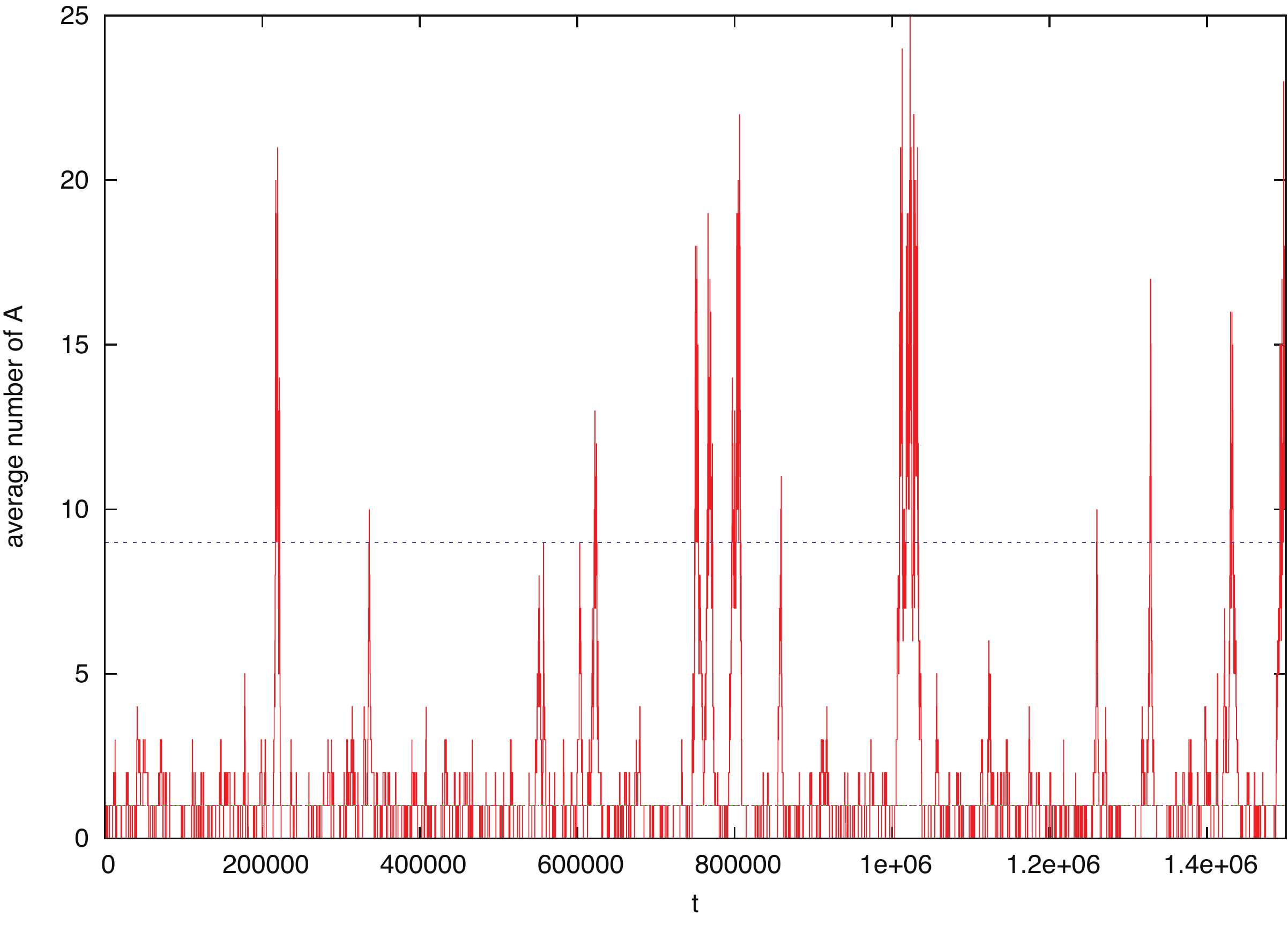} 
  \includegraphics[scale=0.3]{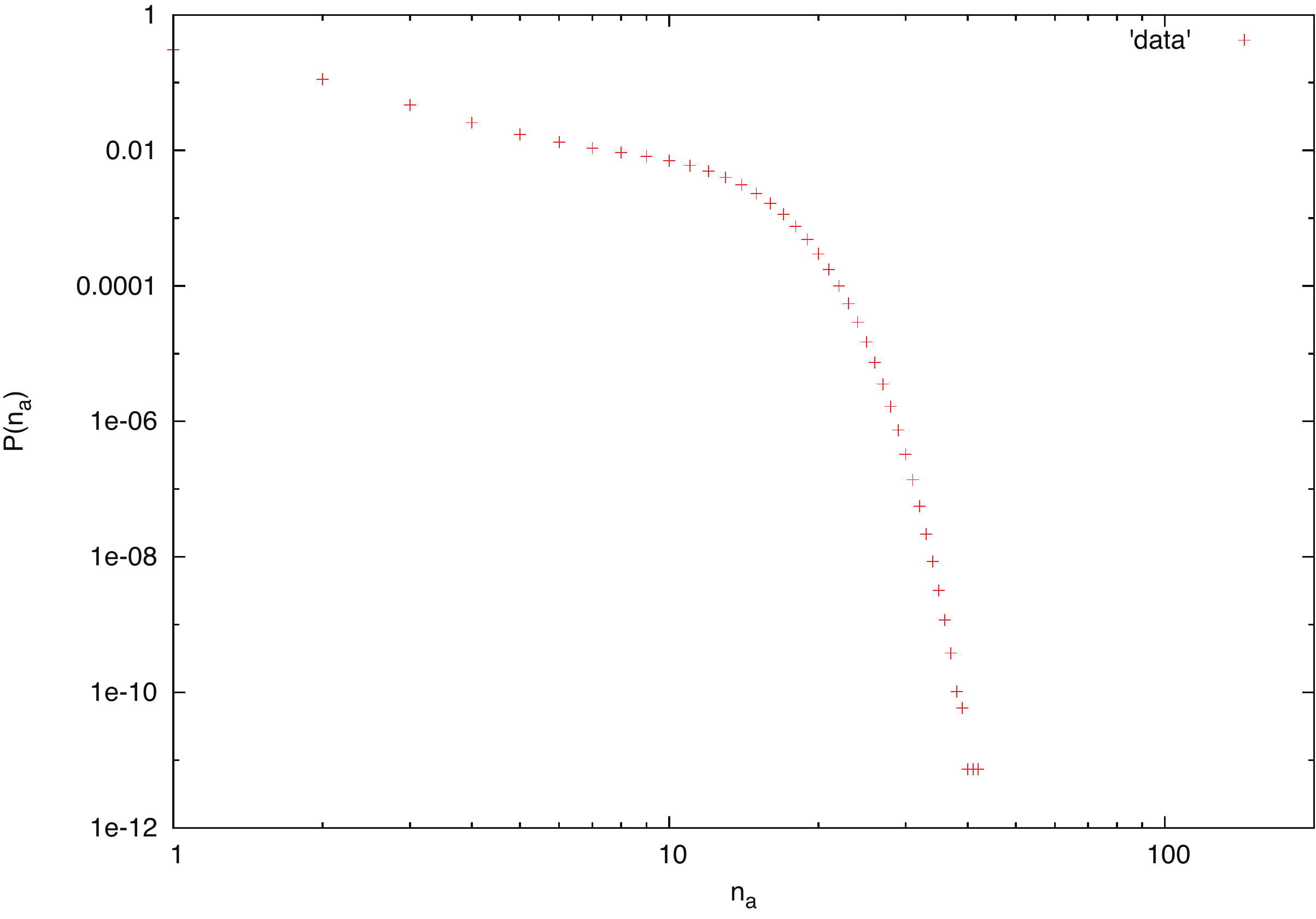} 
  \caption{
  Upper panel: Timeseries for the particle number $\rn$ in the 4-state
  model of Fig.~\ref{fig:g_0123_unconn}.  Dashed lines label the
  minima of the effective potential~(\ref{eq:0123_Kramers_probs_form})
  in the stationary-point expansion.  Lower panel: histogram of the
  stationary distribution for $\rn$ from the simulations, showing
  monotone decrease and a shoulder with a mode around $\rn \approx 15$.
  \label{fig:4states} 
  } 
  \end{center}
\end{figure}

Fig.~\ref{fig:4states} shows a timeseries for the particle number
$\rn$ in a Gillespie simulation, with dashed lines indicating the
minima of the effective potential~(\ref{eq:0123_Kramers_probs_form}).
Some features of the stationary-point approximation are reflected: a
majority of the timeseries remains near $\rn = {\xi}^{\left( 1
\right)} = 1$ with strong mean regression, while excursions with
modest persistence and wider fluctuations occur out to $\rn
\sim {\xi}^{\left( 3 \right)} = 9$.  However, the excursions
do not have the character of fully metastable equilibria.  The log of
the empirical stationary distribution from the simulations is monotone
decreasing, with a visible shoulder (the signature that excursions are
persistent) with a mode around $\rn \approx 15$.

Eq.~(\ref{eq:0123_p_occ_Kramers}) gives, for the occupation
probabilities of the two states, approximate values
\begin{align}
  p_{\bar{9}}
& \approx 
  0.20 
\nonumber \\ 
  p_{\bar{1}} = 
  \left( 1 - p_{\bar{9}} \right)
& \approx  
  0.80 , 
& 
\label{eq:0123_Kramers_probs_eval}
\end{align}
in which $p_{\bar{9}}$ corresponds roughly to the empirical mixing
coefficient $c_{\rm eff}$ in Eq.~(\ref{eq:eff_mixture_coeffs}).  The
Kramers formula captures the larger weight on ${\left< {\Psi}_Y \! 
\left( \rn \right) \right>}^{\left( 1 \right)}$, but over-estimates
the admixture of ${\left< {\Psi}_Y \!  \left( \rn \right) \right>}^{\left( 3
\right)}$ by about a factor of three.

This CRN was also studied by Anderson \textit{et
  al.}~\cite{Anderson:Lyapunov:15}.  Using the fact that the steady
state probability $\rho_{\rm ss}$ is known, they showed that the
non-equilibrium potential, defined here as $-\log(\rho_{\rm ss})$,
converges to the Lyapunov function for the corresponding deterministic
dynamics, in an appropriate scaling limit.  The resulting Eq.~(38)
in~\cite{Anderson:Lyapunov:15} is our
Eq.~(\ref{eq:0123_Kramers_probs_form}) obtained as a large-deviation
function.

\subsection{Two species, cross-catalysis, and loss of factorability}
\label{sec:two_species_bistable}

The final model we will develop shows the loss, for $\delta > 0$, of
the factorability which characterizes the steady states of
deficiency-0 CRNs under the ACK theorem.  We retain the properties
already developed, of deviation from Poisson statistics, and the
capacity for multistability, by simply changing the autocatalytic
feedback in the model of Fig.~\ref{fig:g_0123_unconn} to a
\textit{cross-catalytic} feedback between two symmetric chemical
species.

\begin{figure}[ht]
  \begin{center} 
  \includegraphics[scale=0.6]{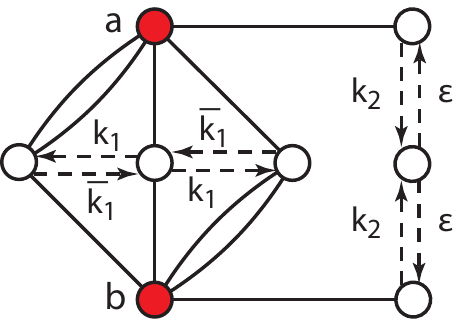}
  \caption{
  This CRN uses two species in a cross-catalytic configuration to
  produce the same potential for bistability as the network of
  Fig.~\ref{fig:g_0123_unconn} produces through 1-species
  autocatalysis.  In this CRN $\delta = 2$, and the scaling behavior
  of each species separately is similar in many respects to 1-species
  scaling of the network from Fig.~\ref{fig:g_0123_unconn}.
    \label{fig:two_species_bistable} 
  }
  \end{center}
\end{figure}

The resulting multistable network, for two species ${\rm A}$ and ${\rm
B}$, is shown in Fig.~\ref{fig:two_species_bistable}, and its reaction
scheme is given by 
\begin{align}
  \varnothing
& \xrightleftharpoons[k_2]{\epsilon}
  {\rm A} & 
  {\rm A} + {\rm B} 
& \xrightleftharpoons[{\bar{k}}_1]{k_1}
  2 {\rm A} + {\rm B} 
\nonumber \\ 
  \varnothing
& \xrightleftharpoons[k_2]{\epsilon}
  {\rm B} & 
  {\rm B} + {\rm A} 
& \xrightleftharpoons[{\bar{k}}_1]{k_1}
  2 {\rm B} + {\rm A} .
\label{eq:cubic_2spec_scheme}
\end{align}
The (now vector-valued) rate equation takes the form 
\begin{align}
  \frac{\partial n_a}{\partial \tau}
& = 
  \epsilon - 
  k_2 n_a + 
  n_b 
  \left( 
    k_1 n_a - 
    {\bar{k}}_1 n_a^2 
  \right) 
\nonumber \\
  \frac{\partial n_b}{\partial \tau}
& = 
  \epsilon - 
  k_2 n_b + 
  n_a 
  \left( 
    k_1 n_b - 
    {\bar{k}}_1 n_b^2 
  \right) . 
\label{eq:rate_ab_cross}
\end{align}

The cross-catalytic CRN from Fig.~\ref{fig:two_species_bistable}
has Liouville operator
\begin{align}
  \mathcal{L} 
& = 
  \left( 1 - a^{\dagger} \right) 
  \left[
    \left( 
      \epsilon - 
      k_2 a 
    \right) + 
    \left( b^{\dagger} \! b \right) 
    \left( a^{\dagger} \! a \right) 
    \left( 
      k_1  - 
      {\bar{k}}_1 a 
    \right) 
  \right] 
\nonumber \\
& \phantom{=} 
  \mbox{} + 
  \left( 1 - b^{\dagger} \right) 
  \left[
    \left( 
      \epsilon - 
      k_2 b 
    \right) + 
    \left( a^{\dagger} \! a \right) 
    \left( b^{\dagger} \! b \right) 
    \left( 
      k_1  - 
      {\bar{k}}_1 b 
    \right) 
  \right] . 
\label{eq:L_two_species_bistable}
\end{align}
We introduce pairs of raising and lowering operators $\left(
a^{\dagger} , a \right)$, $\left( b^{\dagger} , b \right)$, for the
species ${\rm A}$ and ${\rm B}$ respectively.  The convention we adopt
for ordering the components of the (somewhat complicated) vectors
$\psi \! \left( a , b \right)$ and ${\psi}^{\dagger} \! \left(
a^{\dagger} , b^{\dagger} \right)$ is given in
Eq.~(\ref{eq:psi_vecs_two_species_bistable}), 
and the corresponding stoichiometric decomposition of the Liouville
operator is given in Eq.~(\ref{eq:L_stoich_two_species_bistable}).

\subsubsection{Mean-field solutions and scaling regimes}

The truncated factorials $Y_a^{\underline{j_a}}$ and
$Y_b^{\underline{j_b}}$ that govern the scaling regimes in the
steady-state moment hierarchy are provided in
Eq.~(\ref{eq:proj_vecs_two_species_bistable}).  The lowest-order terms
in the descaled form~(\ref{eq:Glauber_moment_fact_hats_alt}) of the
moment equation are
\begin{equation}
  \left( 
  \frac{k_a}{{\xi}_a}
  Y_a 
  e^{
    \left( k_a - 1 \right) \, \partial / \partial Y_a 
  } + 
  \frac{k_b}{{\xi}_b}
  Y_b 
  e^{
    \left( k_b - 1 \right) \, \partial / \partial Y_b 
  }
  \right) 
  \hat{\mathbb{A}}
  \left< 
    {\hat{\Psi}}_{Y} \! \left( \rn \right) 
  \right>
\label{eq:two_species_low_moments}
\end{equation}

The two vanishing conditions for $Y_a \hat{\mathbb{A}}$ and $Y_b
\hat{\mathbb{A}}$ (coming from the projectors for the two $s$-flows of
this network) are
\begin{align}
  \left( 
    \epsilon - k_2 {\xi}_a
  \right) + 
  {\xi}_a {\xi}_b
  \left( 
    k_1 - {\bar{k}}_1 {\xi}_a
  \right) 
& = 
  0 
\nonumber \\ 
  \left( 
    \epsilon - k_2 {\xi}_b
  \right) + 
  {\xi}_a {\xi}_b
  \left( 
    k_1 - {\bar{k}}_1 {\xi}_b
  \right) 
& = 
  0 . 
\label{eq:two_species_low_moments_van}
\end{align}
These are solved at ${\xi}_a = {\xi}_b = \xi$, with $\xi$ again
satisfying the mean-field equation~(\ref{eq:0123_y1A_vanishing}) for
the single-species model of Sec.~\ref{sec:g_0123_unconn}.

In the other asymptote, the highest-order terms in
Eq.~(\ref{eq:Glauber_moment_fact_hats_alt}) that are not identically
zero in the component-wise product of $Y_a^{\underline{j_a}} \cdot
Y_b^{\underline{j_b}}$ are
\begin{widetext}
\begin{equation}
  \frac{
    k_a k_b
  }{
    2 {\xi}_a {\xi}_b
  }
  \left( 
  \frac{k_a-1}{{\xi}_a}
  Y_a^{\underline{2}} \cdot Y_b 
  e^{- \partial / \partial Y_a} + 
  \frac{k_b-1}{{\xi}_b}
  Y_a \cdot Y_b^{\underline{2}} 
  e^{- \partial / \partial Y_b}
  \right) 
  \hat{\mathbb{A}}
  e^{
    \left( k_a - 1 \right) \, \partial / \partial Y_a + 
    \left( k_b - 1 \right) \, \partial / \partial Y_b 
  } 
  \left< 
    {\hat{\Psi}}_{Y} \! \left( \rn \right) 
  \right>
\label{eq:two_species_high_moments}
\end{equation}
\end{widetext}
The vanishing conditions for these two projectors are
\begin{align}
  {\xi}_a {\xi}_b
  \left( 
    k_1 - {\bar{k}}_1 {\xi}_a
  \right) 
& = 
  0 
\nonumber \\ 
  {\xi}_a {\xi}_b
  \left( 
    k_1 - {\bar{k}}_1 {\xi}_b
  \right) 
& = 
  0 
\label{eq:two_species_high_moments_van}
\end{align}
They are again solved at ${\xi}_a = {\xi}_b = \xi$ but now with $ \xi
= k_1 / {\bar{k}}_1 \equiv K_1$, reproducing the large-$k$ asymptotic
condition from the single-species model of
Sec.~\ref{sec:g_0123_unconn}.

This two-species case may again be solved for mixed moments in the
neighborhood of the diagonal $k_a = k_b$, writing coupled recursion
relations for the ratios of the factorial moments, as shown in
\cite{Krishnamurthy:CRN_moments:17}. In the following section, we
illustrate an alternate solution method using the asymptotic
expansions that we have developed in the earlier sections.

\subsubsection{Polynomial expansion of a solution for the moment
equation in a neighborhood of the diagonal $k_a = k_b$}
\label{sec:two_spec_polyn_exp}

We now illustrate how the representation $\Lambda$ of the generator
for the stochastic process, acting similarly to a Laplacian on the
two-dimensional lattice of moments ${\Phi}_{\left( k_a , k_b
\right)}$, can be approximately solved in a neighborhood of the
diagonal $k_a = k_b$.  The method of solution is to use the symmetry
of the recursion equations under $k_a \leftrightarrow k_b$ to expand
solutions in even powers of $\left( k_a - k_b \right)$, with
coefficient functions of $\left( k_a + k_b \right)$ solved by
asymptotic expansion in a manner similar to that used in the 1-species
models of Sec.~\ref{sec:g_123_conn} and Sec.~\ref{sec:g_0123_unconn}.

We do not have a proof that the radius of convergence of these
solutions covers the entire lattice of $k$ values, but comparisons to
Gillespie simulation show good agreement in neighborhoods of the
diagonal, suggesting that the asymptotic boundary conditions we use
are consistent with those of full solutions.  Existence of approximate
solutions with this form shows a strong breaking of factorability from
the product-Poisson form that is a characteristic of the ACK solution
for deficiency-zero networks.

In the following solutions, two combinations of the rate constants
that will appear repeatedly are given a short-hand:\footnote{The second
term, $\eta$, has already appeared in Eq.~(\ref{eq:c_erg_def}).}
\begin{align}
  \omega 
& \equiv
  \epsilon / k_1
& \eta 
& \equiv 
  k_2 / {\bar{k}}_1 - \omega .
\label{eq:omega_eta_def}
\end{align}

As in the asymptotic solutions for the 1-species models, we begin by
recognizing that the large-$k$ asymptotic form is dominated by the
scaling of the projection operator that is non-zero for the largest
value of $j$ in the sum~(\ref{eq:Glauber_moment_fact_hats_alt}).  This
is the projector given in Eq.~(\ref{eq:two_species_high_moments}).
Therefore we define the descaled moment operator by ${\Phi}_{\left(
k_a, k_b \right)} \equiv K_1^{ k_a + k_b } {\hat{\Phi}}_{\left( k_a,
k_b \right)}$, and look for solutions in the form
\begin{align}
  {\hat{\Phi}}_{\left( k_a, k_b \right)} = 
  \mathcal{N}
  \left[ 
    1 + 
    \eta 
    {\varphi}_{\left( k_a , k_b \right)} 
  \right] , 
\label{eq:Phi_twospec_series_form}
\end{align}
where ${\varphi}_{\left( k_a , k_b \right)} \rightarrow 0$ at large
$k_a$ or $k_b$.\footnote{We justify this assumed scaling by reference
to the large-$k$ limit~(\ref{eq:0123_large_k_sol}) from the similar
1-species model, because the orders of catalysis are more similar to
that case than to the model of Sec.~\ref{sec:g_123_conn} leading to
the soft (logarithmic) divergence of Eq.~(\ref{eq:123_large_k_sol}).}

We introduce diagonal and transverse variables, written as
functions of the vector argument $k$: 
\begin{align}
  \kappa \! \left( k \right)
& \equiv 
  k_a + k_b 
\nonumber \\
  {
    q \! \left( k \right) 
  }^2 
& = 
  {
    \left( k_a - k_b \right)
  }^2 . 
\label{eq:two_spec_kq_def}
\end{align}
In matrix multiplications below, we will often use $\kappa$ and $q^2$
as function names, with the argument $k$ which is the index of
summation suppressed as in usual matrix notation.  We look for
solutions to $\varphi$ in the form of power series,
\begin{equation}
  \varphi = 
  \sum_{\alpha = 0}^{\infty}
  {\varphi}^{\left( \alpha \right)}_{\kappa}
  q^{2 \alpha} . 
\label{eq:varphi_q2_expn}
\end{equation}
Each term is to be chosen so that ${\varphi}^{\left( \alpha
\right)}_{\kappa} \rightarrow 0$ as $\kappa \rightarrow \infty$.  The
functions ${\varphi}^{\left( \alpha \right)}_{\kappa}$ obey recursion
relations similar to those for an infinite sequence of 1-dimensional
moment hierarchies labeled by $\alpha$, except that the vectors in the
sequence are coupled across values of $\alpha$.  Within the solution
for each ${\varphi}^{\left( \alpha \right)}_{\kappa}$, we may treat
$\kappa$ itself as the discrete index of the recursion.  Here as in
the 1-species models, the descaled recursion relation suggests
leading-order asymptotics for ${\varphi}^{\left( \alpha
\right)}_{\kappa}$ in powers of $1/\kappa$, which may be used to seed
numerical solutions.

The conversion from the original lattice ${\hat{\Phi}}_k$ to the
sequence of vectors ${\varphi}^{\left( \alpha \right)}_{\kappa}$ leads
to the following approximation procedure to solve for steady states:
The steady state condition from
Eq.~(\ref{eq:Glauber_moment_fact_hats_alt}) is
\begin{equation}
  0 = 
  \hat{\Lambda}
  \left[ 
    \underline{1} + 
    \eta \varphi 
  \right] = 
  \hat{\Lambda} \underline{1} + 
  \eta 
  \hat{\Lambda} 
  \sum_{\alpha = 0}^{\infty}
  {\varphi}^{\left( \alpha \right)}_{\kappa}
  q^{2 \alpha}, 
\label{eq:two_spec_ss_cond_L}
\end{equation}
where $\underline{1} \equiv \left[ 1 \right] {\left[ 1 \right]}^T$ is
the dyadic matrix of all 1s.  We introduce a zeroth order source term
$s^{\left( 0 \right)}_{\kappa}$ defined by
\begin{equation}
  \hat{\Lambda} \underline{1} = 
  - \eta 
  \frac{\kappa}{K_1} \equiv 
  - \eta s^{\left( 0 \right)}_{\kappa} 
\label{eq:two_spec_s0_def}
\end{equation}
so the steady-state condition is equivalent to the series solution of
an inhomogeneous Laplacian equation
\begin{equation}
  \hat{\Lambda}
  \sum_{\alpha = 0}^{\infty}
  {\varphi}^{\left( \alpha \right)}_{\kappa}
  q^{2 \alpha} = 
  s^{\left( 0 \right)}_{\kappa} , 
\label{eq:varphi_from_s0}
\end{equation}
in which $\hat{\Lambda}$ serves as Laplacian and $s^{\left( 0
\right)}_k$ is the source for the inhomogeneous solution.

\begin{figure*}[ht]
  \begin{center} 
  \includegraphics[scale=0.75]{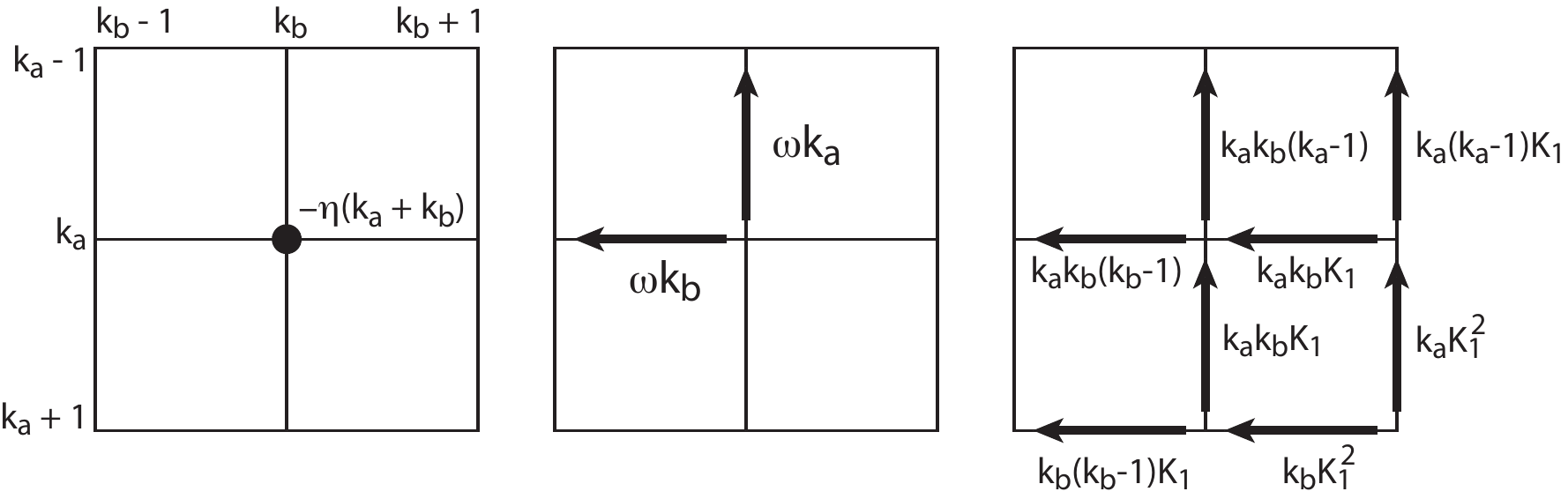} 
  \caption{
  Graphic representation of the action of the generator
  $\hat{\Lambda}$ given by Eq~(\ref{eq:Glauber_moment_fact_hats_alt})
  for the CRN of Fig.~\ref{fig:two_species_bistable}.  The grid
  represents adjacent values of the index pair $\left( k_a , k_b
  \right)$, $\left( k_a \pm 1 , k_b \right)$, $\left( k_a , k_b \pm 1
  \right)$, and $\left( k_a \pm 1 , k_b \pm 1 \right)$, as indicated
  by axis labels in the first panel.  The solid dot in the first panel
  represents multiplication by the moment at $\left( k_a , k_b
  \right)$.  Solid arrows indicate multiplication of moments by the
  projection vector $\left[ \; \mbox{1} \; \; \mbox{-1} \; \right]$,
  acting on the moments at the tip and tail of the arrow.  The labels
  on the lines indicate the function of parameters and $k$ values that
  multiplies each such projector.  Thus the first panel is simply the
  term $-\eta \left( k_a + k_b \right) {\hat{\Phi}}_{\left( k_a , k_b
  \right)}$, etc.  The first and second panels, and the lower-right
  pair of terms in the third panel, come from the terms at order $j_a
  + j_b = 1$ in Eq.~(\ref{eq:Glauber_moment_fact_hats_alt}), given in
  Eq.~(\ref{eq:two_species_low_moments}).  The terms running
  anti-diagonally through the center in the third panel come from the
  terms at order $j_a + j_b = 2$, and the terms in the upper-left
  corner of the third panel come from the terms at order $j_a + j_b =
  3$, shown in Eq.~(\ref{eq:two_species_high_moments}).  The large-$k$
  asymptotic Poisson $\hat{\Phi} \rightarrow \mathcal{N}
  \underline{1}$ is annihilated identically by both panels with
  arrows.
  \label{fig:two_spec_L_diagram} 
  } 
  \end{center}
\end{figure*}

The form of $\hat{\Lambda}$ can be described graphically in terms of
difference operators acting across adjacent positions on the lattice
of $k$ values, as shown in Fig.~\ref{fig:two_spec_L_diagram}.
$\hat{\Lambda}$ acts non-trivially on $q^{2 \alpha}$ as well as on
${\varphi}^{\left( \alpha \right)}_{\kappa}$, so
Eq.~(\ref{eq:varphi_from_s0}) induces connections across orders in
$\alpha$, and we relegate the details of a solution by successive
approximations to App.~\ref{sec:two_spec_bistable_polyn_app}.

\subsubsection{Properties of steady states in the 2-species model}

\begin{figure*}[ht]
  \begin{center} 
  \includegraphics[scale=0.29]{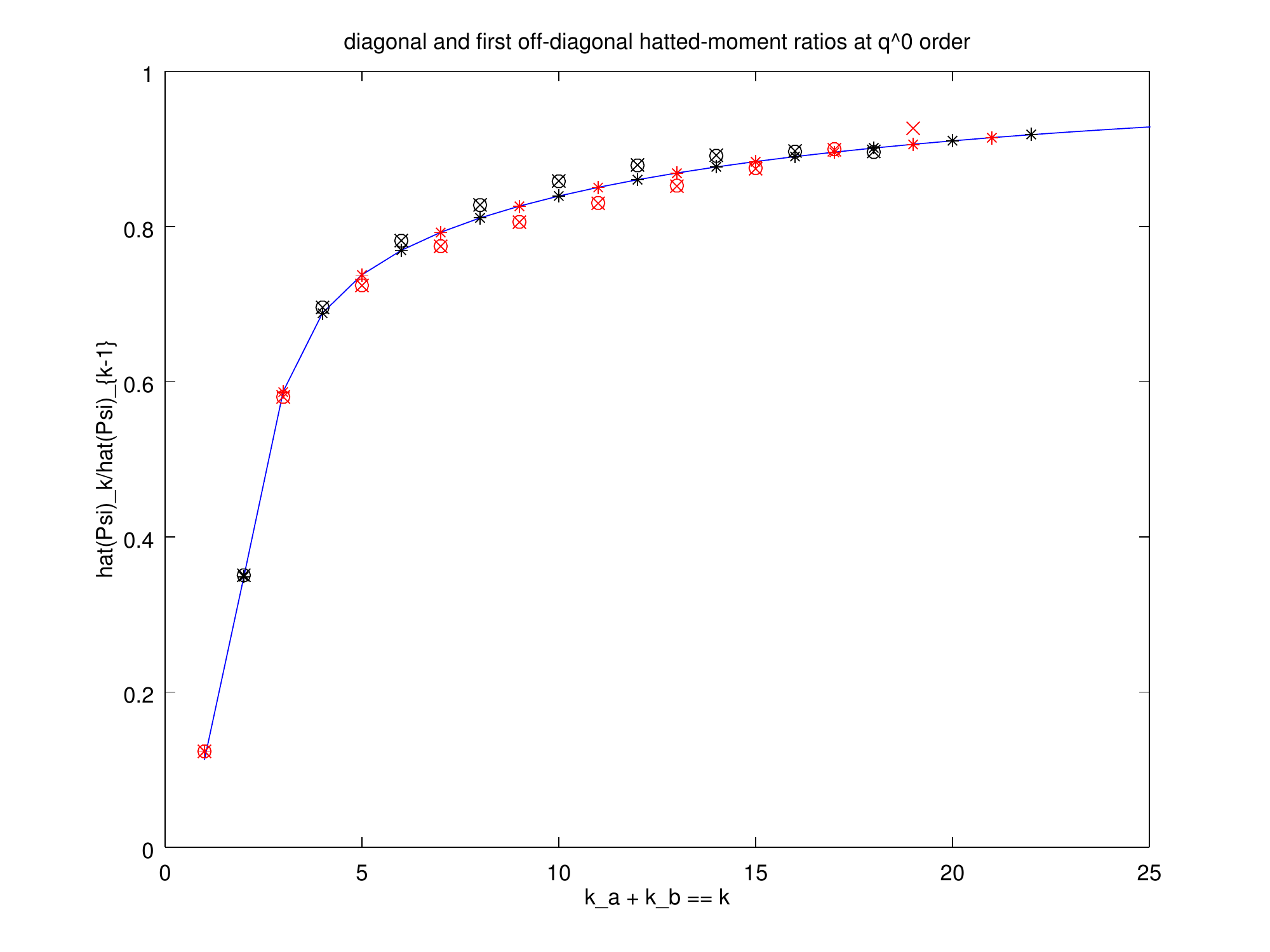} 
  \includegraphics[scale=0.29]{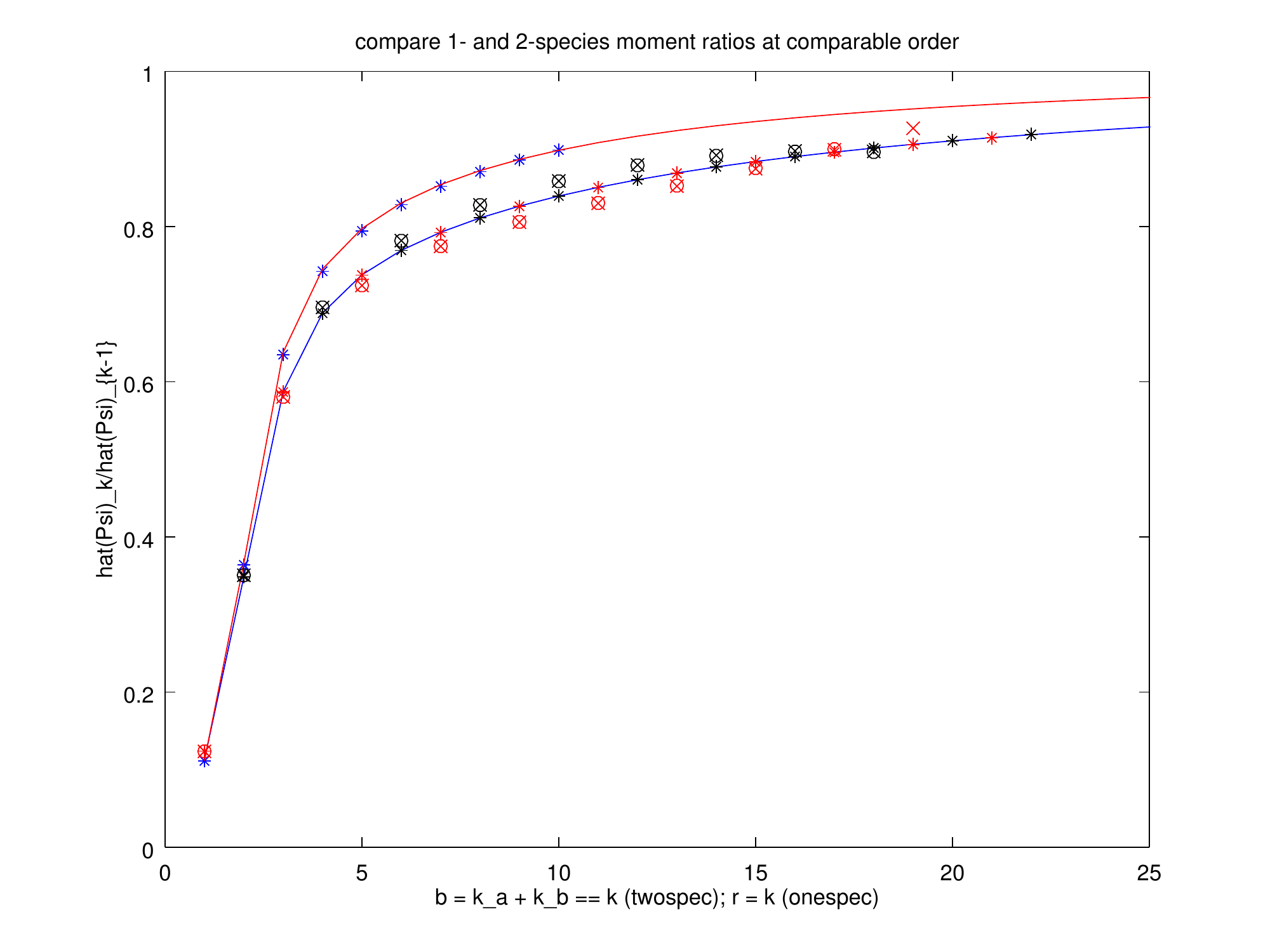} 
\includegraphics[scale=0.29]{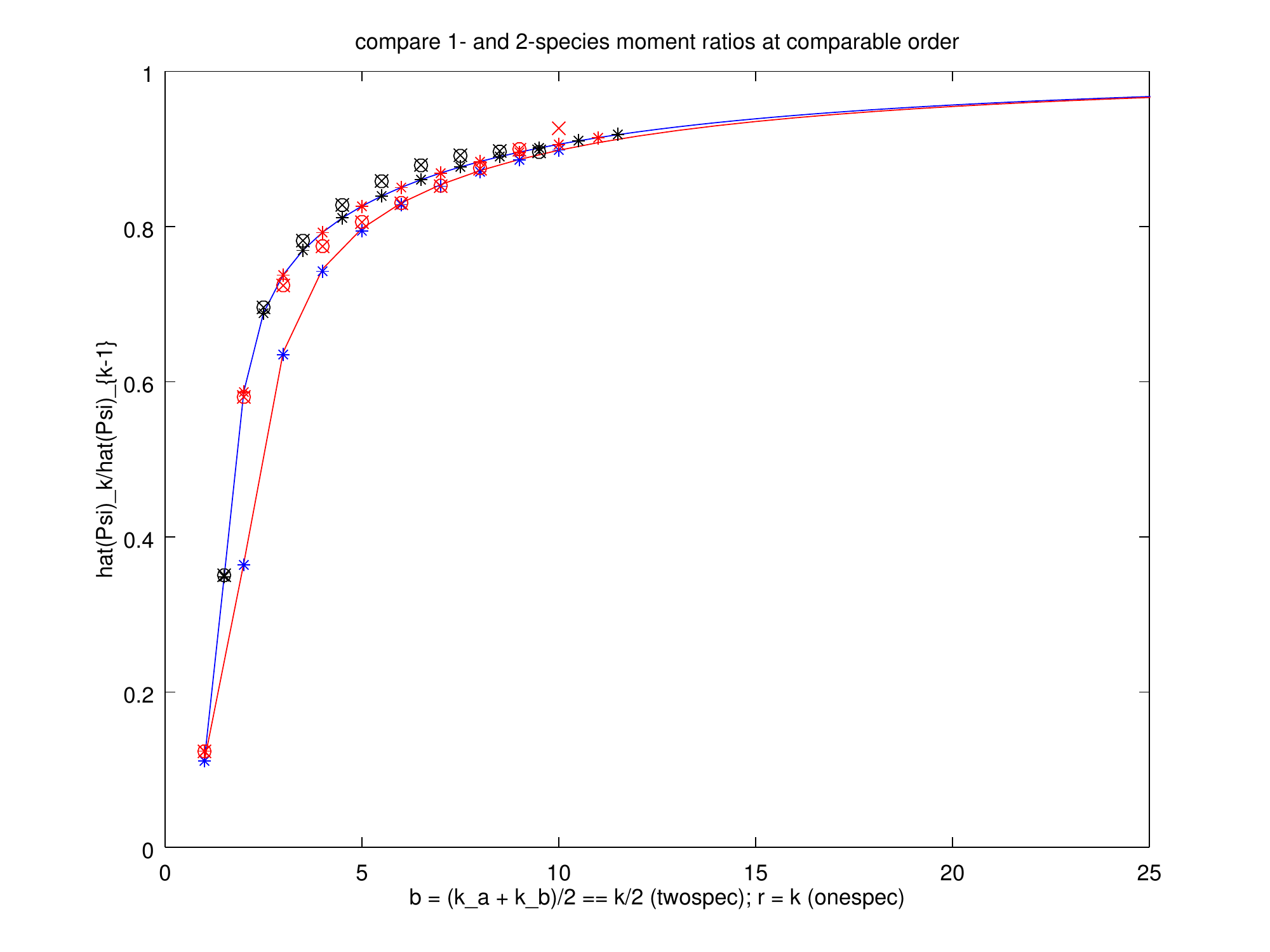} 
  \caption{
  Comparison of Gillespie simulations (asterisks) for the CRN model of
  Fig.~\ref{fig:two_species_bistable}, to the leading solution
  ${\varphi}^{\left( 0 \right)}_{\kappa}$ (blue curve) from
  Eq.~(\ref{eq:varphi_q2_expn}).  First panel shows ratios
  ${\hat{\Phi}}_{\left( k_a , k_b \right)} / {\hat{\Phi}}_{\left( k_a
  -1 , k_b \right)} = {\hat{\Phi}}_{\left( k_a , k_b \right)} /
  {\hat{\Phi}}_{\left( k_a , k_b -1 \right)}$ for $k_a = k_b$ (black
  crosses), and ${\hat{\Phi}}_{\left( k_a -1 , k_b \right)} /
  {\hat{\Phi}}_{\left( k_a -1 , k_b -1 \right)} = {\hat{\Phi}}_{\left(
  k_a , k_b -1 \right)} / {\hat{\Phi}}_{\left( k_a -1 , k_b -1
  \right)}$ for $k_a = k_b$ (red crosses).  Second and third panels show
  comparisons between these adjacent-moment ratios in the 2-species
  model and moment ratios ${\hat{\Phi}}_k / {\hat{\Phi}}_{k-1}$ in the
  1-species model of Fig.~\ref{fig:g_0123_unconn} (red curve) that has an
  equivalent MFT solution.  Second panel plots $k_a + k_b$ directly
  against one-species $k$, showing that for the lowest moments total
  particle number controls similar moment values in the two models.
  Third panel plots $\left( k_a + k_b \right) / 2$ (equivalently,
  $k_a$ or $k_b$) against 1-species $k$, showing that the transition
  to scaling dominated by the autocatalytic reactions is governed
  independently by $k_a$ and $k_b$.
  \label{fig:mom_rats_dense_q0only} 
  } 
  \end{center}
\end{figure*}

The major features of the steady-state solution in this model, which
we have verified against Gillespie simulations, are the following:

\begin{trivlist}

\item \textbf{Order of terms versus $\alpha$:} The na{\"{\i}}ve
scaling dimensions implied for ${\varphi}^{\left( \alpha
\right)}_{\kappa}$ by Eq.~(\ref{eq:Glauber_moment_fact_hats_alt})
suggest that these functions should decay at large $\kappa$ with
increasing powers of $1/\kappa$.  Numerically, this appears to be
borne out, with indeed the entire series ${\varphi}^{\left( \alpha
\right)}_{\kappa}$ decreasing in magnitude with increased $\alpha$.  
In addition to the on-diagonal terms (where $q^2 \equiv 0$), which are
defined entirely in terms of ${\varphi}^{\left( 0 \right)}_{\kappa}$,
terms adjacent to the diagonal, which should be dominated by
${\varphi}^{\left( \alpha \right)}_{\kappa}$ at low orders in
$\alpha$, are well approximated by the solution ${\varphi}^{\left( 0
\right)}_{\kappa}$ across the whole range of $k_a = k_b$.

\item \textbf{Scaling of finite-order approximations along rays of
    $\left| q \right| / \kappa$:}
  The measure of error -- non-zero values of $\partial {\Phi}_k /
  \partial \tau$ -- appears roughly constant along rays of fixed
  $\left| k_a - k_b \right| / \left( k_a - k_b \right) \equiv \left| q
  \right| / \kappa$ at finite orders of approximation in
  ${\varphi}^{\left( \alpha \right)}_{\kappa}$.  Stabilizing the
  asymptotic expansion independently at each order of
  ${\varphi}^{\left( \alpha \right)}_{\kappa}$ becomes increasingly
  difficult as $\alpha$ increases, due to cross-level feedback and the
  successive-approximation algorithm we use for solution.  Thus we
  obtain an approximate solution only through order ${\varphi}^{\left(
      5 \right)}_{\kappa}$.

\item \textbf{Comparison of the 2-species cross-catalytic and
1-species autocatalytic models:} The ratios of adjacent moments in the
value $k_a + k_b$ -- which require at minimum comparing on-diagonal
and first-off-diagonal moments -- are shown in
Fig.~\ref{fig:mom_rats_dense_q0only} and compared to the corresponding
sequence of ratios derived from the moment solutions in
Fig.~\ref{fig:0123_mid_soln}.  We find that the mean value $\left<
{\rn}_a \right> \equiv \left< {\rn}_b \right>$ in the 2-species model
is very close to the mean value $\left< \rn \right>$ from the
1-species model, as suggested by the equivalence of their mean-field
forms, even though both models differ significantly from the
MFT-approximation, which is the solution $\xi$ to
Eq.~(\ref{eq:0123_y1A_vanishing}).  Moreover, the second moments
$\left< {\rn}_a {\rn}_b \right>$ remain close to the 1-species
expectation $\left< \rn \left( \rn -1 \right) \right>$, and again
different in both cases from the MFT prediction.  At higher $k$, a
different behavior is seen: the transition to scaling dominated by the
term~(\ref{eq:two_species_high_moments}) is governed in the 2-species
model by $k_a$ and $k_b$ comparable to 1-species $k$, and not by the
sum $k_a + k_b$.  This is expected by comparing the forms of the two
Liouville operators~(\ref{eq:L_0123})
and~(\ref{eq:L_two_species_bistable}).

\end{trivlist}

\subsubsection{Breaking of the factorability of the ACK theorem
through cross-catalysis} 
\label{sec:two_spec_factor}

The solution scheme defined in Sec.~\ref{sec:two_spec_polyn_exp} and
worked out in App.~\ref{sec:upward_prop_sol} suggests that the moment
hierarchy near the diagonal is well approximated (at least at large
$k$) by a function of $k_a + k_b$, which is a strong deviation from
the factorability that would be produced by the ACK theorem for a
deficiency-0 network.  To study the failure of factorability more
directly than through the numerical approximation scheme of
App.~\ref{sec:upward_prop_sol}, we may alternatively approximate the
large-$k$ behavior of the moment hierarchy by a sum of products of
powers of $1 / k_a$ and $1 / k_b$, solved fully for $\hat{\Lambda}
\hat{\Phi} = 0$ order-by-order in $1/k$.  The two expansions do not
have the same asymptotics along the boundaries $k_a = 0$ and $k_b =
0$, but they can be made to satisfy the same boundedness criteria at
large $k$ in a neighborhood of $k_a = k_b$.

An expansion of the solution to $\hat{\Lambda} \hat{\Phi} = 0$ to
second order in $1 / k$ is given by 
\begin{widetext}
\begin{align}
  {
    \Phi
  }_{
    \left( k_a , k_b \right) 
  }
& \approx 
  \mathcal{N}
  K_1^{
    k_a + k_b
  }
  \left[
    1 + 
    \frac{
      \eta
    }{
      k_a + 
      \left( K_1 - \eta \right) / 2
    } + 
    \frac{
      \eta
    }{
      k_b + 
      \left( K_1 - \eta \right) / 2
    } + 
    \frac{
      \eta 
      \left( \eta - K_1 \right) 
    }{
      \left[ 
        k_a + 
        \left( K_1 - \eta \right) / 2
      \right]
      \left[ 
        k_b + 
        \left( K_1 - \eta \right) / 2
      \right]
    } 
  \right] 
\nonumber \\
& \approx 
  \mathcal{N}
  K_1^{
    k_a + k_b
  }
  \left[
    \left( 
      1 + 
      \frac{\eta}{k_a + K_1/2} + 
      \frac{\eta^2 / 2}{{\left( k_a + K_1/2 \right)}^2} 
    \right) 
    \left( 
      1 + 
      \frac{\eta}{k_b + K_1/2} + 
      \frac{\eta^2 / 2}{{\left( k_b + K_1/2 \right)}^2} 
    \right) 
  \right.
\nonumber \\
& 
  \qquad \qquad \qquad \qquad 
  \mbox{} - 
  \left.
    \frac{
      \eta K_1 
    }{
      \left( k_a + K_1/2 \right) 
      \left( k_b + K_1/2 \right) 
    } 
  \right] , 
\label{eq:two_spec_large_k_sol}
\end{align}
\end{widetext}
which we have checked numerically cancels the error term $\partial
\hat{\Phi} / \partial \tau$ to the correct order of $1/k$.\footnote{
  Eq.~(\ref{eq:two_spec_large_k_sol}) may be compared to
  Eq.~(\ref{eq:0123_large_k_sol}) for the one-species model, and also
  to the leading-order scaling estimate for the term
  ${\varphi}^{\left( 0 \right)}_{\kappa}$ from
  Eq.~(\ref{eq:init_phi_asymptotics}). 
  Along the diagonal $k_a = k_b$,
  Eq.~(\ref{eq:two_spec_large_k_sol}) becomes
  \begin{displaymath}
  {
    \Phi
  }_{
    \left( k_a , k_b \right) 
  } \approx 
  \mathcal{N}
  K_1^{
    k_a + k_b
  }
  \left[
    \frac{
      4 \eta \left( k_a + k_b \right)
    }{
      {
        \left( 
          k_a + k_b + K_1 - \eta 
        \right)
      }^2
    } . 
  \right]    
  \end{displaymath}
  The leading behavior differs from
  Eq.~(\ref{eq:init_phi_asymptotics}) by a factor $\left( 4/3 \right)$
  multiplying $\left( K_1 - \eta \right)$, which is consistent with
  the fact that the scaling solutions in
  App.~\ref{sec:upward_prop_sol} only propagate the effects of
  $\hat{\Lambda}$ upward in a hierarchy of powers; feedbacks down the
  hierarchy are absorbed in higher-order correction terms that have
  the same large-$k$ order as corrections in the multiplier of $\left(
    K_1 - \eta \right)$.}  The first departure from factorability
occurs in the second-order term with numerator $-\eta K_1$, showing
where the $\delta$-flow contributions create correlated fluctuations
that would be ruled out in a ($\delta = 0$)-network.  

\subsubsection{Generalizing to a larger number of species} 
\label{sec:outlook}

Our constructions apply to CRNs with arbitrary numbers of species, but
the foregoing models show how the character of solution methods
changes with increasing numbers.  Because the
generator~(\ref{eq:Glauber_moment_fact_prod}) has finitely-many terms
for any finite CRN, for one-species problems, the number of
undetermined boundary data that must be sampled to search for stable
asymptotic expansions is always finite.  Moreover, the large-$k$ limit
may be extended to improve the precision of approximations from
coarser seed functions.  In two or more dimensions, unknown boundary
data can exist along all surfaces of co-dimension 1 or more, in which
one or more $k_p = 0$.  If an asymptotic bounding surface is moved
outward for the nonzero values of $k_p$, new unknown values are added
to the set that must be sampled along small-$k_p$ boundaries.
Therefore increasingly much of the information in a solution must come
from boundary-condition data, compared to the constraint in the scalar
condition $\Lambda \Phi = 0$.

Other problems of convex analysis, analogous to the Feinberg
deficiency-zero argument, are also left as questions for future work.
Is there a systematic way to represent the number of distinct scaling
regions controlled by terms in the
sum~(\ref{eq:Glauber_moment_fact_prod}) over $j$?  Do large-$k$
asymptotic conditions on the vector $\xi$ of coherent-state parameters
always possess unique solutions?  When are they underdetermined, and
in these cases do the solutions from $s$-flow conditions extend
outward indefinitely?

Despite leaving several detailed questions to be addressed, we
emphasize that the finite rank of the operator $\Lambda$ reduces the
solving of a moment hierarchy to all orders, to a problem of equal
complexity to solving a Laplacian diffusion equation, generally with
beyond-nearest-neighbor couplings.  This is a simpler and less-costly
problem than direct simulation, especially for high-order moments in
systems with large particle numbers.

\section{Conclusions}

The new observations and results of this work may be grouped under the
following four main topics: 
\begin{trivlist}

\item \textbf{A shift in emphasis from topology to dynamics:} In this
  article we have bypassed the use of deficiency as a topological
  index to categorize networks, and focused instead on the contrast
  between mean-regressing and non-mean-regressing flows on the complex
  graph, which is the dynamical property that causes deficiency to be
  important.  The dynamical distinction, which we express in the
  stoichiometric representation~(\ref{eq:Ls_reps}) of the
  stochastic-process generator, continues to be definable in terms of
  the images and kernels of $\mathbb{A}$ and $Y \mathbb{A}$, even in
  ($\delta > 0$)-networks where it cannot be associated with a
  deficiency-0 sub-network.  As shown in
  Sec.~\ref{sec:leading_Poiss_arg}, the unique role of the
  mean-regressing flows as the determinants of the first-moment
  conditions, which persists to all orders in deficiency-0 networks,
  persists in a more limited form as the leading term in a
  $1/n$-expansion for the determination of moments in more general
  networks.

\item \textbf{The Doi operator algebra, Laplace transforms, and an
expanded role for Poisson basis distributions:} The Doi operator
algebra used to express the generators of the stochastic process is
the tool that allows us to associate the $s$-flows and $\delta$-flows
in the stoichiometric representation with corresponding
product-Poisson distributions, the origin and meaning of which are the
same as those of the unique steady-state distributions in the
Anderson-Craciun-Kurtz theorem.  In this way not only the linear
algebra of first moments in the stoichiometric subspace, but the
Poisson family of distributions as basis functions, extends directly
from deficiency-0 to deficiency-nonzero cases.  This simplification
and clarification results from working with the Laplace transform and
the Liouville operator: the elementary projection operators in terms
of which $\mathcal{L}$ naturally decomposes, which annihilate
particular Poisson distributions, describe collective motions that
recursively relate all orders in the moment hierarchy.

\item \textbf{The manifestation of fundamental symmetries in the
    generator of the stochastic process:} The Doi operator algebra
  also exposes symmetries of the generators~(\ref{eq:L_psi_from_A})
  for stochastic CRNs that are obscured in the combination of index
  shifts and number-dependence of rates in the master
  equation~(\ref{eq:T_psi_from_A}), and masked entirely in the
  asymmetric form of the Mean-Field mass-action rate
  equations~(\ref{eq:Glauber_1st_moment},\ref{eq:s_projectors_k1}).
  We use the simplifications this formalism affords to both derive the
  moment hierarchy as well as develop approximations to solve them.

\item \textbf{Locality of scaling regions and connection to the
    Poisson approximations:} The finiteness of the generator acting on
  the moment hierarchy is the feature that allows \textit{scaling
    regimes} to be defined locally in different asymptotic ranges of
  moments, and that makes Poisson-form basis distributions good
  approximations to the moment recursion relations in such regions.
  Despite the fundamental underlying complexity of CRNs -- remember
  that the search for non-locally defined topological properties such
  as shortest paths or feedback cycles can be
  NP-hard~\cite{Andersen:NP_autocat:12} -- the convergence of both
  high- and low-$k$ asymptotic expansions toward the matching region
  buffers the strength with which different regions are coupled, and
  allows the convergence toward the Poisson basis elements to be
  locally governed.  n this way, Poisson distributions which are among
  the lowest-information distributions, serve as a basis for the
  evaluation of the moments in systems with the potential for very
  high information capacity.  As a by-product, we also obtain a
  systematic approach to moment closure, which does not have any of
  the problems of ad-hocness and unphysical
  results~\cite{Schnoerr:Survey:15} prevalent in existing schemes.

\end{trivlist}

\subsection*{Acknowledgments}

DES thanks the Physics Department at Stockholm University for support
during visits in 2014 and 2016 when the bulk of this work was carried
out.  We also acknowledge Dan Rockmore, Scott Pauls, and Greg Leibon
for hospitality and conversations on related topics in 2010, Artur Wachtel in 2015 and Nathaniel Virgo of ELSI in 2016.

\vfill
\eject

\appendix 

\section{Supporting algebra for CRN examples}
\label{sec:app_CRN_algebra}

This appendix provides explicit forms for transfer matrices, Liouville
operators, and truncated factorials of the stoichiometric matrix, for
the CRN models in the main text.

\subsection{Forms of the transfer matrices appearing in master
equations} 
\label{sec:transfer matrices}

Transfer matrices are given below for the indicated CRN models: 

\begin{widetext}
For the graph of Fig.~\ref{fig:g_1_and_3_to_2} the evolution equation
for the density that defines the transfer matrix has the form 
\begin{equation}
  {\dot{\rho}}_{\rn} = 
  \left[
    \left( e^{-\partial / \partial \rn} -1 \right) 
    \alpha \rn + 
    \left( e^{\partial / \partial \rn} -1 \right) 
    \beta
    \rn \left( \rn -1 \right) \left( \rn -2 \right)
  \right]
  {\rho}_{\rn} . 
\label{eq:ME_g_123_NR}
\end{equation}
This equation illustrates in the simplest form how Poisson steady
states come to be ruled out when the ACK theorem no longer applies.
The shift operator acts by single units $\rn \rightarrow \rn \pm 1$,
but the numerical factors in the rates for particle creation and
annihilation differ by second-order terms in $\rn$, which cannot be
absorbed in any Poisson distribution. 

For the weakly reversible graph of Fig.~\ref{fig:g_123_conn}, the
master equation adds a term affecting fluctuations though it preserves
the first-moment rate equation: 
\begin{equation}
  {\dot{\rho}}_{\rn} = 
  \left\{
    \left( e^{-\partial / \partial \rn} -1 \right) 
    \left[ 
      \alpha \rn + 
      \epsilon \rn \left( \rn - 1 \right)
    \right] + 
    \left( e^{\partial / \partial \rn} -1 \right) 
    \left[ 
      \epsilon \rn \left( \rn - 1 \right) + 
      \beta 
      \rn \left( \rn -1 \right) \left( \rn -2 \right)
    \right]
  \right\}
  {\rho}_{\rn} . 
\label{eq:ME_g_123_R}
\end{equation}

The master equation for the CRN of Fig.~\ref{fig:g_0123_unconn} is 
\begin{equation}
  {\dot{\rho}}_{\rn} = 
  \left\{
    \left( e^{-\partial / \partial \rn} -1 \right) 
    \left[ 
      \epsilon + 
      k_1 \rn \left( \rn - 1 \right)
    \right] + 
    \left( e^{\partial / \partial \rn} -1 \right) 
    \left[ 
      k_2 \rn + 
      {\bar{k}}_1
      \rn \left( \rn -1 \right) \left( \rn -2 \right)
    \right]
  \right\}
  {\rho}_{\rn}
\label{eq:ME_g_0123_R}
\end{equation}

For the CRN of Fig.~\ref{fig:two_species_bistable}, $\rn$ becomes a
two-component index to $\rho$, and the master equation becomes 
\begin{align}
  {\dot{\rho}}_{\rn}
& = 
  \left\{
    \left( e^{-\partial / \partial {\rn}_a} -1 \right) 
    \left[ 
      \epsilon + 
      k_1 {\rn}_b {\rn}_a
    \right] + 
    \left( e^{\partial / \partial {\rn}_a} -1 \right) 
    \left[ 
      k_2 {\rn}_a + 
      {\bar{k}}_1
      {\rn}_b {\rn}_a \left( {\rn}_a-1 \right) 
    \right] 
  \right.
\nonumber \\
& \phantom{=}
  \mbox{} + 
  \left. 
    \left( e^{-\partial / \partial {\rn}_b} -1 \right) 
    \left[ 
      \epsilon + 
      k_1 {\rn}_a {\rn}_b
    \right] + 
    \left( e^{\partial / \partial {\rn}_b} -1 \right) 
    \left[ 
      k_2 {\rn}_b + 
      {\bar{k}}_1
      {\rn}_a {\rn}_b \left( {\rn}_b-1 \right) 
    \right] 
  \right\}
  {\rho}_{\rn}
\label{eq:ME_two_species_bistable}
\end{align}

\subsection{Stoichiometric representation forms for Liouville
operators of the models}

The stoichiometric decompositions of Liouville operators not
explicitly written in the main text are provided here: 

\subsubsection{The 1-species, 4-complex model of
Sec.~\ref{sec:g_0123_unconn}} 
\label{sec:g_0123_unconn_app}

The diagonalized form of the Liouville operator~(\ref{eq:L_0123}),
corresponding to master equation~(\ref{eq:ME_g_0123_R}) is
\begin{align}
  \mathcal{L} 
& = 
  \begin{array}{c}
    \left[ 
      \begin{array}{cccc}
        1 & a^{\dagger} & {a^{\dagger}}^2 & {a^{\dagger}}^3  
      \end{array}
    \right] \\ 
    \phantom{} \\ 
    \phantom{} \\ 
    \phantom{}
  \end{array} 
  \frac{1}{
    {\epsilon}^2 + k_2^2 + k_1^2 + {\bar{k}}_1^2
  }
  \left\{ 
    \left( 
      \left( 
        {\epsilon}^2 + k_2^2
      \right) 
      \left[
        \begin{array}{r}
          1 \\
          - 1 \\
          0 \\
          0 
        \end{array}
      \right] + 
      \left( 
        k_1^2 + {\bar{k}}_1^2
      \right) 
      \left[
        \begin{array}{r}
          0 \\
          0 \\
          1 \\
          - 1 
        \end{array}
      \right] 
    \right) 
    \begin{array}{c}
      \left[ 
        \begin{array}{cccc}
          \epsilon & -k_2 & k_1 & -{\bar{k}}_1
        \end{array}
      \right] \\ 
      \phantom{} \\ 
      \phantom{} \\ 
      \phantom{}
    \end{array} + 
  \right.
\nonumber \\
& \phantom{=} 
  \left.
    \left[
      \begin{array}{r}
        1 \\
        - 1 \\
        - 1 \\
        1 
      \end{array}
    \right] 
    \left(
      \begin{array}{c}
        \left( k_1^2 + {\bar{k}}_1^2 \right) 
        \left[ 
          \begin{array}{cccc}
            \epsilon & -k_2 & 0 & 0 
          \end{array}
        \right] - 
        \left( {\epsilon}^2 + k_2^2 \right) 
        \left[ 
          \begin{array}{cccc}
            0 & 0 & k_1 & -{\bar{k}}_1 
          \end{array}
        \right] \\ 
        \phantom{} \\ 
        \phantom{}
      \end{array}
    \right) 
  \right\}
  \left[
    \begin{array}{c}
      1 \\
      a \\
      a^2 \\
      a^3
    \end{array}
  \right] 
\nonumber \\
\label{eq:Lv_one_species_bistable}
\end{align}
in which the top line is the $s$-flow and the bottom line is the
$\delta$-flow.  

\subsubsection{The 2-species, cross-catalytic model of 
Sec.~\ref{sec:two_species_bistable}} 
\label{sec:two_species_bistable_app}

For the 2-species model of Fig.~\ref{fig:two_species_bistable}, the
vector of creation and annihilation operators with respect to which we
will write $\mathcal{L}$ in matrix form is:
\begin{align}
  {\psi}^{\dagger} 
& = 
  \left[
    \begin{array}{cccccc}
      1 & a^{\dagger} & b^{\dagger} & 
      a^{\dagger} b^{\dagger} & 
      {a^{\dagger}}^2 b^{\dagger} & 
      a^{\dagger} {b^{\dagger}}^2 
    \end{array}
  \right]
\nonumber \\
  {\left( \psi \right)}^T
& = 
  \left[
    \begin{array}{cccccc}
      1 & a & b & ab & a^2 b & ab^2 
    \end{array}
  \right]
\label{eq:psi_vecs_two_species_bistable}
\end{align}

Then $\mathcal{L}$ from Eq.~(\ref{eq:L_two_species_bistable}),
corresponding to the master
equation~(\ref{eq:ME_two_species_bistable}), in matrix form and also
diagonalized, becomes
\begin{align}
  \mathcal{L}
& = 
  {\psi}^{\dagger}
  \left\{ 
    \frac{1}{
      2 
      \left( 
        2 {\epsilon}^2 + k_2^2 + 2 k_1^2 + {\bar{k}}_1^2
      \right) 
    }
    \left( 
      \left( 
        2 {\epsilon}^2 + k_2^2
      \right) 
      \left[
        \begin{array}{r}
          2 \\
          - 1 \\
          - 1 \\
          0 \\
          0 \\
          0 
        \end{array}
      \right] + 
      \left( 
        2 k_1^2 + {\bar{k}}_1^2
      \right) 
      \left[
        \begin{array}{r}
          0 \\
          0 \\
          0 \\
          2 \\
          - 1 \\
          - 1 
        \end{array}
      \right] 
    \right) 
    \begin{array}{c}
      \left[ 
        \begin{array}{cccccc}
          2 \epsilon & -k_2 & -k_2 & 
          2 k_1 & -{\bar{k}}_1 & -{\bar{k}}_1 
        \end{array}
      \right] \\ 
      \phantom{} \\ 
      \phantom{} \\ 
      \phantom{} \\ 
      \phantom{} \\ 
      \phantom{}
    \end{array} 
  \right.
\nonumber \\ 
& \phantom{=}
  \mbox{} + 
  \left.
    \frac{1}{
      2 
      \left( 
        k_2^2 + {\bar{k}}_1^2
      \right) 
    }
    \left( 
      k_2^2
      \left[
        \begin{array}{r}
          0 \\
          - 1 \\
          1 \\
          0 \\
          0 \\
          0 
        \end{array}
      \right] + 
      {\bar{k}}_1^2
      \left[
        \begin{array}{r}
          0 \\
          0 \\
          0 \\
          0 \\
          - 1 \\
          1 
        \end{array}
      \right] 
    \right) 
    \begin{array}{c}
      \left[ 
        \begin{array}{cccccc}
          0 & -k_2 & k_2 & 
          0 & -{\bar{k}}_1 & {\bar{k}}_1 
        \end{array}
      \right] \\ 
      \phantom{} \\ 
      \phantom{} \\ 
      \phantom{} \\ 
      \phantom{} \\ 
      \phantom{}
    \end{array} 
  \right.
\nonumber \\ 
& \phantom{=}
  \mbox{} + 
  \left.
    \frac{1}{
      2 
      \left( 
        2 {\epsilon}^2 + k_2^2 + 2 k_1^2 + {\bar{k}}_1^2
      \right) 
    }
    \left[
      \begin{array}{r}
        2 \\
        - 1 \\
        - 1 \\
        - 2 \\
        1 \\
        1 
      \end{array}
    \right] 
    \begin{array}{c}
      \left( 
        2 k_1^2 + {\bar{k}}_1^2
      \right) 
      \left[ 
        \begin{array}{cccccc}
          2 \epsilon & -k_2 & -k_2 & 0 & 0 & 0
        \end{array}
      \right] - 
      \left( 
        2 {\epsilon}^2 + k_2^2
      \right) 
      \left[ 
        \begin{array}{cccccc}
          0 & 0 & 0 & 2 k_1 & -{\bar{k}}_1 & -{\bar{k}}_1 
        \end{array}
      \right] \\ 
      \phantom{} \\ 
      \phantom{} \\ 
      \phantom{} \\ 
      \phantom{} \\ 
      \phantom{}
    \end{array} 
  \right.
\nonumber \\ 
& \phantom{=}
  \mbox{} + 
  \left.
    \frac{
      k_2 {\bar{k}}_1
    }{
      2 
      \left( 
        k_2^2 + {\bar{k}}_1^2
      \right) 
    }
    \left[
      \begin{array}{r}
        0 \\
        - 1 \\
        1 \\
        0 \\
        1 \\
        - 1 
      \end{array}
    \right] 
    \begin{array}{c}
      \left[ 
        \begin{array}{cccccc}
          0 & -{\bar{k}}_1 & {\bar{k}}_1 & 
          0 & k_2 & -k_2 
        \end{array}
      \right] \\ 
      \phantom{} \\ 
      \phantom{} \\ 
      \phantom{} \\ 
      \phantom{} \\ 
      \phantom{}
    \end{array} 
  \right\} 
  \psi 
\label{eq:L_stoich_two_species_bistable}
\end{align}
The top line is the $s$-flow projected out by $\left( a^{\dagger} a +
b^{\dagger} b \right)$; the second line is the $s$-flow projected out
by $\left( a^{\dagger} a - b^{\dagger} b \right)$.  The next two lines
are the $\delta$-flows with the same symmetry or antisymmetry. 

The stoichiometric vectors appearing in
Eq.~(\ref{eq:Glauber_moment_fact_hats_alt}), and their products with
$\mathbb{A}$ are
\begin{align}
  Y_a \equiv 
  Y_a^{\underline{1}}
& = 
  \left[
    \begin{array}{cccccc}
      0 & 1 & 0 & 1 & 2 & 1 
    \end{array}
  \right]
& 
  Y_a^{\underline{1}} \mathbb{A}
& = 
  \left[
    \begin{array}{cccccc}
      \epsilon & -k_2 & 0 & 
      k_1 & -{\bar{k}}_1 & 0
    \end{array}
  \right]
\nonumber \\
  Y_b \equiv 
  Y_b^{\underline{1}} 
& = 
  \left[
    \begin{array}{cccccc}
      0 & 0 & 1 & 1 & 1 & 2 
    \end{array}
  \right] 
& 
  Y_b^{\underline{1}} \mathbb{A}
& = 
  \left[
    \begin{array}{cccccc}
      \epsilon & 0 & -k_2 & 
      k_1 & 0 & -{\bar{k}}_1 
    \end{array}
  \right] 
\nonumber \\
  Y_a \cdot Y_b 
& = 
  \left[
    \begin{array}{cccccc}
      0 & 0 & 0 & 1 & 2 & 2 
    \end{array}
  \right] 
& 
  \left( Y_a \cdot Y_b \right) \mathbb{A}
& = 
  \left[
    \begin{array}{cccccc}
      0 & 0 & 0 & 
      2 k_1 & -{\bar{k}}_1 & -{\bar{k}}_1 
    \end{array}
  \right] 
\nonumber \\
  Y_a^{\underline{2}}
& = 
  \left[
    \begin{array}{cccccc}
      0 & 0 & 0 & 0 & 2 & 0 
    \end{array}
  \right] = 
  Y_a^{\underline{2}} \cdot Y_b 
& 
  Y_a^{\underline{2}} \mathbb{A}
& = 
  2 
  \left[
    \begin{array}{cccccc}
      0 & 0 & 0 & 
      k_1 & -{\bar{k}}_1 & 0
    \end{array}
  \right] = 
  \left( Y_a^{\underline{2}} \cdot Y_b \right) \mathbb{A}
\nonumber \\
  Y_b^{\underline{2}}
& = 
  \left[
    \begin{array}{cccccc}
      0 & 0 & 0 & 0 & 0 & 2 
    \end{array}
  \right] = 
  Y_b^{\underline{2}} \cdot Y_a 
& 
  Y_b^{\underline{2}} \mathbb{A}
& = 
  2 
  \left[
    \begin{array}{cccccc}
      0 & 0 & 0 & 
      k_1 & 0 & -{\bar{k}}_1 
    \end{array}
  \right] = 
  \left( Y_b^{\underline{2}} \cdot Y_a \right) \mathbb{A}
\label{eq:proj_vecs_two_species_bistable}
\end{align}
\end{widetext}

\section{Polynomial expansion for solutions to the two-species CRN
of Sec.~\ref{sec:two_species_bistable}}
\label{sec:two_spec_bistable_polyn_app}

This appendix describes the combination of recursive solution, and
successive approximation, used to solve the hierarchical
expansion~(\ref{eq:varphi_q2_expn}) for the lattice of moments in the
two-species model of Sec.~\ref{sec:two_spec_polyn_exp}.  

Here, to simplify notation and improve readability, we will regard
matrices such as $\hat{\Lambda}$ as operators that shift the indices
$\left( k_a, k_b \right)$ in terms ${\varphi}_k^{\left(
\alpha \right)} q^{2 \alpha}$ by means of  discrete index-shift
operators $e^{\partial / \partial k_a}$, $e^{\partial / \partial
k_b}$, as we did for transfer matrices in Eq.~(\ref{eq:T_psi_from_A})
\textit{et seq.}  Because of the exchange symmetry in the dynamical
equations under $k_a \leftrightarrow k_b$, $\hat{\Lambda}$ acts on
${\varphi}_k^{\left( \alpha \right)}$ through a shift of the $\kappa$
value (by integers), and on $q^{2 \alpha}$ through shifts in $\alpha$.
This allows us to treat $\kappa$ as an index shifted by integers,
analogous to $k$ in single-species models.  Where we suppress
subscript $\kappa$ indices, the whole vector is intended.

With these notational conventions, the action of the generator in
Eq.~(\ref{eq:varphi_from_s0}) can be broken down into three terms:
\begin{equation}
  \hat{\Lambda}
  \left( 
    {\varphi}^{\left( \alpha \right)} 
    q^{2 \alpha} 
  \right) = 
  \left( 
    {\hat{\Lambda}}^0
    {\varphi}^{\left( \alpha \right)} 
  \right) 
  q^{2 \alpha} - 
  s^{\left( \alpha + 1 \right)} 
  q^{2 \left( \alpha + 1 \right)} + 
  \sum_{\beta = 0}^{\alpha - 1}
  {\sigma}_{\beta}^{\left( \alpha \right)}
  q^{2 \beta}
\label{eq:L_decomp_L0_s_sig}
\end{equation}
Here ${\hat{\Lambda}}^0$ is a diagonal operator (in $\alpha$) acting
only on ${\varphi}^{\left( \alpha \right)}$, which takes the form
(refer to Fig.~\ref{fig:two_spec_L_diagram}):
\begin{widetext}
\begin{align}
  {
    \left( 
      {\hat{\Lambda}}^{0} {\varphi}^{\left( \alpha \right)}
    \right)
  }_{\kappa} 
& = 
  \left[ 
    \frac{
      \left( {\kappa}^2 - {\theta}_{\alpha > 0} \right)
      \left( \kappa - 2 - 2 \alpha \right) - 
      2 \alpha {\theta}_{\alpha > 0}
    }{
      4 K_1
    } + 
    \frac{\omega}{K_1}
    \left( \kappa - 2 \alpha \right)
  \right] 
  \left( 
    {\varphi}^{\left( \alpha \right)}_{\kappa - 1} - 
    {\varphi}^{\left( \alpha \right)}_{\kappa}
  \right) 
\nonumber \\
& 
  \quad \mbox{} - 
  \left[ 
    \kappa 
    \left( \frac{\eta}{K_1} + 2 \alpha \right) + 
    2 \alpha 
    \left( \frac{\omega}{K_1} - 2 \alpha \right) + 
    \frac{2 \alpha {\kappa}^2}{4K_1} + 
    {\theta}_{\alpha > 0}
    \frac{\kappa - 2}{4 K_1}
  \right]
  {\varphi}^{\left( \alpha \right)}_{\kappa} 
\nonumber \\
& 
  \quad \mbox{} + 
  \left[ 
    \left( \kappa - 2 \alpha \right) 
    \left( \kappa - 1 \right) 
  \right]
  \left( 
    {\varphi}^{\left( \alpha \right)}_{\kappa} - 
    {\varphi}^{\left( \alpha \right)}_{\kappa + 1}
  \right) + 
  \kappa K_1 
  \left( 
    {\varphi}^{\left( \alpha \right)}_{\kappa + 1} - 
    {\varphi}^{\left( \alpha \right)}_{\kappa + 2}
  \right) - 
  2 \alpha K_1 
  {\varphi}^{\left( \alpha \right)}_{\kappa + 1} 
\nonumber \\
\label{eq:L0_indexed_form}
\end{align}
In the same way as the action of $\hat{\Lambda}$ on the constant
background $\underline{1}$ produced the zero-th order source $- \eta
s^{\left( 0 \right)}$ in Eq.~(\ref{eq:varphi_from_s0}), the action on
each order ${\varphi}^{\left( \alpha \right)}$ generates a source term
$s^{\left( \alpha + 1 \right)}$ in Eq.~(\ref{eq:L_decomp_L0_s_sig})
that propagates cross-terms one order upward in $\alpha$, defined by
\begin{equation}
  s^{\left( \alpha + 1 \right)}_{\kappa} = 
  \frac{1}{4 K_1}
  \left[
    \left( \kappa - 2 \right) 
    \left( 
      {\varphi}^{\left( \alpha \right)}_{\kappa - 1} - 
      {\varphi}^{\left( \alpha \right)}_{\kappa} 
    \right) - 
    2 \alpha 
    {\varphi}^{\left( \alpha \right)}_{\kappa - 1} 
  \right] 
\label{eq:s_aplus1_form}
\end{equation}
In addition to an upward-propagating ``source'' term,
Eq.~(\ref{eq:L_decomp_L0_s_sig}) contains ``feedback'' terms, which
propagate cross-terms downward from order $q^{2\alpha}$ to all lesser
orders $q^{2 \beta}$ with $\beta < \alpha$, given by
\begin{align}
  {\sigma}_{\beta}^{\left( \alpha \right)} 
& = 
  \left[
    \kappa 
    \left(
      \begin{array}{c}
        2 \alpha \\
        2 \beta 
      \end{array}
    \right) - 
    \left(
      \begin{array}{c}
        2 \alpha \\
        2 \beta - 1 
      \end{array}
    \right) 
  \right]
  \left( 
    \frac{\omega}{K_1}
    {\varphi}^{\left( \alpha \right)}_{\kappa - 1} + 
    \left[ 
      K_1 - 
      \left( \kappa - 1 \right) 
    \right] 
    {\varphi}^{\left( \alpha \right)}_{\kappa + 1}     
  \right) 
\nonumber \\
& 
  \mbox{} \quad + 
  2^{2 \left( \alpha - \beta \right)}
  \left[
    \frac{
      \kappa \left( \kappa - 2 \right)
    }{
      2 
    } 
    \left(
      \begin{array}{c}
        2 \alpha \\
        2 \beta 
      \end{array}
    \right) - 
    2 \left( \kappa - 1 \right) 
    \left(
      \begin{array}{c}
        2 \alpha \\
        2 \beta - 1 
      \end{array}
    \right) + 
    2 
    \left(
      \begin{array}{c}
        2 \alpha \\
        2 \beta - 2 
      \end{array}
    \right) 
  \right]
  {\varphi}^{\left( \alpha \right)}_{\kappa}
\nonumber \\
& 
  \mbox{} + 
  \frac{1}{4 K_1}
  \left\{
    {\kappa}^2 
    \left[ 
      \left( \kappa - 2 \right) 
      \left(
        \begin{array}{c}
          2 \alpha \\
          2 \beta  
        \end{array}
      \right) - 
      \left(
        \begin{array}{c}
          2 \alpha \\
          2 \beta - 1 
        \end{array}
      \right) 
    \right] - 
    \left[ 
      \left( \kappa - 2 \right) 
      \left(
        \begin{array}{c}
          2 \alpha \\
          2 \beta - 2 
        \end{array}
      \right) - 
      \left(
        \begin{array}{c}
          2 \alpha \\
          2 \beta - 3 
        \end{array}
      \right) 
    \right] 
  \right\}
  {\varphi}^{\left( \alpha \right)}_{\kappa - 1}
\label{eq:feedbacks_genform}
\end{align}
\end{widetext}
In Eq.~(\ref{eq:feedbacks_genform}), referring to the graphical form
of Fig.~\ref{fig:two_spec_L_diagram}, all terms from $j_a + j_b = 1$
are grouped in the first line, all terms from $j_a + j_b = 2$ are
grouped in the second line, and all terms from $j_a + j_b = 3$ are
grouped in the third line.

\subsection{Solution by upward propagation in powers of $q^2$ and
perturbative correction in $1/\kappa$} 
\label{sec:upward_prop_sol}

The decomposition~(\ref{eq:L_decomp_L0_s_sig}) of the steady-state
condition can now be solved by alternating steps of exact cancellation
of source terms at ascending orders in $\alpha$, and successive
approximation to cancel feedback terms which takes the form of a
perturbative expansion in $1 / \kappa$.

Upward-propagation consists of solving a series of Laplacian equations
in the operator ${\hat{\Lambda}}^0$ in terms of the
sources~(\ref{eq:s_aplus1_form}), as 
\begin{equation}
  {\hat{\Lambda}}^{0} {\varphi}^{\left( \alpha \right)} = 
  s^{\left( \alpha \right)} 
\label{eq:L_forward_prop}
\end{equation}
for all $\alpha \ge 0$. 

Examination of the scaling terms in $\kappa$ in
Eq.~(\ref{eq:L0_indexed_form}) suggests that a bounded large-$\kappa$
asymptotic approximation for each order is 
\begin{equation}
  {\varphi}^{\left( \alpha \right)}_{\kappa} = 
  \frac{
    4 
  }{
    {
      \left( \kappa +  4 \left( K_1 - \eta \right) / 3 \right)
    }^{2 \alpha + 1}
  } + 
  \mathcal{O} \! 
  \left( 
    \frac{1}{{\kappa}^{2\alpha + 2}}
  \right) . 
\label{eq:init_phi_asymptotics}
\end{equation}
Solution of Eq.~(\ref{eq:L_forward_prop}) at each order $\alpha$ by
one-dimensional matched asymptotic expansion proceeds as for the
1-species models.  At order $\alpha = 0$, the solution can be stably
extended down to $\kappa = 0$, but for $\alpha > 1$ the asymptotic
approximation~(\ref{eq:init_phi_asymptotics}) is not sufficient to
produce convergence below $\kappa \sim 20$, and a non-trivial matched
expansion is needed to produce valid higher-order corrections in $q^2$
to the low-order moments.

One round of forward propagation (through all orders $\alpha$) will
produce a solution that does not satisfy $\hat{\Lambda}
\sum_{\alpha = 0}^{\infty} {\varphi}^{\left( \alpha \right)} q^{2
\alpha} - s^{\left( 0 \right)} = 0$, but rather 
\begin{equation}
  \hat{\Lambda}
  \sum_{\alpha = 0}^{\infty}
  {\varphi}^{\left( \alpha \right)}
  q^{2 \alpha} - 
  s^{\left( 0 \right)} = 
  \sum_{\alpha = 0}^{\infty}
  \sum_{\beta = 0}^{\alpha - 1}
  {\sigma}_{\beta}^{\left( \alpha \right)}
  q^{2 \beta}
\label{eq:varphi_all_residues}
\end{equation}
The full solution can be approached perturbatively by using the
feedback terms on the right-hand side of
Eq.~(\ref{eq:varphi_all_residues}) as sources for an iterative
correction to the original $\varphi$.

\subsubsection*{A successive-approximation approach to full solutions
${\varphi}^{\left( \alpha \right)}$}

A scaling analysis following from Eq.~(\ref{eq:init_phi_asymptotics})
suggests that the correction terms needed to cancel the residuals in
Eq.~(\ref{eq:varphi_all_residues}) are suppressed by powers of $1 /
\kappa$, and thus that a method of successive approximations should
converge.  

No feedback term ${\sigma}_{0}^{\left( \alpha \right)}$ to $q^0$ order
from $\alpha = 0$ exists, because $\beta \le \alpha - 1$ in the
sum~(\ref{eq:L_decomp_L0_s_sig}).  For all $\alpha
\ge 1$, ${\sigma}_{0}^{\left( \alpha \right)}$ is given by
\begin{align}
  {
    \left( 
      {\sigma}_{0}^{\left( \alpha \right)} 
    \right) 
  }_{\kappa}
& = 
  \left( 
    \frac{\omega \kappa}{K_1} + 
    \frac{{\kappa}^2 \left( \kappa -2 \right)}{4 K_1} 
  \right) 
  {\varphi}^{\left( \alpha \right)}_{\kappa -1} 
\nonumber \\ 
& 
  \quad \mbox{} + 
  \kappa 
  \left[ 
    K_1 - 
    \left( \kappa -1 \right) 
  \right] 
  {\varphi}^{\left( \alpha \right)}_{\kappa +1} + 
  2^{2\alpha} 
  \frac{
    \kappa \left( \kappa -2 \right)
  }{
    2 
  }
  {\varphi}^{\left( \alpha \right)}_{\kappa} . 
\label{eq:sig0_from_varphi}
\end{align}
Since the lowest-order contribution is from ${\varphi}^{\left( 1
\right)}_{\kappa} \sim 4 / {\kappa}^3$, it follows that
${\sigma}_{0}^{\left( \alpha \right)}$ is no larger for any $\alpha$
than the leading term
\begin{equation}
  {
    \left( 
      {\sigma}_{0}^{\left( 1 \right)} 
    \right) 
  }_{\kappa} = 
  \frac{1}{K_1} + 
  \mathcal{O} \! 
  \left( 
    \frac{K_1}{{\kappa}^2}
  \right) + 
  \mathcal{O} \! 
  \left( 
    \frac{1}{\kappa}
  \right) . 
\label{eq:init_phi_asymp_0}
\end{equation}
\emph{one order lower} (in either $K_1$ or $\kappa$, according to the
range of $\kappa$) than the scaling of $s^{\left( 0 \right)}$ in
Eq.~(\ref{eq:two_spec_s0_def}).  The same argument extends to higher
$\beta$ in Eq.~(\ref{eq:varphi_all_residues}); the first term that
contributes at each order scales with two additional powers of
$1/\kappa$, and so is smaller than the corresponding $s^{\left( \alpha
\right)}$ term in the first iteration of Eq.~(\ref{eq:s_aplus1_form}).

We therefore introduce a second-order correction term
${\varphi}^{\left( \alpha \right) \prime}$, satisfying
\begin{equation}
  \hat{\Lambda}
  \sum_{\alpha = 0}^{\infty}
  {\varphi}^{\left( \alpha \right) \prime}
  q^{2 \alpha} = 
  - \sum_{\alpha = 0}^{\infty}
  \sum_{\beta = 0}^{\alpha - 1}
  {\sigma}_{\beta}^{\left( \alpha \right)}
  q^{2 \beta} + 
  \sum_{\alpha = 0}^{\infty}
  \sum_{\beta = 0}^{\alpha - 1}
  {\sigma}_{\beta}^{\left( \alpha \right) \prime}
  q^{2 \beta} , 
\label{eq:varphi_from_sig_alpha}
\end{equation}
where ${\varphi}^{\left( \alpha \right) \prime}$ is solved by upward
propagation in $q^{2 \alpha}$, as in Eq.~(\ref{eq:L_forward_prop}),
but now with an entire tower of sources $-\sum_{\alpha = \beta +
1}^{\infty}{\sigma}^{\left( \alpha \right)}_{\beta}$ at each order
$q^{2 \beta}$, rather than just the zeroth-order source $s^{\left( 0
\right)}$ that was used for ${\varphi}^{\left( \alpha \right)}$, and 
leaving its own residues ${\sigma}^{\left( \alpha \right) \prime}$.
After infinitely many iterations, the residue terms go to zero as a
sequence in powers of $1/\kappa$, and if the sequence converges, the
sum over corrections will be a solution to the original steady-state
condition~(\ref{eq:varphi_from_s0}).  

An equivalent expression for the closed solution (summing over all
orders of perturbative correction), expressed in terms of the
homogeneous operator ${\hat{\Lambda}}^0$, is
\begin{equation}
  \sum_{\alpha = 0}^{\infty}
  \left( 
    {\hat{\Lambda}}^0
    {\varphi}^{\left( \alpha \right)} - 
    s^{\left( \alpha \right)} 
  \right)
  q^{2 \alpha} = 
  - \sum_{\alpha = 0}^{\infty}
  \sum_{\beta = 0}^{\alpha - 1}
  {\sigma}_{\beta}^{\left( \alpha \right)}
  q^{2 \beta} . 
\label{eq:varphi_self_consist}
\end{equation}
Eq.~(\ref{eq:varphi_self_consist}) is similar in form to a
Schwinger-Dyson equation for Green's function solution, in which
$\left( {\hat{\Lambda}}^0 {\varphi}^{\left( \alpha \right)} -
s^{\left( \alpha \right)} \right)$ serves as a ``bare'' Green's
function, which defines a basis for perturbative incorporation of an
``interaction'' term $\sum_{\alpha = \beta + 1}^{\infty}
{\sigma}_{\beta}^{\left( \alpha \right)}$.

\subsection{Numerical evaluations}

We have implemented the above solution method for a
series~(\ref{eq:varphi_q2_expn}) truncated at five successive orders
of approximation ${\varphi}^{\left( {\alpha}_{\rm max}
\right)}$ for ${\alpha}_{\rm max} = 0 , \ldots, 4$.  Under perfect
convergence of the perturbative
recurrence~(\ref{eq:varphi_from_sig_alpha}), the error measure
$\partial {\hat{\Phi}}_{\left( k_a , k_b \right)} / \partial \tau$
would cancel to order $q^{2 \left( {\alpha}_{\rm max} + 1 \right)}$
around the diagonal (order $q^{10}$ for the highest-order
approximation we compute).  We obtain cancellation along the diagonal
to machine precision for the lowest-order correction
${\varphi}^{\left( 0 \right)}$, but instability of the downward-going
asymptotic expansion at higher orders degrades both the accuracy of
${\hat{\Phi}}_{\left( k_a , k_b \right)}$ on the diagonal for $k_a +
k_b \lesssim 20$, and convergence toward the $q^{2 \left(
{\alpha}_{\rm max} + 1 \right)}$ residual.  

\begin{figure}[ht]
  \begin{center} 
  \includegraphics[scale=0.4]{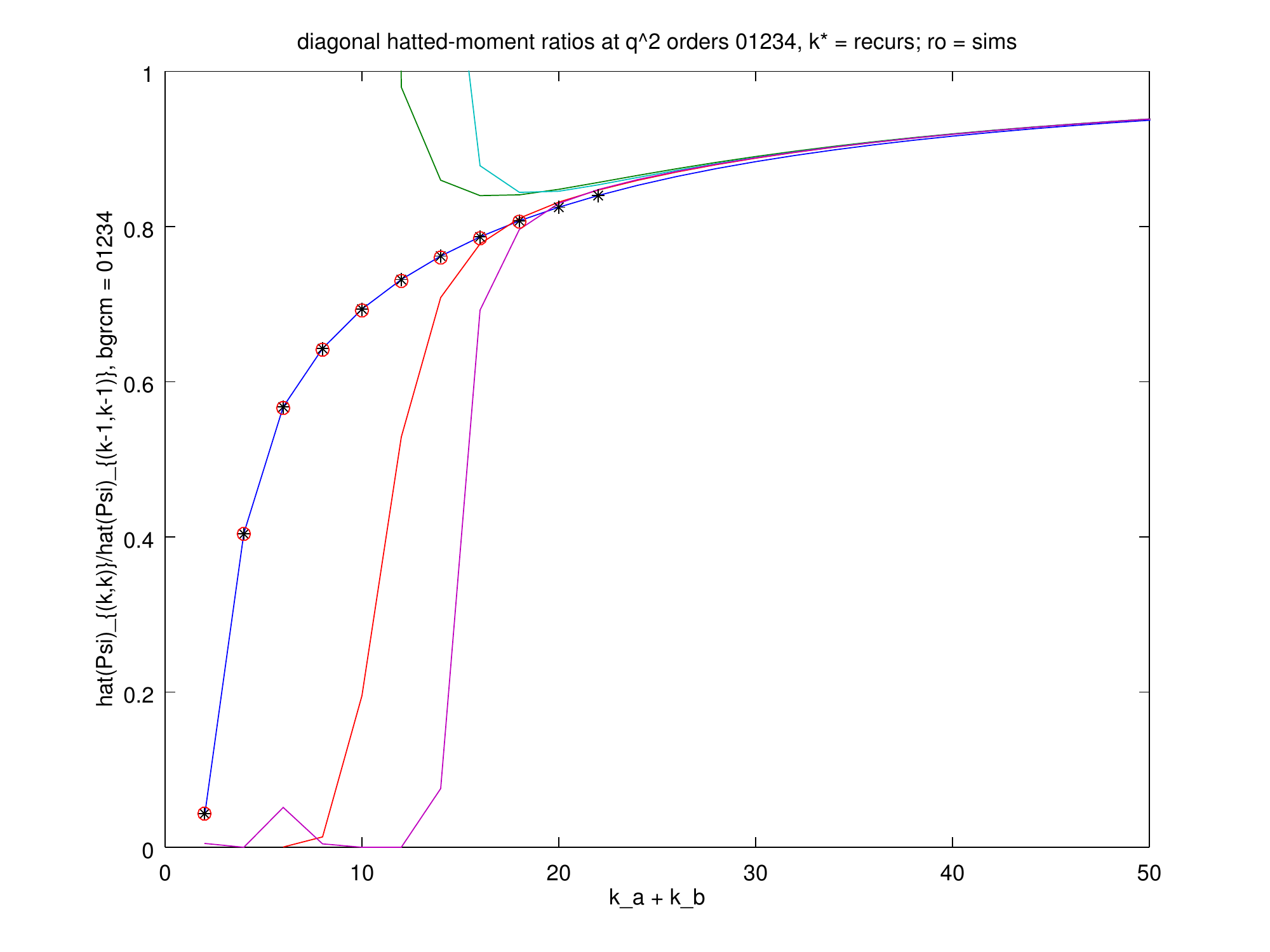} 
  \caption{
  Ratios of moments ${\hat{\Phi}}_{\left( k_a , k_b \right)} /
  {\hat{\Phi}}_{\left( k_a -1 , k_b -1 \right)}$ plotted versus $k_a +
  k_b \equiv k$, along the diagonal $k_a = k_b$.  The
  series~(\ref{eq:varphi_q2_expn}) is truncated at five successive
  orders of approximation ${\varphi}^{\left( {\alpha}_{\rm max}
  \right)}$ for ${\alpha}_{\rm max} = 0 , \ldots, 4$ (color sequence
  bgrcm).  Symbols are the ratios obtained directly from sampled
  moments of a stationary Gillespie simulation, showing that the order
  ${\alpha}_{\rm max} = 0$ provides a good approximation to the
  diagonal moments, as shown in Fig.~\ref{fig:mom_rats_dense_q0only}
  of the main text.  Divergence of higher-order terms -- which occurs
  with opposite sign for odd versus even ${\alpha}_{\rm max}$ --
  reflects instability of the asymptotic expansion downward from large
  $\kappa$.  \textsc{So far I have not implemented a matched
  asymptotic expansion for these terms.}
  \label{fig:mom_rats_diag_01234} 
  } 
  \end{center}
\end{figure}

Fig.~\ref{fig:mom_rats_diag_01234} shows the successive approximations
to the ratio ${\hat{\Phi}}_{\left( k_a , k_b \right)} /
{\hat{\Phi}}_{\left( k_a -1 , k_b -1 \right)}$ along the diagonal $k_a
= k_b$, compared to values obtained from a Gillespie simulation.  The
approximation ${\alpha}_{\rm max} = 0$ already shows good agreement
with simulations.  Divergence of the downward-going asymptotic
expansion for higher-order terms begins around $k_a = k_b \approx 9$,
which is the upper stable solution for the coherent-state mean
number~(\ref{eq:0123_y1A_vanishing}), as predicted in the scaling
analysis of Sec.~\ref{sec:match_AE}.

\begin{figure}[ht]
  \begin{center} 
  \includegraphics[scale=0.4]{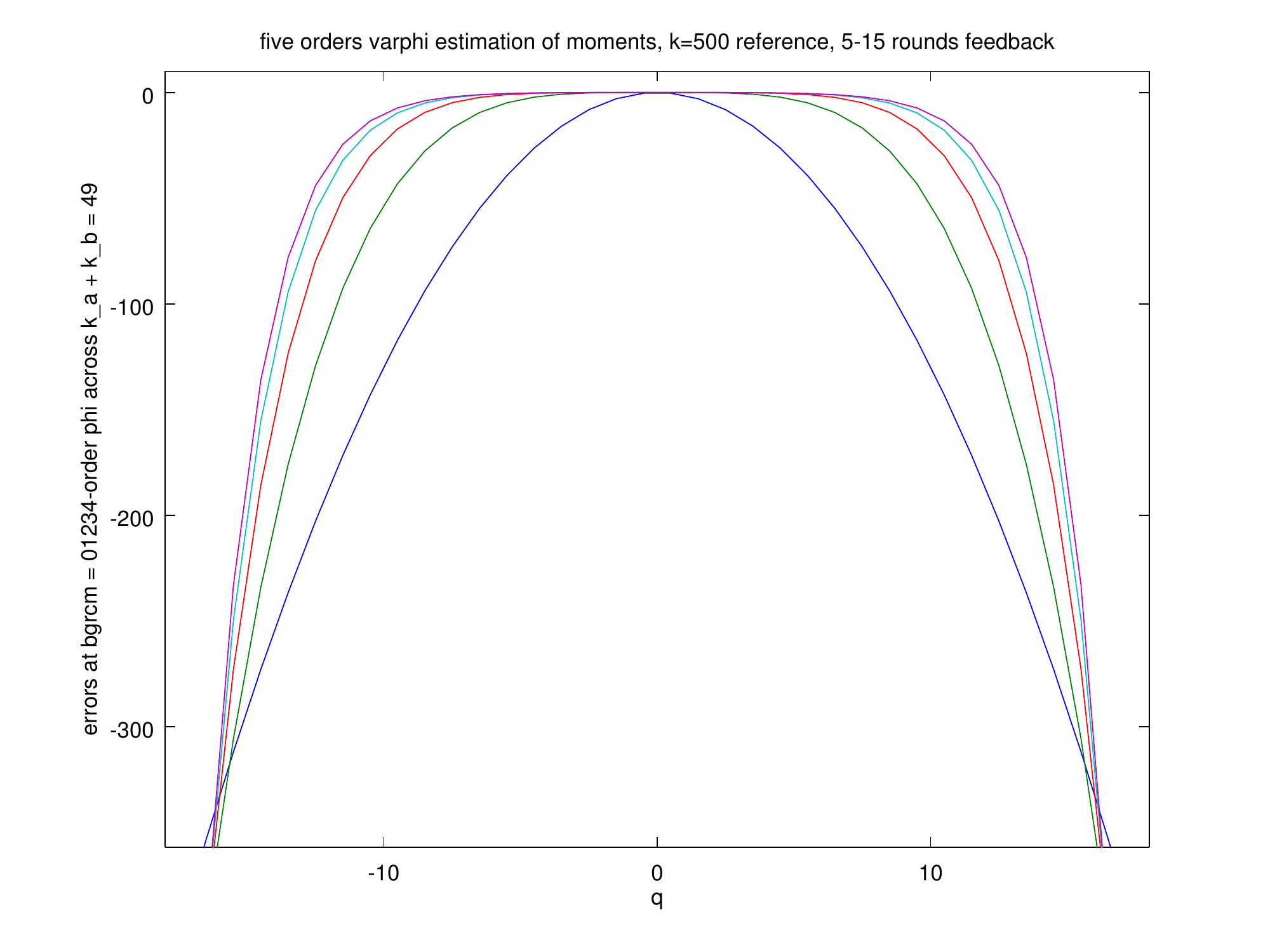} 
  \caption{
  Graph of the residual error measure $\partial {\hat{\Phi}}_{\left(
  k_a , k_b \right)} / \partial \tau$ across the contour $k_a + k_b =
  49$, with the series~(\ref{eq:varphi_q2_expn}) truncated at five
  successive orders of approximation ${\varphi}^{\left( {\alpha}_{\rm
  max} \right)}$ for ${\alpha}_{\rm max} = 0 , \ldots, 4$ (color
  sequence bgrcm).  Under perfect cancellation, the residual error at
  each order ${\alpha}_{\rm max}$ would scale as $q^{2 \left(
  {\alpha}_{\rm max} + 1 \right)}$.  This scaling is very closely
  approximated for ${\alpha}_{\rm max} = 1$, and degrades due to
  imperfect control of asymptotic expansions at higher orders, though
  the approximate behavior is attained.  Crossing of the error curves
  suggests a finite radius of convergence in $q^2$ of the
  series~(\ref{eq:varphi_q2_expn}), for $\left| q \right| / \kappa
  \sim 0.38$.  The asymptotic expansion remains this good or better
  for all larger $\kappa$.
  \label{fig:two_spec_01234_ord_varphi} 
  } 
  \end{center}
\end{figure}

Fig.~\ref{fig:two_spec_01234_ord_varphi} shows the error measure
$\partial {\hat{\Phi}}_{\left( k_a , k_b \right)} / \partial \tau$
(the deviation from a full steady-state condition) across the
anti-diagonal contour $k_a + k_b = 49$, testing the quality of the
convergence of the ${\varphi}^{\left( {\alpha}_{\rm max} \right)}$
approximation.  This contour is in a range $k_a , k_b \gg 9$ where the
asymptotic expansions are still fairly well-controlled.  The
qualitative character of the convergence is well approximated along
the diagonal, but exact cancellation to order $q^{2 \left(
{\alpha}_{\rm max} + 1 \right)}$ degrades at higher ${\alpha}_{\rm
max}$, as both the asymptotic expansion and the recursive
solution~(\ref{eq:varphi_from_sig_alpha}) accumulate numerical errors.
The figure also shows the crossing of error contours at $\left| q
\right| / \kappa \sim 0.38$, suggesting a finite radius of convergence
that does not cover the entire $\left( k_a , k_b
\right)$ lattice.

\begin{figure}[ht!]
  \begin{center} 
  \includegraphics[scale=0.45]{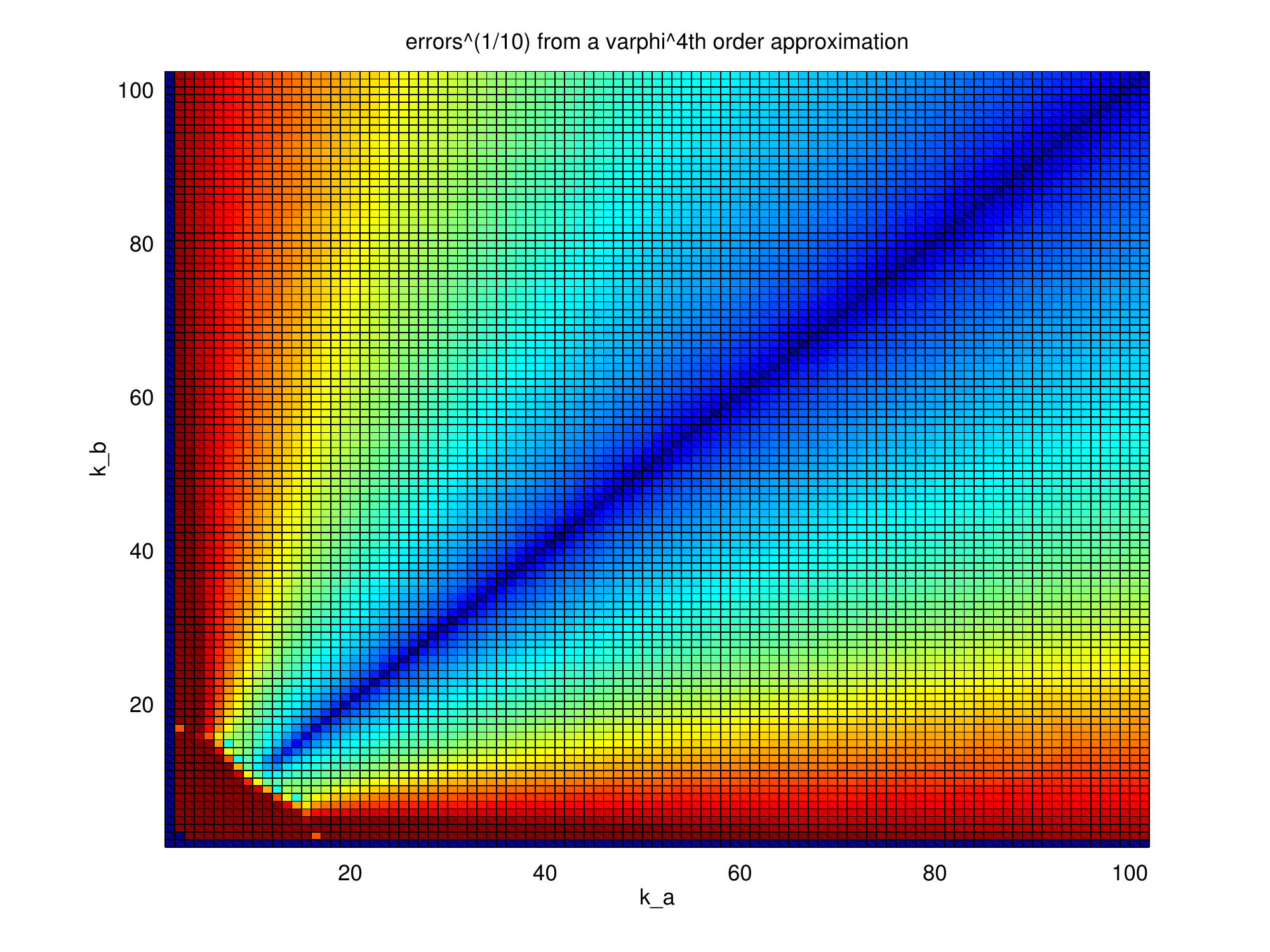} 
  \caption{
  Linear colormap of the error function $\partial {\hat{\Phi}}_{\left(
  k_a , k_b \right)} / \partial \tau$ for ${\alpha}_{\rm max} = 5$,
  evaluated as the left-hand side of Eq.~(\ref{eq:two_spec_ss_cond_L})
  which would equal zero for a stationary distribution.  The deviation
  from zero is raised to the 0.1 power, to produce a linear
  cross-section if the true residuals scale as $q^{10}$.  
  \label{fig:two_spec_error_map_varphi4} 
  } 
  \end{center}
\end{figure}

Fig.~\ref{fig:two_spec_error_map_varphi4} is a contour plot of the
errors $\partial {\hat{\Phi}}_{\left( k_a , k_b \right)} / \partial
\tau$ for ${\alpha}_{\rm max} = 5$.  The anti-diagonal cross section
corresponds to the outermost curve from
Fig.~\ref{fig:two_spec_01234_ord_varphi}, raised to the 0.1 power
appropriate if the total error scales as $q^{10}$.  The region $k_a +
k_b \lesssim 20$ shows the divergence of the asymptotic expansions
already noted in Fig.~\ref{fig:mom_rats_diag_01234}.  In the region
where the asymptotic expansions converge, the errors are roughly
constant along contours of fixed $\left| q \right| /
\kappa$ for large $k$.

\vfill 
\eject 


\end{document}